%% file: langreth_rules.tex
\documentclass[12pt]{iopart}
\usepackage[utf8]{inputenc}

\usepackage{graphicx}
\usepackage{todonotes}
\expandafter\let\csname equation*\endcsname\relax
\expandafter\let\csname endequation*\endcsname\relax

\usepackage{amsmath}
\usepackage{amssymb}
\usepackage{amsthm}
\usepackage{hyperref}
\usepackage{booktabs}
\usepackage{mathrsfs}

\usepackage{color}
\usepackage{perpage} 
\MakePerPage{footnote} 

\graphicspath{{figures/}}

\input{newcommands.tex}

\begin{document}
\date=\today

\title{Contour calculus for many-particle functions}

\author{M. J. Hyrk\"as, D. Karlsson, and R. van Leeuwen}
\address{Department of Physics,
Nanoscience Center P.O.Box 35 FI-40014 University of Jyv\"{a}skyl\"{a}, Finland}
\ead{markku.hyrkas@jyu.fi}
\begin{abstract}
In non-equilibrium many-body perturbation theory, Langreth rules are an efficient way to extract real-time equations from contour ones. However, the standard rules are not applicable in cases that do not reduce to simple convolutions and multiplications. We introduce a procedure for extracting real-time equations from general multi-argument contour functions with an arbitrary number of arguments. This is done for both the standard Keldysh contour, as well as the extended contour with a vertical track that allows for general initial states. This amounts to the generalization of the standard Langreth rules to much more general situations. These rules involve multi-argument retarded functions as key ingredients, for which we derive intuitive graphical rules. We apply our diagrammatic recipe to derive Langreth rules for the so-called double triangle structure and the general vertex function, relevant for the study of vertex corrections beyond the $GW$ approximation.
%
%
\end{abstract}
\submitto{\jpa}
\maketitle

\section{Introduction}
\label{sec:1_introduction}

Many-body perturbation theory (MBPT) is an invaluable asset when studying complex multi-particle phenomena out of equilibrium and at finite temperatures. A crucial step in the development of modern MBPT was its formulation in the language of non-equilibrium Green's functions~\cite{Danielewicz1984,Stefanucci2013}. Within this formulation, quantities are defined on a directed time contour which allows one to derive diagrammatic perturbation theory at finite temperature and out of equilibrium in precisely the same way as for the traditional zero-temperature case~\cite{Fetter2003}. Within the contour formalism, traditional formalisms such as the zero-temperature or the Matsubara formalism, follow from choosing the contour in specific ways and attaching specific time dependencies to the interactions~\cite{Stefanucci2013}.

The contour formalism, however, introduces an additional complexity by the replacement of real-time integrals with contour integrals. This makes objects, such as Green's functions and self-energies, more cumbersome to calculate numerically. An efficient tool was developed by Langreth and Wilkins~\cite{Langreth1972,Langreth1976}, who derived rules for obtaining real-time objects from contour objects. These rules are now commonly referred to as the Langreth rules~\cite{Danielewicz1984,Stefanucci2013}.

The Langreth rules, however, are not applicable when the structure of a contour equation does not reduce to convolutions and products. This situation occurs, for example, when including vertex corrections to the commonly used $GW$ approximation~\cite{Pavlyukh2016}. Furthermore, the Langreth rules are not applicable if the equation contains three-point or higher-order objects, such as the vertex function and the Bethe-Salpeter kernel in the Hedin equations~\cite{Hedin1965}.

In such cases, a direct evaluation of the contour integrals, by splitting the integrals over the various branches, results in an unwieldy amount of terms, many of which add up to yield zero contribution. Moreover, the objects involved often have no physical interpretation. This problem was studied by Danielewicz~\cite{Danielewicz1990} who introduced a way to obtain real-time components in terms of explicitly retarded or advanced objects which have a physical interpretation. Moreover, he derived rules that allow some of the vanishing terms to be discarded from the outset.

The purpose of the present paper is twofold. First, we extend the analysis of Danielewicz~\cite{Danielewicz1990} to the realm of non-equilibrium systems with general initial states, by considering the addition of an vertical time branch to the contour. Furthermore, we present alternative proofs of known results. Second, we extend the Langreth rules to general contour equations, to cover all cases of interest. An important new addition to the original Langreth rules are those for the so-called double triangle graph, appearing in the lowest order vertex corrections beyond the $GW$ approximation. These rules already found an important application in the construction of positive semi-definite spectral functions in non-equilibrium systems~\cite{Hyrkas2018}.

The paper is structured as follows. In Section \ref{sec:2_theoretical_background} we discuss the structure of commonly encountered contour equations, and briefly review the Langreth rules. In Section \ref{sec:3_keldysh_functions} we discuss general properties of contour objects. In Section \ref{sec:4_integrals_over_keldysh_functions} we give an alternative derivation of results of Danielewicz~\cite{Danielewicz1990}, providing rules to extract the real-time part of $n$-point functions. We also provide a different graphical recipe to intuitively obtain the end results. In Section \ref{sec:5_extended_contour} we extend the discussion of Section \ref{sec:4_integrals_over_keldysh_functions} for contours including the Matsubara branch, thereby generalizing the results by Danielewicz and extending our graphical recipe correspondingly. In Section \ref{sec:6_langreth_rules} we derive extended Langreth rules for general diagrams, and in Section \ref{sec:7_rules_for_double_triangle} we apply our formalism to derive Langreth rules for selected diagrams of common interest, that cannot be handled using the standard rules. We conclude in Section \ref{sec:conclusions}.

\section{Theoretical Background and Motivation}
\label{sec:2_theoretical_background}

\subsection{Time-dependent ensemble averages}
\begin{figure}
\includegraphics[width=1.0\textwidth,angle=0]{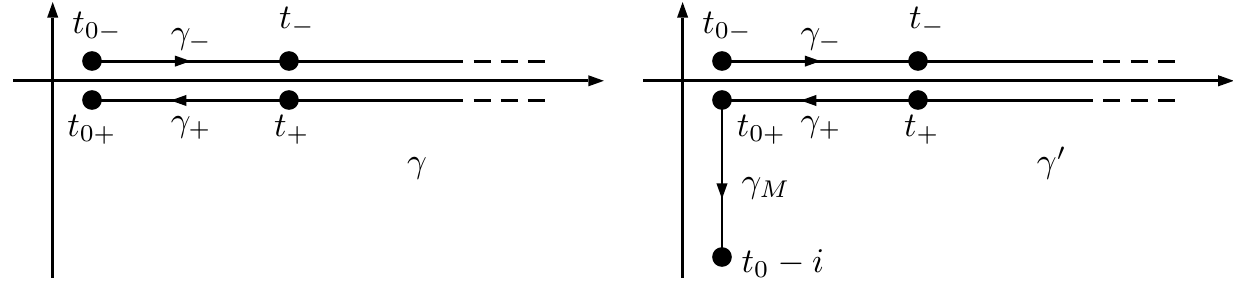}
\caption{The two different contours used in this work. (Left) The Keldysh contour $\gamma$, consisting of a forward branch $\gamma_-$ and a backward branch $\gamma_+$. $t_-(t_+)$ denotes a contour time on the forward (backward) branch at distance $t$ from $t_0$. The contour is ordered such that $t_+$ is later than $t_-$. (Right) The extended contour $\ext{\gamma}$, with an added vertical branch $\gamma_M$. Note that the horizontal branches are shifted vertically for illustrative purposes only; both branches are on the real axis.}  \label{contours}
\end{figure}

The time-dependent ensemble average $O(t)$ of an operator $\Oh(t)$ is given by~\cite{Stefanucci2013}
\begin{equation}  \label{eq:2_expectation_value}
O(t) =  \Tr \left[ \rhoh \, \Uh(t_0,t) \Oh(t) \Uh(t,t_0) \right],
\end{equation}
where $\Uh(t,t_0)$ is the time-evolution operator, $t_0$ the initial time, and $\rhoh$ is the density matrix. The ensemble average can be written in a form convenient for perturbation theory by introducing a directed time contour $\gamma_t$ \cite{Keldysh1965,Kadanoff1962,Stefanucci2013}, that runs from the initial time $t_0$ to $t$, and then back to $t_0$ again. For manipulations, it is convenient to extend the contour to $t = \infty$, which leaves the result unchanged since $\Uh(t,t')\Uh(t',t) = 1$. We denote this contour by $\gamma$, which is often referred to as the Keldysh contour~\cite{Keldysh1965} (see Figure~\ref{contours}, left figure. Note, however, that in the original work by Keldysh, the contour starts at $t_0=-\infty$). The ensemble average of \Eq{eq:2_expectation_value} can then be expressed as \cite{Stefanucci2013}
\begin{equation}\label{eq:2_expectation_value_contour}
O(z) = \trace{\rhoh \, \mathcal{T}_ \gamma
\left \{
\rme^{-i\int_\gamma \dif \bar{z} \, \hat{H}(\bar{z})} \Oh(z)
\right \} }.
\end{equation}
The contour time $z$ is a parameter on the contour $\gamma$. The first part of the contour, from $t_0$ to $\infty$, we call the forward branch $\gamma_-$, and denote times on it by $t_-$. The returning branch we call the backward branch $\gamma_+$ with times $t_+$. The contour-ordering operator $\mathcal{T_\gamma}\{\ldots \}$ orders operators according to the order of their contour-time arguments, with the convention that later contour times (as measured along the contour) are ordered to the left, and that for equal times the existing order is retained~\cite{Stefanucci2013}. The Hamiltonian on the contour is defined through $\Hh(t_\pm) = \Hh(t)$, and likewise for the operators $\Oh$. Consequently, the ensemble average is independent of the branch index, so that $O(t) = O(t_\pm)$.

Since the density matrix $\hat{\rho}$ is a positive semi-definite operator, it can always be written as~\cite{Leeuwen2012,VanLeeuwen2013b,Stefanucci2013,Wagner1991}
\begin{equation}\label{eq:2_density_matrix}
\rhoh = \frac{\rme^{-\Hh_M}}{Z} = \frac{\rme^{-i \int_{t_0}^{t_0 - i} \dif \bar{z} \, \Hh_M(\zb)}}{Z},
\end{equation}
where $\Hh_M(\zb) = \Hh_M$ is time-independent and $Z = \trace{\rme^{-\Hh_M}}$. In general, the operator $\Hh_M$ is an $n$-body operator unrelated to $\Hh$, but in the special case where $\rhoh$ represents the grand canonical ensemble, it is given by $\Hh_M = \beta(\Hh-\mu \Nh)$, where $\mu$ is the chemical potential and $\Nh$ the particle number operator~\cite{Leeuwen2012,VanLeeuwen2013b,Stefanucci2013}. \Eq{eq:2_density_matrix} allows for writing the ensemble average \Eq{eq:2_expectation_value_contour} as
\begin{equation} \label{expValueExtendedContour}
 O(z) = \frac{\trace{\mathcal{T}_{\ext{\gamma}}
 \left \{
 \rme^{-i\int_{\ext{\gamma}} \dif \bar{z} \, \hat{H}(\bar{z})} \Oh(z)
 \right \} }}{Z}
\end{equation}
using an extended contour~\cite{Konstantinov1961,Danielewicz1984,Wagner1991} $\gamma'$ with an vertical Matsubara branch $\gamma_M$ going from $t_0$ to $t_0 - i$ attached to the end and a corresponding extended definition of $\hat{H}(z)$ given by $\hat{H}(z \in \gamma_M) = \hat{H}_M$ (see Figure~\ref{contours}).

Using the extended contour, initial correlations can be treated perturbatively~\cite{Leeuwen2012}. The expansion is performed by expanding all $n$-body terms higher than one in $\Hh(z)$, which allows Wick's theorem to be applied. In our following discussion, we will make use of both contours (see Figure~\ref{contours}).

\subsection{Contour-time structure of diagrammatic expansions}

To discuss the types of structures that arise in many-body perturbation theory, we here focus on the diagrammatic expansion of the single-particle Green's function $G$ when the interaction $\Vh$ is of two-body type. The discussion pertains to both the Keldysh contour and the extended contour.

The simplest structure that appears is the convolution:
\begin{equation}
C(z_1,z_2) = \int_\gamma \dif \zb_1 \, A(z_1,\zb_1) B(\zb_1,z_2),
\end{equation}
which appears, for example, in the Dyson equation. The expressions generally involve also relevant quantum numbers. For clarity, we will suppress arguments other than time arguments.

\begin{figure}
\begin{align*}
\Sigma[G,W] &= \diagram{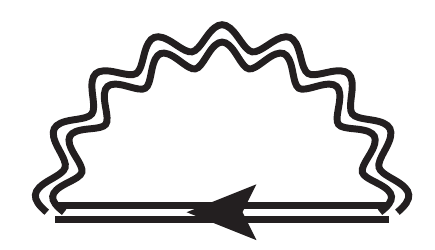} + \diagramcenter{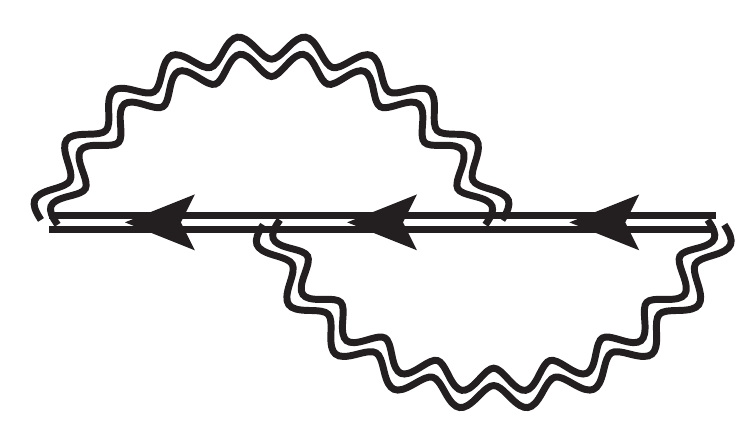} \\
P[G,W] &= \diagramcenter{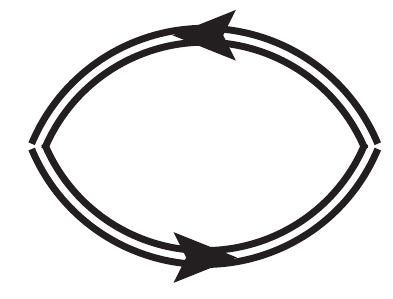} + \diagramcenter{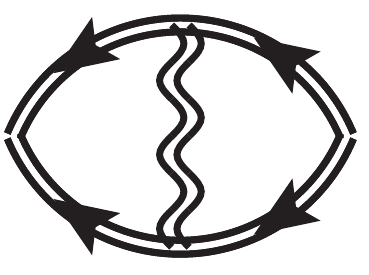}
\end{align*}
\caption{The self-energy to second order in the screened interaction (upper) and the polarization up to first order (lower). The $GW$ approximation amounts to keeping the first diagram in each row. The second diagrams are of the double-triangle structure.}
\label{2_sigmaGW}
\end{figure}

The Langreth rules use the fact that the functions involved can be expressed as
\begin{equation} \label{eq:2_greater_lesser_representation}
A(z_1,z_2) = \theta(z_1, z_2) A^>(t_1,t_2) + \theta(z_2, z_1) A^<(t_1,t_2),
\end{equation}
using real-time greater ($A^>$) and lesser ($A^<$) components for contour orders $z_1 > z_2$ and $z_2 > z_1$ respectively. The Langreth rules for a convolution yield, for the extended contour,
\begin{equation} \label{eq:2_Langreth_rule_M_<>}
A^\lessgtr = B^R \cdot C^\lessgtr + B^\lessgtr \cdot C^A + B^\rceil \star C^\lceil.
\end{equation}
In \Eq{eq:2_Langreth_rule_M_<>}, we have defined the short-hand notation for real-time convolutions as
\begin{equation}
[B \cdot C](t_1,t_2) = \int_{t_0}^\infty \dif t_3 \, B(t_1,t_3) C(t_3,t_2),
\end{equation}
and the star represents the imaginary-time convolution
\begin{equation}
[B^\rceil \star C^\lceil](t_1,t_2) \doteq -i \int_0^1 \dif t_3 \, B^\rceil(t_1,t_3) C^\lceil(t_3,t_2).
\end{equation}
The retarded ($A^R$) and advanced ($A^A$) compositions are defined as
\begin{align}
A^R(t_1, t_2) &= \Theta(t_1 - t_2) \big[ A^>(t_1, t_2) - A^<(t_1, t_2) \big] \\
A^A(t_1, t_2) &= -\Theta(t_2 - t_1) \big[ A^>(t_1, t_2) - A^<(t_1, t_2) \big].
\end{align}
Here, $\Theta(t)$ is 1 for $t>0$, and zero otherwise. We use $\Theta(t_1-t_2)$ to denote the real-time step function, which is to be distinguished from the contour step function $\theta(z_1,z_2)$. The relations between these functions are given by $\theta(t_{1-},t_{2-}) = \Theta(t_1-t_2)$, $\theta(t_{1+},t_{2+}) = \Theta(t_2-t_1)$, $\theta(t_{1-},t_{2+})=0$ and $\theta(t_{1+},t_{2-})=1$.
The retarded and advanced compositions can also be immediately obtained from another Langreth rule:
\begin{equation}
A^{R/A} = B^{R/A} \cdot C^{R/A}.
\end{equation}
All Langreth rules for convolutions are shown in Table \ref{tab:basic_langreth_rules}. The Langreth rules for the Keldysh contour can be obtained simply by taking the terms from the extended contour and putting all imaginary-time convolutions to zero~\cite{Stefanucci2013}. Another structure that appears often in many-body perturbation theory is a product:
\begin{equation}
C(z_1,z_2) = A(z_1,z_2)B(z_2,z_1).
\end{equation}
Notable examples come from the $GW$ approximation~\cite{Hedin1965}. Here, the exchange-correlation self-energy $\Sigma_{xc}(z_1,z_2) = i G(z_1,z_2) W(z_2,z_1)$ is of product form, where $W$ is the screened interaction and $G$ is the Green's function. Another example is the polarization $P(z_1,z_2) = -i G(z_1,z_2) G(z_2,z_1)$ (see Figure~\ref{2_sigmaGW}). These structures can also be treated by Langreth rules, shown in Table \ref{tab:basic_langreth_rules}.

When considering higher-order diagrams, however, structures emerge that are not reducible to these basic types. For example, the diagrams in Figure~\ref{2_sigmaGW}, second order in $W$ for $\Sigma$, and first order for $P$, are not of the chain convolution or product type. These diagrams are examples of a structure which we call the double-triangle,
\begin{equation} \begin{split} \label{eq:2_double_triangle_convolution}
&F(z_1,z_2) = \int_\gamma \dif \zb_3 \dif \zb_4 \, A(z_1,\zb_3) B(\zb_3,z_2) C(\zb_3,\zb_4) D(z_1,\zb_4) E(\zb_4,z_2),
\end{split} \end{equation}
which can be diagrammatically expressed as
\begin{equation}\label{2_double_triangle_convolution}
\bigdiagram{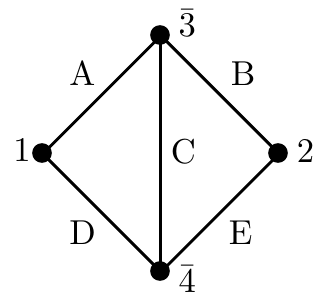}
\end{equation}
Higher-order diagrams will yield increasingly complex structures. For these types of diagrams, the original Langreth rules are not enough, and need to be generalized.

Another example for which the Langreth rules cannot be applied directly is the case in which we have integrals over general $n$-point functions. An example is one of the Hedin equations, which has the structure
\begin{equation} \begin{split} \label{eq:2_hedin_equation}
H (z_1,z_2) = \int_{\gamma} \dif \zb_3 \dif \zb_4 \,
A(z_1,\zb_3) B(z_1,\zb_4) C(\zb_4,z_2,\zb_3)
\end{split} \end{equation}
in which the three-point function $C$ appears. This equation is the Hedin equation for the exact exchange-correlation self-energy $\Sigma_{xc} = H$ if we identify $A = W$ and $B=G$, and $C = i\Lambda$ is the so-called vertex function~\cite{Hedin1965}. It can be diagrammatically expressed as
\begin{equation}\label{2_hedin_convolution}
\bigdiagram{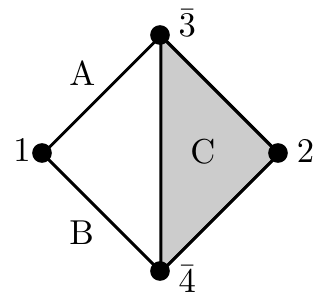}
\end{equation}
In this case the problem arises of expressing the three-point function $C$ on the contour in terms of its real-time components~\cite{Ness2011a}. 
We also mention the Bethe-Salpeter equation for the two-particle Green's function as another important equation to which the Langreth rules cannot be directly applied.

These considerations motivate the generalization of the Langreth rules for more complex expressions than convolutions and products, as well as the consideration of general $n$-point functions. These issues will be discussed in the following sections.

\section{Properties of Contour Functions}
\label{sec:3_keldysh_functions}

To generalize the Langreth rules, we introduce general real-time components that encode the information contained in a contour function. Specifically, we are interested in contour functions that have a diagrammatic representation in terms of Green's functions. In this section, we consider the case of the Keldysh contour, while the extended contour will be treated in Section \ref{sec:5_extended_contour}.

For dealing with functions with an arbitrary number of arguments, we will introduce a convenient notation. We start by defining an ordered set of labels $\Ncal = \{ n_1, \ldots, n_N \}$. By ordered set we mean that different orderings of the elements are considered to be different sets, for example
$
\{ a, b, c \} \neq \{ b, a, c \}
$. A collection of contour or real times corresponding to the labels $\Ncal$ is denoted by
$
z_\Ncal = \{ z_{n_1},\ldots,z_{n_N} \}
$ and similarly for real times $t_\Ncal$. We stress that the contour times $z$ are also real-time numbers, but carry an additional branch index to indicate which branch they are on. The collection of real times $t_\Ncal$ is the collection $z_\Ncal$ with the branch indices removed. We define the contour step function $\theta(z_\Ncal) = 1$ if $z_{n_1} > z_{n_2} > \cdots > z_{n_N}$ on the contour, and zero otherwise. For zero and one argument sets $\Ncal$, we define $\theta(z_\Ncal) = 1$.
Permutations of ordered sets are denoted by
\begin{equation}
P(\Ncal) = \{ P(n_1), \ldots, P(n_N) \}.
\end{equation}
We often sum over all $N!$ permutations belonging to the symmetric group $S_N$ of order $N$, which we denote by $\sum_{P \in S_N}$. With these definitions, we can conveniently define a \emph{Keldysh function} $\con{O}(z_\mathcal{N})$, denoted by a cursive letter, as a function of $N$ contour variables that can be expressed as
\begin{equation} \label{eq:3_keldysh_sum}
\con{O}(z_\mathcal{N}) = \sum_{P \in S_N} \theta(z_{P(\mathcal{N})}) O^{P(\check{\mathcal{N}})}(t_\mathcal{N}),
\end{equation}
with some set of real-time functions $O^{P(\check{\Ncal})}(t_\Ncal)$.
The háček ( $\check{}$ ) above an argument index denotes the position of that argument in the argument list of the relevant function. For example,
\begin{equation}
 O^{\check{b}\check{c}\check{a}} (t_a,t_b,t_c) = O^{231}(t_a,t_b,t_c).
\end{equation}
Therefore, the háček is a mapping from labels to integers, which will make the components independent of the labeling of their arguments. The value of this notation will become clear later. The real-time functions $O^{P(\check{\Ncal})}(t_\Ncal)$ are referred to as the \emph{Keldysh components} of $\con{O}(z_\mathcal{N})$. We call \Eq{eq:3_keldysh_sum} a \emph{Keldysh sum} representation. The concept of a Keldysh function is a useful one, since both Green's functions themselves, as well as diagrams built out of Green's functions, are Keldysh functions. For future use, we define a short-hand notation. We define the formal sum $ L = \sum_j \sigma_j l_j$ as a sum over signs $\sigma_j = \pm$ and integer strings $l_j$ of equal length. Correspondingly, we define a linear combination of Keldysh components
\begin{equation}\label{eq:linearCombination}
 O^{L} = \sum_j \sigma_j O^{l_j}.
\end{equation}
For example,
\begin{equation}
 O^{123 - 231} = O^{123} - O^{231}.
\end{equation}
This notation will be especially useful when we encounter nested commutators.

Keldysh functions depend on branch indices of their arguments only so far as they affect the ordering of the arguments on the contour. The domain of definition of a Keldysh function on the contour can be divided into sub-domains of constant contour order, such that in each sub-domain the function can be described by a real-time function. The Keldysh components are assumed to be defined in all of $\mathbb{R}^N$, although only the values of $O^{P(\check{\Ncal})}$ for which the multiplying step-function yields $1$ contribute to $\con{O}$ through the Keldysh sum. As a special case, a Keldysh function with a single contour argument, $\con{O}(z_{n_1})$ has a single Keldysh component $O^1(t_{n_1}) = \con{O}(t_{n_1\pm})$.

As an example, a two-point Keldysh function $\con{A}(z_a,z_b)$ has the representation
\begin{equation} \label{eq:3_keldysh_sum_example}
\con{A}(z_a,z_b) = \theta(z_a,z_b) A^{\check{a}\check{b}}(t_a,t_b) + \theta(z_b,z_a) A^{\check{b}\check{a}}(t_a,t_b),
\end{equation}
where, as before, $A^{\check{a}\check{b}} = A^{12}$ and $A^{\check{b}\check{a}} = A^{21}$. \Eq{eq:3_keldysh_sum_example} is identical to  \Eq{eq:2_greater_lesser_representation}, with $A^> = A^{12}$ and $A^< = A^{21}$. Thus, Keldysh components are a generalization of greater and lesser components to more variables, where each Keldysh component corresponds to a particular contour-order of the arguments.

We now examine two important properties of Keldysh functions. The first is that products of Keldysh functions are Keldysh functions. This can be seen by multiplying two Keldysh functions, and re-expressing the products of step functions in terms of sums of single step functions of varying argument lists, see, e.g., \Eq{thetafunctions} in \ref{app:decomposition_of_step_functions}. For example, multiplying two two-point functions $\con{C}(z_a,z_b)\con{D}(z_b,z_c)$ involves multiplications of, for example,
\begin{equation}
 \theta(z_a,z_b) \theta(z_c,z_b) = \theta(z_c,z_a,z_b) + \theta(z_a,z_c,z_b).
\end{equation}
The terms on the right-hand side appear in the expansion of a new Keldysh function $\con{E}(z_a,z_b,z_c)$. In general, an $N$-point Keldysh function multiplied with an $M$-point Keldysh function yields an $L$-point Keldysh function  with $L\leq M+N$.

The second property that we will use is that integrating time arguments of a Keldysh function over the contour yields another Keldysh function. This can be seen by directly integrating a Keldysh sum. Let us consider the unit permutation in an $N$-point Keldysh sum, and integrate the variable $z_i$:
\begin{equation} \begin{split} \label{eq:3_integral_example}
&\int_\gamma \dif z_i \, \theta(z_\Ncal) O^{\Ncal}(t_\Ncal) =
\theta(z_{\Ncal \setminus i}) \int_{z_{i+1}}^{z_{i-1}} \dif z_i \, O^{\Ncal}(t_\Ncal).
\end{split} \end{equation}
Here, ${\Ncal \setminus i}$ is the set $\Ncal$ with the element $i$ removed. By considering different branches for $z_{i-1}$ and $z_{i+1}$, and taking into account that the Keldysh component $O^{\Ncal}(t_\Ncal)$ is independent of the branch indices, we can readily see that the integral on the right-hand side is independent of the branch indices as well, and is thus an $(N-1)$--point Keldysh component. In particular, it follows that integrating $N$ arguments from an $N$-point Keldysh function yields zero,
\begin{equation} \label{eq:4_total_integral_lemma}
\int_\gamma \dif z_\Ncal \, \con{\bar{O}}(z_\mathcal{N}) = 0.
\end{equation}

An arbitrary integrand in a Feynman diagram is a polynomial in Keldysh functions, and hence a Keldysh function. Thus, any Feynman diagram is a Keldysh function. As a small remark, a perturbation expansion also includes time-local interactions. For example, the two-point interaction has the form
\begin{equation}
V(z_1,z_2) = v(z_1) \delta(z_1, z_2).
\end{equation}
If these interaction lines connect internal vertices of a diagram, the delta-functions are integrated out and only a multiplication by the coupling constant $v(z_1)$ remains. Since a multiplication by a real-time function trivially leaves a Keldysh function a Keldysh function, any diagram with only internal interaction lines is a Keldysh function. External interaction lines, i.e. interaction lines that connect to an external vertex, can be dealt with by expressing the function as a sum of singular and regular parts.

\begin{figure}
\centering
\includegraphics[width=0.5\textwidth,angle=0]{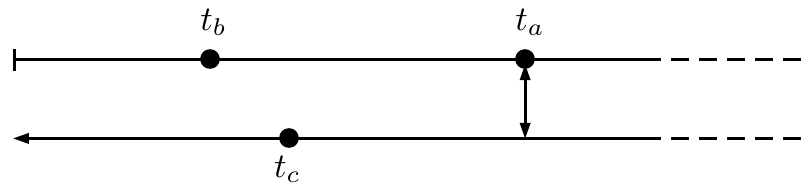}
\caption{If $z_a$ is the argument with the highest real-time value $t_a$, the contour ordering is $cab$, independently on whether $z_a$ is on the forward or the backward branch. }
\label{fig:3_highest_time_symmetry}

\end{figure}

Finally, we point out a symmetry property of Keldysh functions that will be useful. A Keldysh function $\con{O}(z_\mathcal{N})$ is symmetric with respect to the branch index of the parameter with the highest real-time value, i.e.
\begin{equation} \label{eq:3_largest_real_time_symmetry}
\con{O}(\ldots,t_{a+},\ldots) = \con{O}(\ldots,t_{a-},\ldots), \qquad \text{if } t_a > t_{\mathcal{N} \setminus a},
\end{equation}
since the contour order is identical with either $t_{a+}$ or $t_{a-}$ (see figure \ref{fig:3_highest_time_symmetry}).

\section{Integrals over Keldysh Functions}
\label{sec:4_integrals_over_keldysh_functions}

In the previous section we discussed how Keldysh functions can be expressed in terms of real-time Keldysh components. We will make use of this decomposition to transform an integral involving Keldysh functions to equations between the corresponding Keldysh components. The equations we are interested in have the general form
\begin{equation} \label{eq:4_general_integral_equation}
\con{O}(z_\mathcal{E}) = \int_\gamma \dif z_\mathcal{I} \, \con{\bar{O}}(z_\mathcal{N}),
\end{equation}
in which a Keldysh function $\con{\bar{O}}(z_\mathcal{N})$ has some of its arguments integrated over the Keldysh contour. We will typically denote integrands by barred symbols. We call $\mathcal{E} = \{ e_1, \ldots, e_E \} \subset \mathcal{N}$ the external arguments and $\mathcal{I} = \{ i_1,\ldots, i_I \} = \mathcal{N} \setminus \mathcal{E}$ the internal arguments. The function $\con{\bar{O}}$ may have a diagrammatic structure, but in this section we only assume that it is a Keldysh function. The fact that $\con{O}(z_\mathcal{E})$ is a Keldysh function was derived in the last section.

The question is then how to obtain the Keldysh components of $\con{O}$ directly in terms of the Keldysh components of $\con{\bar{O}}$. An elegant result accomplishing this, in case of the Keldysh contour, was derived by Danielewicz~\cite{Danielewicz1990} using an expansion in terms of retarded compositions of Keldysh components. In this section we will provide an alternative derivation, and generalize it to the extended contour in the subsequent section. 

\subsection{Deforming the Contour}

We use the fact that the contour can be truncated, so that it turns back at $t = t_M$, with $t_M$ the largest real-time value of an external argument:
\begin{equation}
\bigdiagram{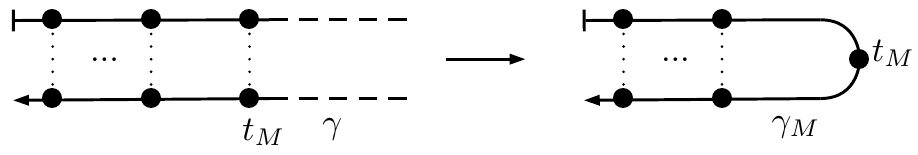}
\end{equation}
This follows from \Eq{eq:4_total_integral_lemma}, since we can consider a contour that starts at $t_M$, goes to infinity and returns to $t_M$, for which the integral over all internal variables vanishes. We denote the truncated contour by $\gamma_M$, and we have
\begin{equation} \label{eq:4_contour_truncation_lemma}
\int_\gamma \dif z_\Ical \, \con{\bar{O}}(z_\Ncal) = \int_{\gamma_M} \dif z_\Ical \, \con{\bar{O}}(z_\Ncal), \qquad t_M \geq t_\Ecal.
\end{equation}

The same idea can be used to deform the contour $\gamma_M$ into several loops. In an integral back and forth on the real-axis, components with the same order will appear in both branches, cancelling each other out. Consider, for example, the contour $\gamma_{ab}$, consisting of two loops $\gamma_a$ and $\gamma_b$:
\begin{equation} \label{fig:4_contour_deformation}
\bigdiagram{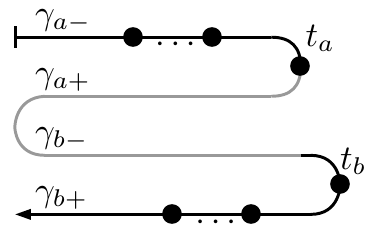}
\end{equation}
where two new branches, going from $t_a$ to $t_0$ and back along the real-axis, have been added between the arguments $a$ and $b$. In other words:
\begin{equation}
\int_\gamma \dif z_\mathcal{I} \, \con{\bar{O}}(z_\mathcal{N}) = \int_{\gamma_{ab}} \dif z_\mathcal{I} \, \con{\bar{O}}(z_\mathcal{N}).
\end{equation}
It follows that additional back-and-forth loops can be freely added to the contour without changing the value of the integral. This idea will be used in the subsequent sections.

\subsection{A Single External Argument}

The situation in which all but one argument of a Keldysh function $\con{\Ob}(z_e,z_\Ical) = \con{\Ob}(z_e,z_{i_1},\ldots,z_{i_I})$ are integrated over,  provides a useful stepping stone to the general result. We therefore consider
\begin{equation}\label{eq:4_single_external_start}
\con{O}(z_e) = O^1(t_e) = \int_{\gamma_e} \dif z_\Ical \ \con{\Ob}(z_e,z_\Ical),
\end{equation}
where the contour $\gamma_e$ has been truncated to turn back at $t_e$. Note that, as was seen in the previous section, a Keldysh function of a single argument is equal to its only Keldysh component. Therefore the right-hand side of \Eq{eq:4_single_external_start} can not depend on the branch index of $z_e$ either. We consider the Keldysh sum
\begin{equation}\label{oneExternalParameter}
\con{\Ob}(z_e,z_\Ical) = \sum_{P \in S_{I+1}} \theta_{P(e i_1 \cdots i_I)} \Ob^{P(\check{e} \check{i}_1 \cdots \check{i}_I)}(t_e, t_\Ical),
\end{equation}
over permutations of the $I+1$ arguments, where we have introduced the short-hand notation for contour step functions $\theta(z_{n_1},\ldots,z_{n_N}) = \theta_{n_1 \cdots n_N}$. We remark that although we denote $z_e$ to be the first in the argument list in \Eq{oneExternalParameter}, the equation holds for any location of $z_e$ in the list, illustrating the usefulness of the háček notation. Our strategy to obtain a real-time expression from \Eq{eq:4_single_external_start} will be to split the $I$ contour integrals into separate contributions from the forward branch $\gamma_-$ and the backward branch $\gamma_+$. For this purpose, it will be convenient to write the sum in \Eq{oneExternalParameter} as a sum over permutations of the $I$ internal arguments, and an additional sum over the position ($j$) of the external argument:
\begin{equation} \label{eq:4_Pj_representation}
\con{\Ob}(z_e,z_\Ical) = \sum_{j=0}^I \sum_{P \in S_I} \theta_{(P,j)} \Ob^{\check{(P ,j)}} (t_e,\tb),
\end{equation}
where we have defined
\begin{align}\label{eq:4_Pj_definition}
(P,j) = P(i_1 \cdots i_j) e P(i_{j+1} \cdots i_I).
\end{align}
For example, $(P,0) = e P(i_1 \cdots i_I)$, $(P,2) = P(i_1 i_2) e P(i_3 \cdots i_I)$. As such, $j$ is also the number of arguments to the left of $e$, while we have $I-j$ arguments to the right of $e$. This means that the arguments $z_{P(i_1)}, \ldots, z_{P(i_j)}$ are on the backward branch, while $z_{P(i_{j+1})}, \ldots, z_{P(i_I)}$ are on the forward branch. We illustrate the situation in the figure below.
\begin{equation}
\bigdiagram{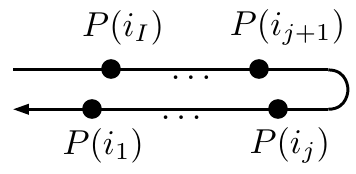}
\end{equation}

When \Eq{eq:4_Pj_representation} is inserted into \Eq{eq:4_single_external_start}, each integral is restricted to either the forward or the backward branch. For $j=0$, there are no arguments later than $z_e$, which means that all $I$ integrals are over the forward branch. For $j=1$, one integral is over the backward branch, and $I-1$ over the forward branch. \Eq{eq:4_single_external_start} can then be written as
\begin{equation} \begin{split} \label{eq:4_integral_split}
O^1(t_e) &= \sum_{j=0}^I  \sum_{P \in S_I}  \int_{\gamma_e^+} \dif z_{P(i_1 \cdots i_j)} \int_{\gamma_e^-} \dif z_{P(i_{j+1} \cdots i_I)} \\
&\underbrace{\theta_{P(i_1 \cdots i_j)}}_{\text{backward branch}} \underbrace{\theta_{P(i_j)e}}_{=1} \underbrace{\theta_{eP(i_{j+1})}}_{=1} \underbrace{\theta_{P(i_{j+1} \cdots i_I)}}_{\text{forward branch}} \Ob^{\check{(P,j)}} (t_e,t_\Ical).
\end{split} \end{equation}
The two step-functions in the middle always equal one, since the integral domains only extend to $t_e$. We also define, that step-functions with one or zero arguments always yield unity, which ensures that \Eq{eq:4_integral_split} is well defined for all values of $j$.

For integrals over the forward branch, the contour integrals can be converted into real-time integrals via the replacement $\int _{\gamma_e} \dif z \to \int_{t_0}^{t_e} \dif t$, and the contour step functions can be converted into real-time step functions, using $\theta(t_{1-},t_{2-}) = \Theta(t_1 - t_2)$.
For the backward branch, the replacement $\int _{\gamma_e} \dif z \to -\int_{t_0}^{t_e} \dif t$ yields an additional minus sign due to the reversal of the integration direction, yielding an extra factor of $(-1)^j$ from the backward-branch integrations. The conversion of contour step functions to real-time step functions is reversed as compared to the forward branch, since $\theta(t_{1+},t_{2+}) = \Theta(t_2 - t_1)$. We now use the short-hand notation $\Theta(t_{n_1},\ldots,t_{n_N}) = \Theta_{n_1 \cdots n_N}$, allowing \Eq{eq:4_integral_split} to be written in the form
\begin{equation} \begin{split} \label{eq:4_integral_split_2}
O^1(t_e) &= \sum_{j=0}^I  \sum_{P \in S_I}
 \int_{t_0}^\infty \dif t_\Ical (-1)^j \Theta_{e P(i_j \cdots i_1)} \Theta_{e P(i_{j+1} \cdots i_I)} \Ob^{\check{(P,j)}} (t_e,t_\Ical),
\end{split} \end{equation}
where we have placed $t_e$ back into the step-functions, and extended the integral to infinity.
%

To elucidate the structure of \Eq{eq:4_integral_split_2} we give examples for one ($I=1$) and two ($I=2$) integrations.

For $I=1$, the sum over permutations in \Eq{eq:4_integral_split_2} yields one term, and the $j$-sum yields two terms:
\begin{equation}
O^1(t_e) = \int_{t_0}^\infty \dif t_{i_1} \  \left (
 \Theta_{e i_1} \Ob^{\check{e}\,\check{i}_1}(t_e,t_{i_1}) - \Theta_{e i_1} \Ob^{\check{i}_1 \check{e}}(t_e,t_{i_1}) \right) =
 \int_{t_0}^\infty \dif t_{i_1}  \Theta_{e i_1} \Ob^{[\check{e}, \check{i}_1]}(t_e,t_{i_1}),
\end{equation}
where we have defined the commutator $[\check{e}, \check{i}_1] = \check{e}\check{i}_1 - \check{i}_1\check{e}$ and used the short-hand notation introduced in \Eq{eq:linearCombination}. Defining the retarded composition of two arguments as
\begin{equation}
 \Ob^{R(\check{e},\check{i}_1)}(t_e,t_{i_1}) = \Theta_{e i_1} \Ob^{[\check{e}, \check{i}_1]}(t_e,t_{i_1}).
\end{equation}
we obtain the compact expression
\begin{equation}
O^1(t_e) = \int_{t_0}^{\infty} \dif t_{i_1} \Ob^{R(\check{e},\check{i}_1)}(t_e,t_{i_1}).
\end{equation}

Let us now consider $I=2$, and the equation
\begin{align}
\con{O}(z_e) = \int_{\gamma_e} \dif z_{i_1} \dif z_{i_2} \con{\Ob}(z_e,z_{i_1},z_{i_2}),
\end{align}
\Eq{eq:4_integral_split_2} gives after writing out the $j$ sum:
\begin{equation} \begin{split} \label{eq:4_I_2_step_one}
&O^1(t_e) =
 \int_{t_0}^\infty \dif t_{i_1} \dif t_{i_2} \
 \sum_{P \in S_2} \\
 &\times
 \big [
 \underbrace{\Theta_{e P(i_1 i_2)} \Ob^{\check{e} \check{P(i_1)} \check{P(i_2)}}}_{j=0} -
 \underbrace{\Theta_{e P(i_1)} \Theta_{e P(i_2)} \Ob^{\check{P(i_1)} \check{e} \check{P(i_2)}}}_{j=1} +
 \underbrace{\Theta_{e P(i_2 i_1)} \Ob^{\check{P(i_1)} \check{P(i_2) \check{e}}}}_{j=2}
 \big ].
\end{split} \end{equation}
We will break the derivation from this point into three steps, so that it can be easily compared with the derivation of the general result, that follows the same steps.
\begin{enumerate}
\item
First the step-functions in the $j=1$ term in \Eq{eq:4_I_2_step_one} are written in the form $\Theta_{e P(i_1)} \Theta_{e P(i_2)} = \Theta_{e P(i_1 i_2)} + \Theta_{e P(i_2 i_1)}$.
\item
Next we relabel permutations in the two terms containing $\Theta_{e P(i_2 i_1)}$, so that $P(i_1)$ and $P(i_2)$ are swapped. We can then factor out the step function $\Theta_{e P(i_1 i_2)}$ to obtain
\begin{equation} \begin{split} \label{eq:4_I_2_step_two}
&O^1(t_e) =
 \int_{t_0}^\infty \dif t_{i_1} \dif t_{i_2} \
 \sum_{P \in S_2} \\
 &\times
 \Theta_{e P(i_1 i_2)} \big [
 \Ob^{\check{e} \check{P(i_1)} \check{P(i_2)}}
 - \Ob^{\check{P(i_1)} \check{e} \check{P(i_2)}} -  \Ob^{\check{P(i_2)} \check{e} \check{P(i_1)}}
 + \Ob^{\check{P(i_2)} \check{P(i_1) \check{e}}}
 \big ].
\end{split} \end{equation}
\item
Finally we observe that the sum of terms in the square brackets in \Eq{eq:4_I_2_step_two} can be written as
\begin{equation} \begin{split}
\Ob^{\check{e} \check{P(i_1)} \check{P(i_2)}}
 - \Ob^{\check{P(i_1)} \check{e} \check{P(i_2)}} -  \Ob^{\check{P(i_2)} \check{e} \check{P(i_1)}}
 + \Ob^{\check{P(i_2)} \check{P(i_1) \check{e}}} = \Ob^{[\check{e}, \check{P(i_1)}, \check{P(i_2)}]},
\end{split} \end{equation}
where $[a,b,c] = [[a,b],c]$ is a nested commutator (see \ref{app:structure_of_nested_commutators} for a discussion on properties of nested commutators). Defining a generalized retarded composition as
\begin{equation}
 \Ob^{R(\check{e},\check{i}_1 \check{i}_2)}(t_e,t_{i_1},t_{i_2}) =
 \sum_{P \in S_2}
 \Theta_{e P(i_1 i_2)} \Ob^{[\check{e}, \check{P(i_1)}, \check{P(i_2)}]},
\end{equation}
then leads to the compact result
\begin{equation}
 O^1(t_e) = \int_{t_0}^\infty \dif t_{i_1} \dif t_{i_2} \Ob^{R(\check{e},\check{i}_1 \check{i}_2)}(t_e,t_{i_1},t_{i_2}).
\end{equation}

\end{enumerate}

We now show that the general case of $I = N$ can be rewritten in terms of generalized retarded compositions. The proof proceeds by the same three steps as used above for the $I=2$ case.
\begin{enumerate}
\item
The product of step functions in \Eq{eq:4_integral_split_2} is written as a sum of step-functions as
\begin{equation}
\Theta_{e P(i_j \cdots i_1)} \Theta_{e P(i_{j+1} \cdots i_I)} = \sum_{T \in \Tcal_{I,j}} \Theta_{e T(P(i_1)) T(P(i_2)) \cdots T(P(i_I)))} = \sum_{T \in \Tcal_{I,j}} \Theta_{e\, T \circ P(\Ical)}.
\end{equation}
Here the set $\Tcal_{I,j}$ contains every permutation of the arguments $\Ical$ for which the subsets $\{ i_j, \ldots, i_1 \}$ and $\{ i_{j+1}, \ldots, i_I \}$ remain in the same relative order as given by the step-functions on the left-hand side. Graphically this corresponds to every permutation that can result from projecting the forward and backward branches vertically to the same axis (see figure \ref{fig:4_T_permutations}).
\begin{figure}
\centering
\includegraphics[scale=1]{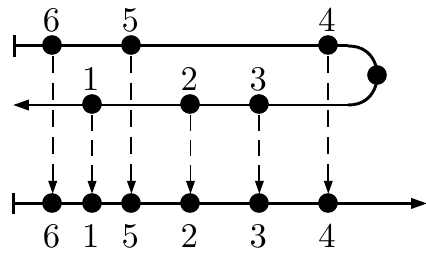}
\caption{An example of a permutation belonging to the set $\Tcal_{6,3}$, permuting $123456$ to $432516$. Note that the relative order of $456$ is retained, while the order of $123$ is inverted. Moving the arguments on the upper contour without changing their order generates all the permutations in $\Tcal_{6,3}$. \label{fig:4_T_permutations}}
\end{figure}

The structure of these permutations is derived in detail in \ref{app:decomposition_of_step_functions}.
\Eq{eq:4_integral_split_2} now takes the form
\begin{align} \label{eq:4_general_step_one}
 O^1(t_e) =
 \sum_{j=0}^I
  \sum_{P \in S_I}
  \sum_{T \in \Tcal_{I,j}}
 \int_{t_0}^\infty \dif t_\Ical \ (-1)^j
 \Theta_{e\, T \circ P(\Ical)} \Ob^{\check{(P,j)}} (t_e,t_\Ical).
\end{align}
\item
Since we sum over all permutations $P$ in the group $S_I$, we can equivalently sum over all permutations $U = T \circ P$, where $U \in S_I$. Inserting this relation into \Eq{eq:4_general_step_one}, and reordering, yields
\begin{align} \label{eq:4_general_step_two}
 O^1(t_e) =
  \sum_{U \in S_I}
  \int_{t_0}^\infty \dif t_\Ical \
 \Theta_{e\, U(\Ical)}
 \sum_{j=0}^I
\sum_{T \in \Tcal_{I,j}}
 (-1)^j
 \Ob^{\check{(T^{-1} \circ U ,j)}} (t_e,t_\Ical).
\end{align}
\item
Finally we will show that the sum over components of $\Ob$ in \Eq{eq:4_general_step_two} corresponds to a nested commutator:
\begin{equation}\label{eq:4_general_step_three}
 \sum_{j=0}^I \sum_{T \in \Tcal_{I,j}} (-1)^j \Ob^{\check{(T^{-1} \circ U ,j)}} (t_e,t_\Ical) =
 \Ob^{[\ech,U(\ich_1),U(\ich_2),\cdots,U(\ich_I)]} (t_e,t_\Ical).
\end{equation}

It is sufficient to consider \Eq{eq:4_general_step_three} for the identity permutation $U(i_1 \cdots i_I) = i_1 \cdots i_I$, as the general case simply follows from relabeling. Let us also focus on a specific $j$ term.
Using the definition of $(T^{-1} ,j)$, \Eq{eq:4_Pj_definition}, we have
\begin{equation} \label{eq:4_general_step_four}
 \sum_{T \in \Tcal_{I,j}} (-1)^j (T^{-1} ,j) =
 \sum_{T \in \Tcal_{I,j}} (-1)^j
 T^{-1} (i_1 \cdots i_j) \, e \, T^{-1} (i_{j+1} \cdots i_I).
\end{equation}
The permutations $T^{-1}$ are the inverse of those considered in step (i), and can thus be represented by inverting the arrows in figure \ref{fig:4_T_permutations}, see figure \ref{fig:4_T_inverse_permutations}.
\begin{figure}
\centering
\includegraphics[scale=1]{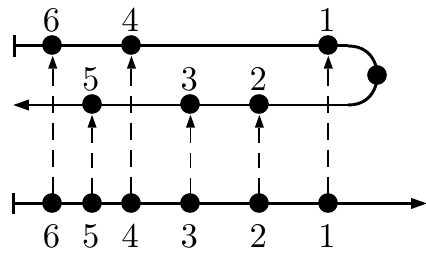}
\caption{An example of an inverse of a permutation belonging to the set $\Tcal_{6,3}$, permuting $123456$ to $532146$. Since $j = 3$, three arguments are moved to the backward branch. Inverses of all the permutations in $\Tcal_{6,3}$ are generated by all the possible ways to place three arguments on the backward branch, and three on the forward branch. Thus there are $6!/3!$  permutations in $\Tcal_{6,3}$. \label{fig:4_T_inverse_permutations}}
\end{figure}
Thus permutation $T^{-1}$, when $T \in \Tcal_{I,j}$, permutes $1\cdots I$ in such a way that the first $j$ arguments are in decreasing order and the remaining $I - j$ arguments are in increasing order. These are exactly the permutations generated by $[e,1,2,\ldots,I]_j$, i.e. the part of the nested commutator with $j$ elements to the left of $e$ (see the discussion in \ref{app:structure_of_nested_commutators}). Furthermore, since a nested commutator generates a minus sign for each element to left of $e$, it has the sign $(-1)^j$, matching that in \Eq{eq:4_general_step_four}. We thus have
\begin{equation}\label{eq:partialResultCommutator}
 \sum_{T \in \Tcal_{I,j}} (-1)^j
 T^{-1} (i_1 \cdots i_j) \, e \, T^{-1} (i_{j+1} \cdots i_I) =
 [e,i_1,\cdots,i_I]_j.
\end{equation}
A more rigorous derivation of \Eq{eq:partialResultCommutator} is found by comparing the definitions of the permutations $T^{-1}(i_1 \cdots i_I)$, found in \Eq{eq:definingTcal1}, to the permutations $Q$ given in \Eq{eq:definingQcal}. The definitions are identical, and thus \eqref{eq:partialResultCommutator} follows directly from \Eq{eq:defNestedCommutatorQ}.

The full nested commutator is given by $[e,i_1,\cdots,i_I] = \sum_j [e,i_1,\cdots,i_I]_j$.  Summing over $j$ in \Eq{eq:4_general_step_three} thus yields the total nested commutator $[e,i_1,\cdots,i_I]$, which proves \Eq{eq:4_general_step_three} for the identity permutation. The proof for other permutations follows in the same way from an initial relabeling.

Having proven \Eq{eq:4_general_step_three}, we now insert it into \Eq{eq:4_general_step_two}. By defining the retarded composition as
\begin{equation} \label{eq:4_fully_retarded_definition}
 \Ob^{R(\ech,\check{\Ical})} (t_e,t_\Ical) =
 \sum_{P \in S_I} \Theta_{e P(i_1 i_2 \cdots i_I)} \Ob^{[\ech,P(\ich_1),P(\ich_2),\cdots,P(\ich_I)]} (t_e, t_\Ical),
\end{equation}
we obtain the final result
\begin{equation} \label{eq:4_single_argument_result}
 O^1(t_e) = \int_{t_0}^\infty \dif t_\Ical \, \Ob^{R(\ech,\check{\Ical})} (t_e, t_\Ical).
\end{equation}
The function $\Ob^{R(\ech,\check{\Ical})}$ is called a retarded composition of Keldysh components.
\end{enumerate}

We call the ordered union $\{ \ech,\check{\Ical} \}$ a retarded set. Its first argument $\ech$ we call the top argument, and the arguments in $\check{\Ical}$ the retarded arguments. The retarded composition is non-zero only when the top argument has a higher real-time value than any of the retarded arguments. It follows from its definition in \Eq{eq:4_fully_retarded_definition} that the retarded composition is symmetric with respect to permutations of the retarded arguments, so that:
\begin{equation} \label{eq:4_retarded_composition_symmetry}
\Ob^{R(\ech,P(\check{\Ical}))} (t_e,t_\Ical) = \Ob^{R(\ech,\check{\Ical})} (t_e,t_\Ical),
\end{equation}
for any permutation $P \in S_I$. Note also that \Eq{eq:4_fully_retarded_definition} is not unique in satisfying \Eq{eq:4_single_argument_result}. In particular one could, if desired, symmetrize \Eq{eq:4_fully_retarded_definition} with respect to the internal times without changing the integral in \Eq{eq:4_single_argument_result}. We use the definition in \Eq{eq:4_fully_retarded_definition} for its simplicity.

\subsection{Multiple External and Internal Arguments}

We now have all the pieces required to lay out the proof for the case with an arbitrary number of both internal $I$ and external $E$ arguments. Let us return to the equation
\begin{equation} \label{eq:4_1_general_integral_equation}
O(z_\mathcal{E}) = \int_\gamma \dif z_\mathcal{I} \, \bar{O}(z_\mathcal{N}),
\end{equation}
Let us take as an example the Keldysh component $O^{1\cdots E}$. We deform the contour $\gamma$ into $E$ loops, obtaining
\begin{equation} \label{eq:4_integral_split_3}
\int_\gamma \dif z = \Big( \int_{\gamma_{e_1}} + \ldots + \int_{\gamma_{e_E}} \Big) \dif z.
\end{equation}
For example for two external arguments we have two loops
\begin{equation}\label{4_contour_truncation_2}
\bigdiagramcenter{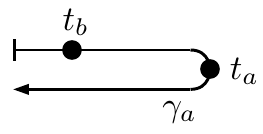} \rightarrow \bigdiagramcenter{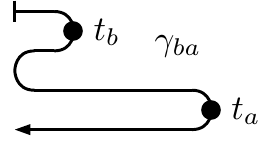}
\end{equation}
Substituting \Eq{eq:4_integral_split_3} for each integral in \Eq{eq:4_1_general_integral_equation} results in a sum containing each possible distribution of the internal arguments among the loops. We write this as
\begin{equation} \label{eq:4_1_M_sum}
\int_\gamma \dif z_\mathcal{I} = \sum_\mathcal{I} \int_{\gamma_{e_1}} \dif z_{\mathcal{I}_{e_1}} \ldots \int_{\gamma_{e_E}} \dif z_{\mathcal{I}_{e_E}},
\end{equation}
where the sum is over every possible way to split the set $\mathcal{I}$ into the subsets $\mathcal{I}_{e_1},\ldots,\mathcal{I}_{e_E}$, while retaining the relative order of the indices within each subset. $\int_{\gamma_{e_i}} \dif z_{\mathcal{I}_{e_i}}$ denotes integrating the arguments $z_{\mathcal{I}_{e_i}}$ over the loop $\gamma_{e_i}$.

In each term of the $\Ical$ sum, we can handle the integral over each loop $\gamma_{e_i}$ separately, because the arguments on the other loops are always earlier or later in contour time. Performing the integral over $\gamma_{e_i}$, we obtain a composition in which $t_{\mathcal{I}_{e_i}}$ are retarded with respect to $t_{e_i}$, leading to
\begin{equation} \begin{split} \label{eq:4_1_multi_parameter_result}
& O^{\check{e}_1\cdots \check{e}_E}(t_\mathcal{E}) =
\sum_\mathcal{I} \int_{t_0}^\infty \dif t_\mathcal{I} \, \bar{O}^{R(\check{e}_1,\check{\mathcal{I}}_1) \cdots R(\check{e}_E,\check{\mathcal{I}}_E)}(t_\mathcal{N}).
\end{split} \end{equation}
We note that since $z_\mathcal{E}$ is ordered as $(e_1,\cdots,e_E)$, it follows that $\check{e}_1\cdots \check{e}_E = 1 \cdots E$. All other Keldysh components can be obtained from \Eq{eq:4_1_multi_parameter_result} by permuting $1,\ldots,E$ on both sides. The integrand in \Eq{eq:4_1_multi_parameter_result} is a general type of multi-retarded composition defined as
\begin{equation} \label{eq:4_retarded_definition}
 \bar{O}^{R(\check{e}_1,\check{\mathcal{I}}_1) \cdots R(\check{e}_E,\check{\mathcal{I}}_E)}(t_\mathcal{N}) =
 \sum_{ P_1 \in S_{I_1}} \Theta_{e_1 P_1(\Ical_1)}
 \cdots
 \sum_{ P_E \in S_{I_E}} \Theta_{e_E P_E(\Ical_E)}
 \bar{O}^{[\check{e}_1,P_1(\Ical_1)] \cdots [\check{e}_E,P_E(\Ical_E)]}(t_\mathcal{N}),
\end{equation}
where $I_k$ is the number of elements in the set $\Ical_k$, and we define $\Theta_{e P(\Ical)} = \Theta_{e P(i_1) \cdots P(i_I)}$ for $\Ical = \{ i_1,\ldots,i_I \}$. We furthermore define the nested commutator
\begin{equation}
 [e,P(\Ical)] = [e,P(i_1),\ldots,P(i_I)],
\end{equation}
see \ref{app:structure_of_nested_commutators} for further details. Note that in \Eq{eq:4_retarded_definition} some of the sets $\Ical_i$ can be empty, in which case one substitutes $R(\check{e}_i,\emptyset) \rightarrow \check{e}_i$, and correspondingly $[\check{e}_i,\emptyset] \rightarrow \check{e}_i$. For example for one internal argument all the sets are empty except one, and \Eq{eq:4_1_multi_parameter_result} reduces to
\begin{equation}
O^{\check{e}_1 \cdots \check{e}_E}(t_\mathcal{E}) = \int_{t_0}^\infty \dif t_i \sum_{k=1}^E \bar{O}^{\check{e}_1 \cdots R(\check{e}_k,\check{i}) \cdots \check{e}_E}(t_\mathcal{N}).
\end{equation}
The definition of a multi-retarded composition subsumes both the completely retarded compositions defined in \Eq{eq:4_fully_retarded_definition} (when $E = 1$) and the Keldysh components (when $E = N$ and thus all $\Ical_i$ are empty).

Let us now illustrate \Eq{eq:4_1_multi_parameter_result} with an example. We apply the equation to a four-point function when we integrate two arguments:
\begin{equation} \label{eq:4_1_expansion_example}
\con{D}(z_a,z_d) = \int_\gamma \dif z_b \dif z_c \,  \con{\bar{D}}(z_a,z_b,z_c,z_d).
\end{equation}
In \Eq{eq:4_1_expansion_example}, the internal arguments constitute the set $\mathcal{I} = \{ b, c \}$, and the external arguments are in $\mathcal{E} = \{ a,d \}$. The four possible divisions of $\mathcal{I}$ into two subsets are
$\{\mathcal{I}_1, \mathcal{I}_2 \} =
\{
\{b,c\}, \emptyset \}$, and
$\{\{b\},\{c\}\}$, and
$\{\{c\}, \{b\} \}$, and
$\{\emptyset, \{b,c\} \}$.

The component $D^{12}_{ad}$ is given by \Eq{eq:4_1_multi_parameter_result}:
\begin{equation} \begin{split} \label{eq:4_1_expansion_example_result}
D^{12}_{ad} &= \int \bar{D}^{R(\check{a},\check{b}\check{c})\check{d}}_{a\bar{b}\bar{c}d} + \int \bar{D}^{R(\check{a},\check{b})R(\check{d},\check{c})}_{a\bar{b}\bar{c}d} + \int \bar{D}^{R(\check{a},\check{c})R(\check{d},\check{b})}_{a\bar{b}\bar{c}d} + \int \bar{D}^{\check{a}R(\check{d},\check{b}\check{c})}_{a\bar{b}\bar{c}d} \\
&= \int \bar{D}^{R(1,23)4}_{a\bar{b}\bar{c}d} + \int \bar{D}^{R(1,2)R(4,3)}_{a\bar{b}\bar{c}d} + \int \bar{D}^{R(1,3)R(4,2)}_{a\bar{b}\bar{c}d} + \int \bar{D}^{1R(4,23)}_{a\bar{b}\bar{c}d},
\end{split} \end{equation}
where the integrals are over the barred arguments (real-time integrals are from $t_0$ to $\infty$ unless otherwise stated). This equation illustrates the usefulness of the háček notation, as for example the argument $z_d$ denotes the second argument of $D$ but the fourth argument of the integrand $\bar{D}$.
The other component $D^{21}_{ad}$ is obtained by reordering the retarded sets so that $d$ comes after $a$ in the contour order:
\begin{equation} \begin{split} \label{eq:4_lesser_component}
D^{21}_{ad} &= \int \bar{D}^{4R(1,23)}_{a\bar{b}\bar{c}d} + \int \bar{D}^{R(4,3)R(1,2)}_{a\bar{b}\bar{c}d} + \int \bar{D}^{R(4,2)R(1,3)}_{a\bar{b}\bar{c}d} + \int \bar{D}^{R(4,23)1}_{a\bar{b}\bar{c}d}.
\end{split} \end{equation}
The retarded compositions appearing in the real-time integrals can be calculated from the Keldysh components using the equation for multi-retarded compositions, \Eq{eq:4_retarded_definition}:
\begin{equation}
\bar{O}^{R(1,23)4}_{abcd} = \Theta_{abc} \bar{O}^{[[1,2],3]4}_{abcd} + \Theta_{acb} \bar{O}^{[[1,3],2]4}_{abcd}
\end{equation}
\begin{equation}
\bar{O}^{R(1,2)R(4,3)}_{abcd} = \Theta_{ab} \Theta_{dc} \bar{O}^{[1,2][4,3]}_{abcd}.
\end{equation}

\subsection{The Diagrammatic Representation}
\label{sec:diagrammatic_recipe}

The previous discussion of obtaining different real-time components was mostly algebraical. In this section we describe a diagrammatic recipe for obtaining the components. The recipe includes four steps.
\begin{enumerate}
\item Write the initial contour equation in diagrammatic form.
\item Choose a Keldysh component.
\item Represent the various terms using diagrammatic rules.
\item Convert the result into a real-time expression.
\end{enumerate}

We will use as an example the equation
\begin{equation} \label{eq:4_2_diag_expansion_example}
\con{D}(z_a,z_d) = \int_\gamma \dif z_b \dif z_c \,  \con{\bar{D}}(z_a,z_b,z_c,z_d).
\end{equation}
The steps are performed as follows.

\begin{enumerate}
\item The contour functions are depicted by collections of vertices representing their arguments. We use filled circles for external arguments and empty circles for internal arguments. For \Eq{eq:4_2_diag_expansion_example} we draw
\begin{equation} \label{eq:4_diag_expansion_example_1}
\con{D}_{ad} = \int \con{\bar{D}}_{a\bar{b}\bar{c}d} \quad \rightarrow \quad \bigdiagramcenter{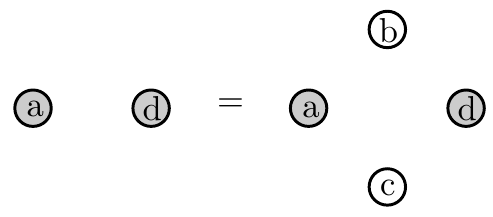}
\end{equation}

\item We choose the Keldysh component by drawing a contour through the external vertices to determine their ordering:
\begin{equation}
\bigdiagramcenter{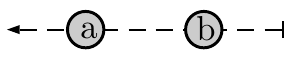} \quad : \quad \text{$z_a$ comes after $z_b$ on the contour.}
\end{equation}
The contour must be drawn in the same way on both sides of the equation. Choosing the component $D^{21}_{ad} = D^{\dch\ach}_{ad}$ we obtain from \Eq{eq:4_diag_expansion_example_1}
\begin{equation} \label{eq:4_diag_expansion_example_2}
\bigdiagramcenter{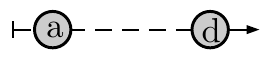} = \bigdiagramcenter{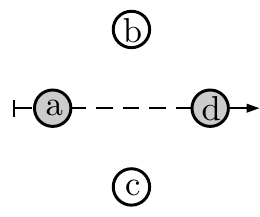}
\end{equation}

\item Each internal vertex is assigned to a retarded set with an external vertex, marked by a double circle, as the top argument. We denote a retarded set diagrammatically by circling the vertices in it:
\begin{equation}
\bigdiagramcenter{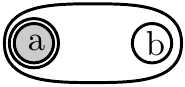} \quad : \quad R(\ach, \bch) \quad \text{($t_b$ is retarded with respect to $t_a$)}
\end{equation}
If assigning the internal vertices to retarded sets can be done in multiple ways, each option generates a separate diagram. Thus the right hand side of \Eq{eq:4_diag_expansion_example_2} becomes
\begin{equation}
\bigdiagramcenter{4_contour_ordered_example_b} = \bigdiagramcenter{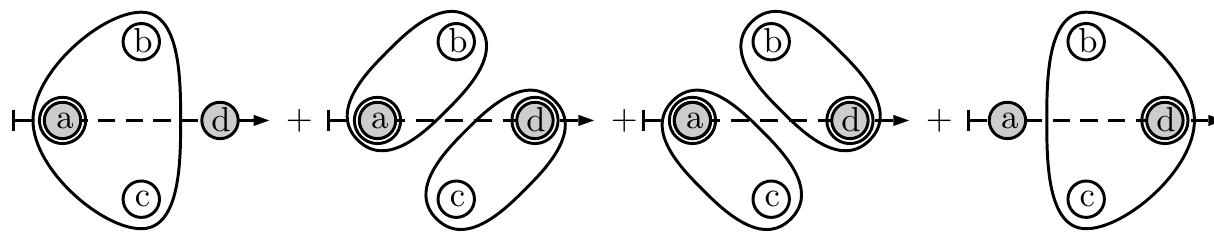}
\end{equation}

\item
Each diagram is converted to a single retarded composition using two rules:
\begin{enumerate}
\item For every circled set of vertices, one obtains a retarded set:
\begin{equation}
\bigdiagramcenter{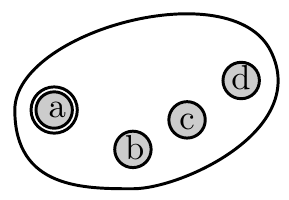} \quad \rightarrow \quad R(\check{a},\check{b}\check{c}\check{d})
\end{equation}
\item The retarded sets are ordered according to their order on the contour in the diagram.
\begin{equation}
\bigdiagramcenter{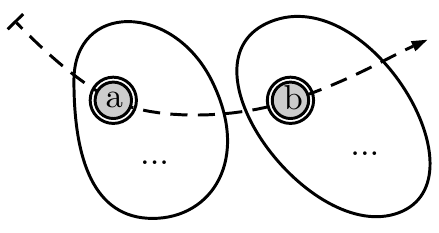} \quad \rightarrow \quad R(\check{b},\cdots)R(\check{a},\cdots) \quad (z_b > z_a)
\end{equation}
\end{enumerate}

For the example case we then obtain
\begin{equation} \begin{split}
&\bigdiagramcenter{4_contour_ordered_example_a} = \bigdiagramcenter{4_retarded_expansion_2} \\
&\qquad D^{\check{d}\check{a}}_{ad} \;\qquad = \quad \bar{D}^{\check{d}R(\check{a},\check{b}\check{c})}_{a\bar{b}\bar{c}d} \;\quad\quad + \quad \bar{D}^{R(\check{d},\check{c})R(\check{a},\check{b})}_{a\bar{b}\bar{c}d} \;\;\; + \quad \bar{D}^{R(\check{d},\check{b})R(\check{a},\check{c})}_{a\bar{b}\bar{c}d} \;\;\; + \quad \bar{D}^{R(\check{d},\check{b}\check{c})\check{a}}_{a\bar{b}\bar{c}d},
\end{split} \end{equation}
where we integrate over all barred arguments. Writing the argument positions explicitly we obtain the result we obtained algebraically in \Eq{eq:4_lesser_component}.
\end{enumerate}
The diagrammatic procedure described above can be applied to obtain any Keldysh components of a general $n$-point Keldysh function.

\subsection{Obtaining Retarded Compositions}

In the previous section we discussed a diagrammatic recipe for obtaining real-time Keldysh components, the result being expressed in terms of retarded compositions. It is possible to generalize the recipe to directly obtain these retarded compositions~\cite{Danielewicz1990}. Diagrammatically, this fits elegantly with what has been presented so far. Here we introduce the diagrammatic recipe, and refer to \ref{app:expansion_for_retarded_compositions} for a derivation. 

As an example, we consider a case with five external variables and two internal ones:
\begin{equation} \label{eq:4_retarded_diagrammatics_example}
\con{E}(z_a,z_b,z_c,z_d,z_e) = \int_\gamma \dif z_f \dif z_g \, \con{\bar{E}}(z_a,z_b,z_c,z_d,z_e,z_f,z_g),
\end{equation}
and calculate the retarded composition $E^{R(1,2)R(3,45)} = E^{R(\check{a},\check{b})R(\check{c},\check{d}\check{e})}$.
We perform the same four steps as for Keldysh components:
\begin{enumerate}
\item
The diagrammatic equation for \Eq{eq:4_retarded_diagrammatics_example} is
\begin{equation}
\bigdiagram{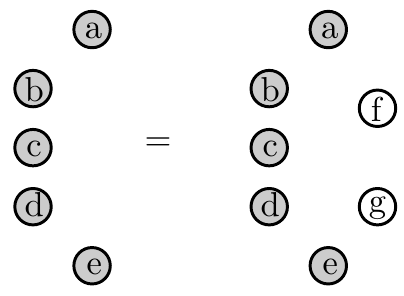}
\end{equation}

\item
To choose the desired composition $E^{R(\check{a},\check{b})R(\check{c},\check{d}\check{e})}$ we encircle the retarded sets $R(\check{a},\check{b})$ and $R(\check{c},\check{d}\check{e})$ separately, and order the sets by drawing a contour through $a$ and $c$:
\begin{equation}
\bigdiagram{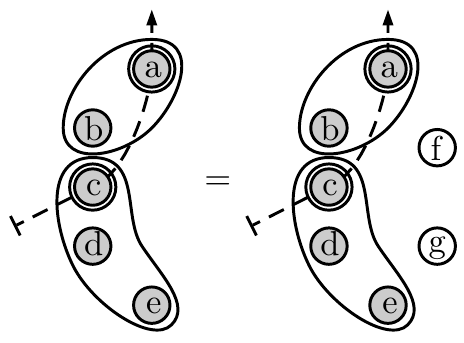}
\end{equation}

\item
The internal vertices are now assigned to retarded sets. If an external vertex is already within a retarded set, the internal vertices are simply added to the pre-existing set. Each different possibility generates a diagram, and we obtain
\begin{equation}
\bigdiagram{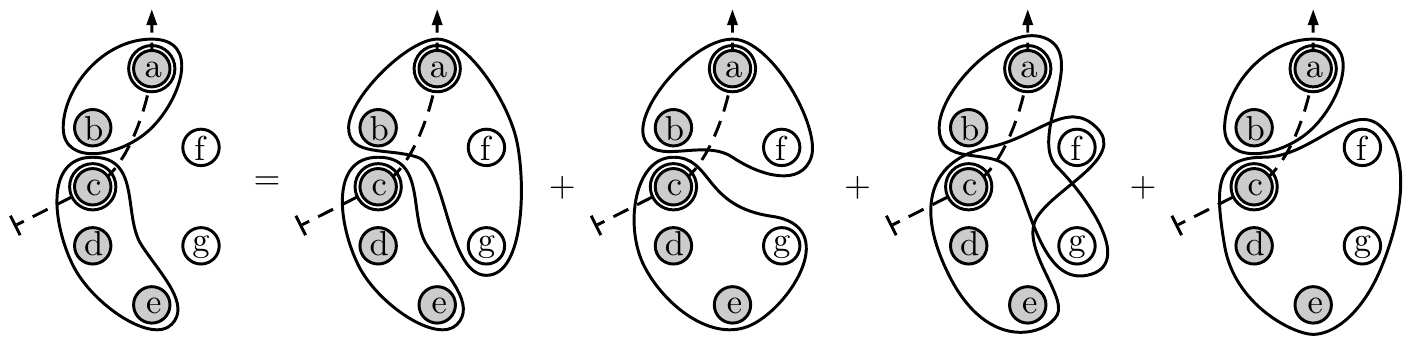}
\end{equation}
\item
We can then directly read off the retarded representation
\begin{equation} \begin{split} \label{eq:4_retarded_example_result}
&E^{R(\ach,\bch)R(\cch,\dch\ech)}_{abcde} \\
&= \int \bar{E}^{R(\ach,\bch\fch\gch)R(\cch,\dch\ech)}_{abcde\bar{f}\bar{g}} + \int \bar{E}^{R(\ach,\bch\fch)R(\cch,\dch\ech\gch)}_{abcde\bar{f}\bar{g}} + \int \bar{E}^{R(\ach,\bch\gch)R(\cch,\dch\ech\fch)}_{abcde\bar{f}\bar{g}} + \int \bar{E}^{R(\ach,\bch)R(\cch,\dch\ech\fch\gch)}_{abcde\bar{f}\bar{g}}.
\end{split} \end{equation}
\end{enumerate}

As seen in the example studied above, a retarded composition is obtained by summing over all different distributions of the internal vertices into retarded sets. In general, if
\begin{equation}
\con{O}(z_\mathcal{E}) = \int_\gamma \dif z_\mathcal{I} \, \con{\bar{O}}(z_\mathcal{N}),
\end{equation}
then a general multi-retarded composition with $H$ retarded sets, as defined in \Eq{eq:4_retarded_definition}, will be given by
\begin{equation} \label{eq:4_retarded_result}
O^{R(\check{h}_1,\check{\mathcal{H}}_1)\cdots R(\check{h}_H,\check{\mathcal{H}}_H)}(t_\mathcal{E}) = \sum_\mathcal{I} \int \dif t_\mathcal{I} \, \bar{O}^{R(\check{h}_1,\check{\mathcal{H}}_1 \cup \check{\mathcal{I}}_1) \cdots R(\check{h}_H,\check{\mathcal{H}}_H \cup \check{\mathcal{I}}_H)}(t_\mathcal{N}),
\end{equation}
where $\{ h_1,\ldots,h_H \}$ and $\mathcal{H}_0, \ldots, \mathcal{H}_H$ constitute a non-overlapping cover of $\Ecal$.
A proof of \Eq{eq:4_retarded_result} is presented in \ref{app:expansion_for_retarded_compositions} (\Eq{eq:4_retarded_result} is equivalent to Eq (4.18) in Danielewicz~\cite{Danielewicz1990}). In our example \Eq{eq:4_retarded_example_result}, $E^{R(\ach,\bch)R(\cch,\dch\ech)}_{abcde}$ corresponds to $h_1 = a, \mathcal{H}_1 = \{b\}$,  and $ h_2 = c$, $\mathcal{H}_2 = \{d,e\}$. The possible internal sets $\mathcal{I}$ that are summed over are
$\{\mathcal{I}_1, \mathcal{I}_2 \} =
\{
\{f,g\}, \emptyset \}$, and
$\{\{f\},\{g\}\}$, and
$\{\{g\}, \{f\} \}$, and
$\{\emptyset, \{f,g\} \}$.
 Note that when the left-hand side of \Eq{eq:4_retarded_result} is a Keldysh component, $\mathcal{H}_i = \emptyset$, the equation reduces to \Eq{eq:4_1_multi_parameter_result}.

\section{The Extended Contour}
\label{sec:5_extended_contour}

\subsection{Matsubara-Restricted Keldysh Functions}

We now generalize the discussion of the previous two sections to the case of the extended contour. For a Keldysh function on the extended contour, $\con{O}(z_\mathcal{N})$ the arguments can take values on the Matsubara branch ($z = t_0 - i t$) as well as on the horizontal Keldysh branches ($z = t_\pm$). Out of the total set of arguments $\Ncal$, we can select a set $\Mcal$ of arguments on the Matsubara branch $\gamma_M$, and $\Kcal$ arguments on the Keldysh branch. Naturally, $\Ncal = \Kcal \cup \Mcal$. Selecting a time $z_m = t_0 - i t_m$ on the Matsubara branch yields a real-time function of $t_m$. For the remaining arguments on the Keldysh contour, the function is still a Keldysh function. Any such function in which we restrict the domain we refer to as an \emph{Matsubara-restricted Keldysh function} (MK-function) $\con{O}^{M(\check{\Mcal})}(t_{\Mcal} \cup z_\Kcal)$, defined by
\begin{equation}
 \con{O}^{M(\check{\Mcal})}(t_{\Mcal} \cup z_\Kcal) =
 \sum_{Q\in S_M} \Theta(t_{Q(\Mcal)})
 \sum_{P\in S_K} \theta(z_{P(\Kcal)}) O^{M(Q(\check{\Mcal})) P(\check{\Kcal})}(t_\Ncal).
\end{equation}
Note that the union set designation $t_{\Mcal} \cup z_\Kcal$ is just for convenience of notation and does not imply that the arguments $t_M$ are earlier in the argument list. We have used the fact that all the Matsubara arguments are later in contour time than any of the arguments on the Keldysh contour. When no argument is on the Matsubara branch, $\mathcal{M} = \emptyset$, we obtain a Keldysh function of a type we considered previously $\con{O}^{\mathcal{M}(\emptyset)}(z_\Ncal) = \con{O}(z_\Ncal)$. The equation above can be rewritten as
\begin{equation}\label{eq:extendedKeldyshFunction}
 \con{O}^{M(\check{\Mcal})}(t_{\Mcal} \cup z_\Kcal) =
 \sum_{P\in S_K} \theta(z_{P(\Kcal)}) O_S^{M(\check{\Mcal}) P(\check{\Kcal})}(t_\Ncal),
\end{equation}
where we have defined the symmetrized Matsubara  function
\begin{equation}
 O_S^{M(\check{\Mcal}) P(\check{\Kcal})}(t_\Ncal) =
 \sum_{Q\in S_M} \Theta(t_{Q(\Mcal)}) O^{M(Q(\check{\Mcal})) P(\check{\Kcal})}(t_\Ncal).
\end{equation}
The symmetrized functions are symmetric under exchanging the order of the Matsubara labels in the super index.

For example, for a four-point Keldysh function $\con{C}(z_a,z_b,z_c,z_d)$, if $z_a,z_c$ are on the Matsubara branch and $z_b,z_d$ on the Keldysh branch, the symmetrized Matsubara function is
\begin{equation}
 O_S^{M(\check{a}\check{c} ) P(\check{b}) P(\check{d})}(t_a,t_b,t_c,t_d) =
 \sum_{Q\in S_2} \Theta(t_{Q(a)} - t_{Q(c)}) O^{M(Q(\check{a}) Q(\check{c})) P(\check{b})P(\check{d})}(t_a,t_b,t_c,t_d).
\end{equation}
The MK function can be written according to \Eq{eq:extendedKeldyshFunction} as
\begin{equation}
 \con{O}^{M(\check{a}\check{c})}(t_a,z_b,t_c,z_d) =
 \sum_{P\in S_2} \theta(z_{P(b)},z_{P(d)}) O_S^{M(\check{a}\check{c}) P(\check{b})P(\check{d})}(t_a,t_b,t_c,t_d).
\end{equation}
In the following, we only work with the symmetrized functions, and will therefore drop the sub-index $S$ from the general expansion \Eq{eq:extendedKeldyshFunction}. Each of the components in this expansion we will still call a Keldysh component.

Since MK functions are Keldysh functions, multiplications and convolutions between MK functions are again MK functions. Also, retarded compositions will appear naturally from integrations, as will be seen below. Retarded compositions of MK functions have the form
\begin{equation}
O^{M(\check{\mathcal{M}})R(\check{h}_1,\check{\mathcal{H}}_1)\cdots R(\check{h}_H,\check{\mathcal{H}}_H)}(t_\mathcal{N}),
\end{equation}
where $\{ h_1,\ldots,h_H \}, \mathcal{M}, \mathcal{H}_0, \ldots, \mathcal{H}_H$ constitute a non-overlapping cover of $\Ncal$.

\subsection{Integrals over the Extended Contour}

Let us now consider an integral over the extended contour:
\begin{equation} \label{eq:5_extended_contour_start}
\con{O}(z_\mathcal{E}) = \int_{\ext{\gamma}} \dif z_\mathcal{I} \, \con{\bar{O}}(z_{\mathcal{N}}).
\end{equation}
Our basic strategy is to split the integrations as $\int_{\ext{\gamma}} = \int_{\gamma_M} + \int_\gamma$ and in each integral replace the integrand with a suitable MK function. This breaks the right-hand side of \Eq{eq:5_extended_contour_start} into terms containing real-time integrals over the Matsubara branch and integrals over the Keldysh contour, for which results were derived previously.

Take for example the equation
\begin{equation}
\con{D}_{ad} = \int_{\ext{\gamma}} \dif z_b \dif z_c \, \con{\bar{D}}_{abcd} = \int_{\gamma'} \con{\bar{D}}_{a\bar{b}\bar{c}d},
\end{equation}
where we introduced the notation that we integrate over barred variables. Suppose now that $a$ and $d$ are on the Keldysh branches. The substitution $\int_{\ext{\gamma}} = \int_{\gamma_M} + \int_\gamma$ generates four terms, in each of which we can replace $\con{\bar{D}}$ by an MK function. We obtain
\begin{equation} \label{eq:4_integral_example}
\con{D}_{ad} = \int_{\gamma'} \con{\bar{D}}_{a\bar{b}\bar{c}d} = \int \con{\bar{D}}^{M(\check{b}\check{c})}_{a\bar{b}\bar{c}d} + \int \con{\bar{D}}^{M(\check{b})}_{a\bar{b}\bar{c}d} + \int \con{\bar{D}}^{M(\check{c})}_{a\bar{b}\bar{c}d} + \int \con{\bar{D}}_{a\bar{b}\bar{c}d}, \quad a,d\in \Kcal,
\end{equation}
where the unspecified integrals are over the Matsubara branch if the barred argument is part of the Matsubara set, and over the Keldysh contour otherwise. Integrals over the Matsubara branch take the form
\begin{equation}
\int_{\gamma_M} \dif z_i \, \con{O}(z_i) = -i \int_0^1 \dif t_i \, \con{O}(t_0 - i t_i) = -i \int_0^1 \dif t_i \, O^{M(\ich)}(t_i).
\end{equation}
Thus the $\int$-sign in \Eq{eq:4_integral_example} contains an implicit factor $(-i)^M$, with $M$ the number of arguments in the Matsubara set. Diagrammatically we denote \Eq{eq:4_integral_example} by
\begin{equation}
\bigdiagram{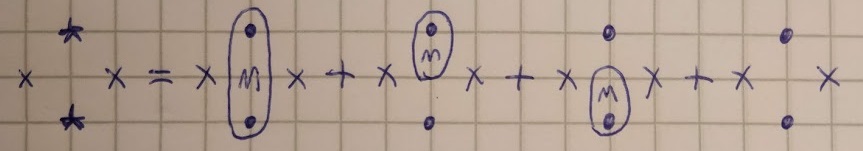}
\end{equation}
where internal vertices that we integrate over the extended contour $\gamma'$ are denoted by squares, and vertex labels in a Matsubara set are circled by a dashed line. Once we choose the component, the retarded set representation can be obtained for each of these terms, by regarding the Matsubara vertices as spectators. This yields, for example for the component $D^{12}_{ad} = D^{\check{a}\check{d}}_{ad}$ and the second term in \Eq{eq:4_integral_example},
\begin{equation}
\int \con{\bar{D}}^{M(\check{b})\check{a}\check{d}}_{a\bar{b}\bar{c}d} = \int \bar{D}^{M(\check{b})R(\check{a},\check{c})\check{d}}_{a\bar{b}\bar{c}d} + \int \bar{D}^{M(\check{b})\check{a}R(\check{d},\check{c})}_{a\bar{b}\bar{c}d},
\end{equation}
which we write diagrammatically as
\begin{equation}
\bigdiagram{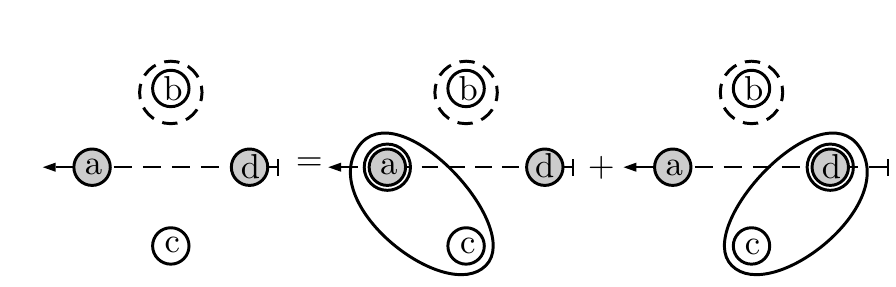}
\end{equation}

The diagrammatic recipe on the extended contour can be described using the same four steps as for the Keldysh contour, given in section \ref{sec:diagrammatic_recipe}. The only difference is, that in step iii) there are additional diagrams resulting from placing internal vertex labels in the Matsubara set.

Real-time components with one or more external arguments on the Matsubara branch can be obtained through the same steps. As an example, $D^{M(\check{a})\check{d}}_{ab}$ is
\begin{equation}
\bigdiagram{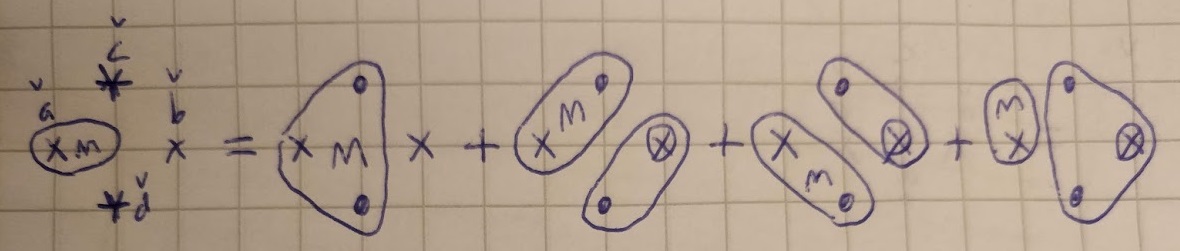}
\end{equation}
which corresponds to the equation
\begin{equation}
D^{M(\check{a})\check{d}}_{ad} = \int {\bar{D}}^{M(\check{a}\check{b}\check{c})\check{d}}_{a\bar{b}\bar{c}d} + \int {\bar{D}}^{M(\check{a}\check{b})R(\check{d},\check{c})}_{a\bar{b}\bar{c}d} + \int {\bar{D}}^{M(\check{a}\check{c})R(\check{d},\check{b})}_{a\bar{b}\bar{c}d} + \int {\bar{D}}^{M(\check{a})R(\check{d},\check{b}\check{c})}_{a\bar{b}\bar{c}d}.
\end{equation}

The general equations on the extended contour are closely related to the equations derived above on the Keldysh contour. For a Keldysh component of an MK function with no arguments on the Matsubara branch we have
\begin{equation} \begin{split} \label{eq:5_extended_contour_result}
O^{\check{e}_1\cdots \check{e}_E}(t_\mathcal{E}) = \sum_\mathcal{I} \int \dif t_\mathcal{I} \, \bar{O}^{M(\check{\mathcal{I}}_0) R(\check{e}_1,\check{\mathcal{I}}_1) \ldots R(\check{e}_E,\check{\mathcal{I}}_E)}(t_\mathcal{N}),
\end{split} \end{equation}
which is identical to the result on the Keldysh contour (\Eq{eq:4_1_multi_parameter_result}), except that the sum also covers the additional set $\mathcal{I}_0$, containing the internal arguments on the Matsubara branch. A general retarded composition is given by a similarly modified version of \Eq{eq:4_retarded_result}:
\begin{equation} \label{eq:5_extended_retarded_result}
O^{M(\check{\mathcal{H}}_0)R(\check{h}_1,\check{\mathcal{H}}_1)\cdots R(\check{h}_H,\check{\mathcal{H}}_H)}(t_\mathcal{E}) = \sum_\mathcal{I} \int \dif t_\mathcal{I} \, \bar{O}^{M(\check{\mathcal{H}}_0 \cup \check{\mathcal{I}}_0)R(\check{h}_1,\check{\mathcal{H}}_1 \cup \check{\mathcal{I}}_1) \cdots R(\check{h}_H,\check{\mathcal{H}}_H \cup \check{\mathcal{I}}_H)}(t_\mathcal{N}).
\end{equation}
The above equation is the most general form of a multi-retarded composition we consider. The \Eq{eq:5_extended_retarded_result} reduces to  \Eq{eq:5_extended_contour_result} by setting $\mathcal{H}_i = \emptyset$ for $i = 0,\ldots,H$, and to \Eq{eq:4_retarded_result} by setting $\mathcal{H}_0 = \mathcal{I}_0 = \emptyset$. Note that if all the external parameters in \Eq{eq:5_extended_retarded_result} are in the Matsubara set $\Ical_0$, meaning that only $\mathcal{H}_0$ is non-empty, then all integrals reduce to integrals over the set $\Ical_0$, i.e. all integrals are over the Matsubara branch.

\section{Deriving Langreth Rules}
\label{sec:6_langreth_rules}

In the previous section we discussed a general integral equation of the form
\begin{equation} \label{eq:6_integral_equation}
\con{O}(z_\mathcal{E}) = \int_{\ext{\gamma}} \dif z_\mathcal{I} \, \con{\bar{O}}(z_{\mathcal{N}}),
\end{equation}
and obtained expressions for the retarded compositions of $\con{O}$ in terms of the retarded compositions of Keldysh components of $\con{\bar{O}}$. In practice $\con{\bar{O}}$ typically has some structure in terms of functions of fewer arguments, being for example a product of Green's functions appearing in a perturbation expansion, so that it takes the form
\begin{equation}\label{eq:multiplication}
\con{\bar{O}}(z_\mathcal{N}) = \prod_i \con{\bar{O}}_i(z_{\mathcal{N}_i}).
\end{equation}
where $\Ncal_i \subset \Ncal$. We call $\con{\bar{O}}_i$ the sub-functions of $\con{\bar{O}}$. By a \emph{Langreth rule} we mean an equation that expresses a retarded composition of $\con{O}$ in terms of retarded compositions of the sub-functions $\con{\bar{O}}_i$.

For example, let us take the convolution:
\begin{equation}
\bigdiagramcenter{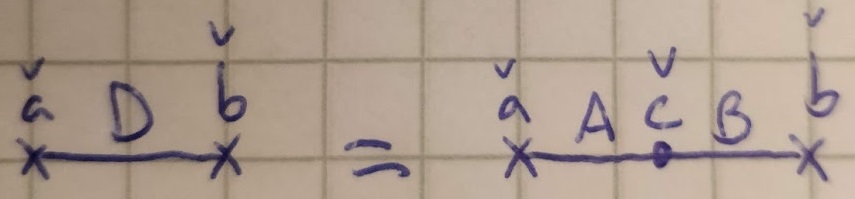} \qquad\qquad \con{D}_{ab} 
 = \int_{\gamma'} \con{A}_{a\bar{c}}\con{B}_{\bar{c}b}.
\end{equation}
Suppose we are interested in the component $D^{12}$. The diagrammatic recipe gives
\begin{equation} \begin{split} \label{eq:6_example_1}
D^{\ach\bch}_{ab} 
&=
\int [\con{A}_{a\bar{c}} \con{B}_{\bar{c}b}]^{M(\check{c})\check{a}\check{b}} + \int [\con{A}_{a\bar{c}} \con{B}_{\bar{c}b}]^{R(\check{a},\check{c})\check{b}} + \int [\con{A}_{a\bar{c}} \con{B}_{\bar{c}b}]^{\check{a}R(\check{b},\check{c})},
\end{split} \end{equation}
where on the second line we have used the háček notation to avoid confusion between argument numbering in the sub-functions and the total function. The recipe, as laid out so far, thus leads to expressions containing retarded compositions of products of sub-functions. The issue faced in this section is how to evaluate such compositions directly in terms of the components of the sub-functions.

Before proceeding further, we will introduce some terminology:
\begin{itemize}

\item An $n$-point subfunction is diagrammatically represented by a polygon of $n$ vertices representing the $n$ arguments. For example, $n=1$ is a dot, $n=2$ is a line, and $n=3$ is a triangle.

\item Two subfunctions sharing the same argument correspond to a diagram in which two polygons share the same vertex.

For example the expression $\con{A}_{abcd} \con{B}_{cdfgh} \con{C}_{ijk} \con{D}_{be} \con{E}_{ef}$ converts to the diagram
\begin{equation}
\bigdiagramcenter{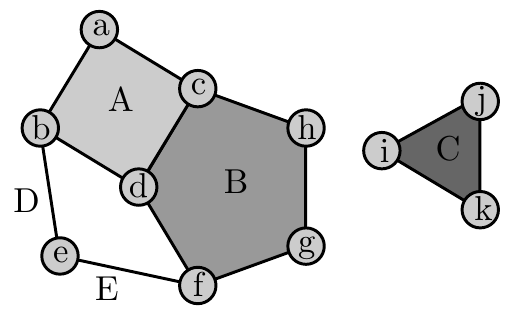}
\end{equation}

\item Two vertices are \emph{directly connected} if they belong to the same polygon, and therefore the corresponding arguments appear in the same sub-function.

\item Two vertices are \emph{connected} if there is a path consisting of polygon edges connecting them. Likewise the corresponding arguments of subfunctions are said to be connected.

\item A set of vertices is \emph{connected}, if each two vertices in the set are either directly connected or connected by an edge path that does not leave the set. If this is not the case, the set of vertices is \emph{disconnected}. The same nomenclature is used for arguments.

\end{itemize}

\subsection{Arguments on the Matsubara branch}

The goal of this section is to express an MK function in terms of its sub-functions. Generally, an MK function $\con{bar{O}}^{M(\check{\Mcal})}(t_\Mcal \cup z_\Kcal)$ for some $\Mcal \subset \Ncal$, can be worked out as follows. For sub-function $\con{\bar{O}}_i$ we define $\Mcal_i = \Mcal \cap \Ncal_i$ and $\Kcal_i = \Ncal_i \setminus \Mcal_i$, one then has
\begin{equation} \begin{split} \label{eq:6_matsubara_sets_general}
\con{bar{O}}^{M(\check{\Mcal})}(t_\Mcal \cup z_\Kcal)
&= \prod_i \con{\bar{O}}^{M(\check{\Mcal}_i)}_i(t_{\Mcal_i} \cup z_{\mathcal{K}_i}).
\end{split} \end{equation}

In practice one can simply move the Matsubara set superscript from the total function to each individual sub-function, as in
\begin{equation} \begin{split}
[\con{A}_{a\bar{c}}\con{B}_{\bar{c}b}]^{M(\check{c})} = \con{A}^{M(\check{c})}_{a\bar{c}} \con{B}^{M(\check{c})}_{\bar{c}b},
\end{split} \end{equation}
while dropping from the Matsubara set any arguments that do not appear in the relevant sub-function. Note that since $\con{A}^{M(\check{c})}(z_a,t_c)$ and $\con{B}^{M(\check{c})}(t_c,z_b)$ are Keldysh functions of a single argument, they are equal to the MK-components $A^{M(\cch)\ach}(t_a,t_c)$ and $B^{M(\cch)\bch}(t_c,t_b)$ respectively. For example, if we take a product with vertex structure
\begin{equation}
\bigdiagramcenter{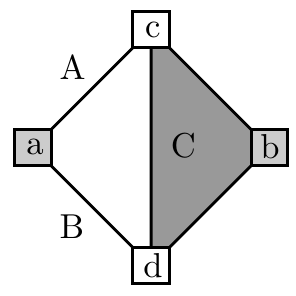} \qquad\qquad \int_{\gamma'} \con{A}_{a\bar{c}} \con{B}_{a\bar{d}} \con{C}_{\bar{d}b\bar{c}},
\end{equation}
and calculate the MK function $[\cdots]^{M(\check{c})}$, the integrand is
\begin{equation}
[\con{A}_{ac} \con{B}_{ad} \con{C}_{dbc}]^{M(\check{c})} = A^{M(\check{c})\check{a}}_{ac} \con{B}_{ad} \con{C}^{M(\check{c})}_{dbc}.
\end{equation}
Diagrammatically the above equation corresponds to splitting the diagram into two separate pieces:
\begin{equation} \label{eq:6_hedin_cutting_example}
\bigdiagramcenter{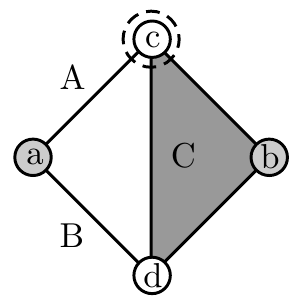} = \bigdiagramcenter{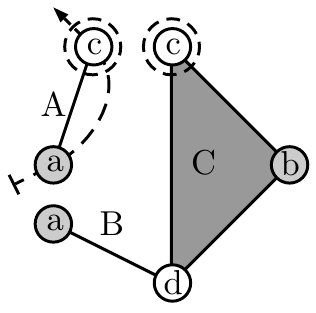}
\end{equation}

Thus any sub-function with some of its arguments in a Matsubara set, reduces to the corresponding MK function. If no more than one argument is outside a Matsubara set, the sub-function reduces immediately to a real-time MK component. The real-time components can then be separated from the diagram, leaving behind a simpler diagram on the Keldysh contour. For example, in \Eq{eq:6_hedin_cutting_example}, $A$ can be separated in the integrand, and what remains is essentially a convolution between two-point functions, since $\con{C}_{dbc}^{M(\check{c})}(z_a,z_b,t_c)$ is a two-point Keldysh function with an additional real-time argument. The problem on the extended contour is then essentially reduced to the problem on the Keldysh contour, which we will discuss in the next sections.

\subsection{Arguments on the Keldysh branch}

We take now all the arguments on the Keldysh contour, and consider Keldysh components of products. As an example we consider $C_{abc}^{\check{a}\check{b}\check{c}} = [\con{A}_{ac} \con{B}_{cb}]^{\check{a}\check{b}\check{c}}$, which is, by definition, given by the real-time function that describes the product $\con{A}_{ac} \con{B}_{cb}$ in the $z_a > z_b > z_c$ subspace. In this subspace the product takes the form
\begin{equation}
\con{A}_{ac} \con{B}_{cb} = [\underbrace{\theta_{ac}}_{=1} A^{\check{a}\check{c}}_{ac} + \underbrace{\theta_{ca}}_{=0} A^{\check{c}\check{a}}_{ac}] [\underbrace{\theta_{cb}}_{=0} B^{\check{c}\check{b}}_{cb} + \underbrace{\theta_{bc}}_{=1} B^{\check{b}\check{c}}_{cb}] = A^{\check{a}\check{c}}_{ac} B^{\check{b}\check{c}}_{cb},
\end{equation}
and therefore
\begin{equation} \label{eq:6_keldysh_example_1}
C_{abc}^{\check{a}\check{b}\check{c}} = A^{\check{a}\check{c}}_{ac} B^{\check{b}\check{c}}_{cb}.
\end{equation}

The general case can be handled analogously. Writing the right-hand side of \Eq{eq:multiplication} as a Keldysh sum we obtain
\begin{equation}
\con{O}(z_\Ncal) = \prod_i \sum_{P_i \in S_{N_i}} \theta_{P_i(\Ncal_i)} \bar{O}^{P_i(\check{\Ncal_i})}_i(t_{\Ncal_i}).
\end{equation}
Suppose we wish to obtain the Keldysh component $\con{O}^{P(\check{\Ncal})}(t_\Ncal)$. It can be obtained from the above equation by choosing $z_{P(n_1)} > z_{P(n_2)} > \ldots > z_{P(n_N)}$. The left hand side then reduces to the desired Keldysh component. On the right-hand side, most of the step-function vanish. The only step functions that remain, and attain the value $1$, are those for which $P_i(\Ncal_i) = P(\Ncal)\setminus \Ncal_i^c $, where we defined the complementary set $\Ncal_i^c = \Ncal \setminus \Ncal_i$. The permutation $P$ therefore picks out a unique permutation $P_i$ for each set $\Ncal_i$, and we have
\begin{equation} \label{eq:componentCalculus}
\con{O}^{P(\check{\Ncal})}(t_\Ncal) =
\prod_i \bar{O}^{P_i(\check{\Ncal}_i)}_i(t_{\Ncal_i}), \quad  P_i(\Ncal_i) = P(\Ncal)\setminus \Ncal_i^c.
\end{equation}
As an example, let us now take the term $C_{abc}^{\check{b}\check{a}\check{c}} = [\con{A}_{ac} \con{B}_{cb}]^{\check{b}\check{a}\check{c}}$. In this case, $P(\{ a,b,c\}) = \{b,a,c\}$, $\Ncal_1 = \{a,c\}$, $\Ncal_1^c = \{b\}$, and $\Ncal_2 = \{c,b\}$, $\Ncal_2^c = \{a\}$. Therefore, $P_1(\Ncal_1) = \{ b,a,c\} \setminus \{ b\} = \{ a,c\}$, and likewise $P_2(\Ncal_2) = \{b,c\}$. In this particular case, $P_1$ is the identity permutation and $P_2$ is the transposition. Equation \eqref{eq:componentCalculus} applied to our example yields
\begin{equation} \label{eq:6_keldysh_example_2}
C_{abc}^{\check{b}\check{a}\check{c}} = A^{\check{a}\check{c}}_{ac} B^{\check{b}\check{c}}_{cb}.
\end{equation}
In practice, only those components of the sub-functions remain, in which the argument labels are in the same relative order as they are for the full product function. Thus the correct result can be obtained simply by moving the full string of super indices to each subfunction, and then removing all the labels that are not part of the argument list of the particular subfunction.

Retarded compositions of products can be worked out by writing the compositions in terms of Keldysh components by using \Eq{eq:4_retarded_definition}. For example, the integrand in the second term in \Eq{eq:6_example_1} can be written as
\begin{equation} \begin{split}
[\con{A}_{a {c}} \con{B}_{{c}b}]^{R(\check{a},\check{c})\check{b}} &= \Theta_{ac} \left( [\con{A}_{a{c}} \con{B}_{{c}b}]^{\check{a}\check{c}\check{b}} - [\con{A}_{a{c}} \con{B}_{{c}b}]^{\check{c}\check{a}\check{b}} \right) \\
&= \Theta_{ac} \left( A^{\check{a}\check{c}}_{a{c}} - A^{\check{c}\check{a}}_{a{c}}\right) B^{\check{c}\check{b}}_{{c}b} \\
&= A^{R(\check{a},\check{c})}_{a{c}} B^{\check{c}\check{b}}_{{c}b}.
\end{split} \end{equation}
The third term in \Eq{eq:6_example_1} can be handled likewise. We can then express the component $D^{12}_{ab}$ in \Eq{eq:6_example_1} as
\begin{equation} \begin{split} \label{eq:6_example_1_result}
D^{12}_{ab} &=
\int[\con{A}_{a\bar{c}} \con{B}_{\bar{c}b}]^{M(\check{c})\check{a}\check{b}} + \int[\con{A}_{a\bar{c}} \con{B}_{\bar{c}b}]^{R(\check{a},\check{c})\check{b}} + \int[\con{A}_{a\bar{c}} \con{B}_{\bar{c}b}]^{\check{a}R(\check{b},\check{c})} \\
&=
\int A^{M(\check{c})\check{a}}_{a\bar{c}} B^{M(\check{c})\check{b}}_{\bar{c}b} + \int A^{R(\check{a},\check{c})}_{a\bar{c}} B^{\check{c}\check{b}}_{\bar{c}b} + \int A^{\check{a}\check{c}}_{a\bar{c}} B^{R(\check{c},\check{b})}_{\bar{c}b}.
\end{split} \end{equation}

The above equation is an example of a known Langreth rule, \Eq{eq:2_Langreth_rule_M_<>}.   The procedure laid out above can in principle be used to derive Langreth rules from arbitrary equations. Below we will derive additional rules that make these calculations less cumbersome. 

\subsection{The Vanishing of Retarded Compositions on Disconnected Sets}

Specifying the structure of a Keldysh function in terms of sub-functions typically introduces new symmetries on top of those of a general Keldysh function, since there are permutations of super indices that do not change the components of the subfunctions. It is an advantage of the retarded set representation that these symmetries can be employed to directly discard certain terms of the representation. As we will show, this feature is consequence of the definition of retarded compositions in terms of nested commutators. We will therefore begin by considering commutator expressions.
%
%
%
%
%

In the previous section we noted that for our example two different permutations $P(\Ncal)$ give the same result, see equations \Eq{eq:6_keldysh_example_2} and \Eq{eq:6_keldysh_example_1}:
\begin{equation} \begin{split} \label{eq:exampleOrdering}
[\con{A}_{ac}\con{B}_{cb}]^{\check{a}\check{b}\check{c}} &= A^{\check{a}\check{c}}_{ac} B^{\check{b}\check{c}}_{cb} \\
[\con{A}_{ac}\con{B}_{cb}]^{\check{b}\check{a}\check{c}} &= A^{\check{a}\check{c}}_{ac} B^{\check{b}\check{c}}_{cb}.
\end{split} \end{equation}
Other permutations of the superindices, like $acb$ and $bca$, will not produce the same result.
The situation can be illustrated by the following diagram, in which higher points on the vertical axis denote later contour times:
\begin{equation}
\bigdiagram{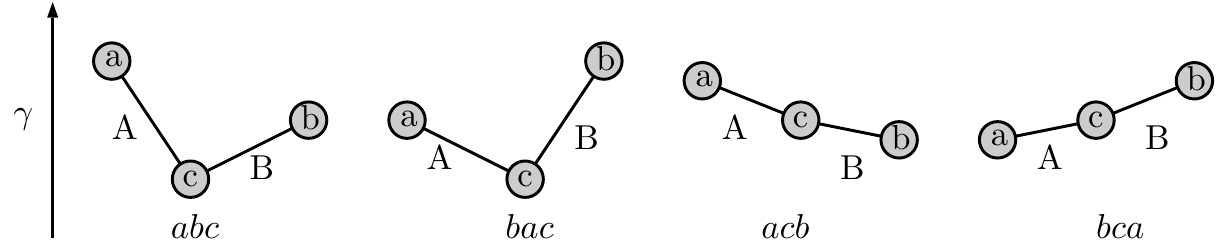}
\end{equation}
For the permutations $abc$ and $bac$, the order within sub-functions $A$ and $B$ remains the same when the order of $a$ and $b$ is exchanged. On the other hand, exchanging $a$ and $b$ by going from $abc$ to $bca$ does change the internal order of $A$ and $B$, as the ordering with respect to $c$ is changed. Thus the symmetry appears only when $a$ and $b$ are neighbouring in the contour order. Furthermore, if there was a sub-function connecting $a$ and $b$, its internal order would be reversed by any exchange of $a$ and $b$, breaking the symmetry.

Generally, if for a function $\con{O}(z_\Ncal)$ the arguments $z_a$ and $z_b$ are next to each other on the contour, exchanging their positions only changes their relative contour order. Furthermore, if the vertices $a$ and $b$ are not directly connected, it follows that in \Eq{eq:componentCalculus} none of the sets $\Ncal_i$ contains both $a$ and $b$, and consequently changing their relative order does not change the order in any $\Ncal_i$. We can therefore state a rule:
\begin{equation}
O^{\check{X} \check{a}\check{b} \check{Y}} = O^{\check{X} \check{b}\check{a} \check{Y}},\quad \text{when $a$ and $b$ are not directly connected},
\end{equation}
where $\check{X}$ and $\check{Y}$ are strings of the remaining superindices. This rule can be used repeatedly, so that for example
\begin{equation} \label{eq:6_vanishing_rule_2}
O^{\check{X} \check{a} \check{c} \check{b} \check{Y}} = O^{\check{X} \check{c} \check{a} \check{b} \check{Y}} = O^{\check{X} \check{c} \check{b} \check{a} \check{Y}},\quad \text{when $a$ is not directly connected to either $b$ or $c$.}
\end{equation}
Note that $b$ and $c$ may be directly connected, since their relative order is not changed in \Eq{eq:6_vanishing_rule_2}. Continuing in this way, one arrives at a general rule for a string of arguments $a_1 \ldots a_A $, or a linear combinations $Z$ of such strings:
\begin{equation}
O^{\check{X} [\check{Z},\check{b}] \check{Y}} =
O^{\check{X} \check{Z}\check{b} \check{Y}} -
O^{\check{X} \check{b}\check{Z} \check{Y}}
= 0,\quad \text{when $b$ is not directly connected to any  $a_i$}.
\end{equation}
%
If we choose $Z$ to be the nested commutator $Z = [a_1,\ldots,a_{A-1}]$ and we denote $b=a_A$, we obtain
%
\begin{equation} \begin{split} \label{eq:6_nested_commutator_pre_rule}
O^{\check{X} [\check{a}_1,\ldots,\check{a}_A] \check{Y}} = 0,\quad &\text{ when $a_A$ is not directly connected to any other $a_i$}
\end{split} \end{equation}
Furthermore, if we pick some argument $a_k$ from the nested commutator, and write $O^{\check{X} [\check{a}_1,\ldots,\check{a}_A] \check{Y}} = O^{\check{X} [[\check{a}_1,\ldots,a_k],\ldots,a_A] \check{Y}}$, when the outermost nested commutator is expanded, each of the resulting terms will vanish according to rule \eqref{eq:6_nested_commutator_pre_rule}, unless $a_k$ is directly connected to some argument $a_l$ with $l < k$. This is true for any $a_k$, which allows us to state a more general rule:
\begin{equation} \begin{split} \label{eq:6_nested_commutator_rule}
\bar{O}^{\check{X} [\check{a}_1,\ldots,\check{a}_A] \check{Y}} = 0,\quad &\text{unless every argument $a_k$ is directly connected} \\ &\text{to at least one argument to its left in the commutator.}
\end{split} \end{equation}
Thus every argument in a nested commutator must be directly connected to an argument to its left, and this argument must in turn be directly connected to another, until the leftmost argument $a_1$ is reached. Therefore rule \eqref{eq:6_nested_commutator_rule} requires that each argument is connected to the leftmost argument by a chain of direct connections that links to the left in the commutator.

We are now in position to work out the condition for a retarded composition to vanish. In the definition of a retarded composition
\begin{equation} \label{eq:6_retarded_definition}
\bar{O}^{\check{X} R(\check{a}_1,\check{a}_2 \cdots \check{a}_A) \check{Y}}(t_\Ncal) = \sum_{P\in N_{A-1}} \Theta_{a_1 a_{P(2)} \cdots a_{P(A)}} \bar{O}^{\check{X} [\check{a}_1,\check{a}_{P(2)},\ldots,\check{a}_{P(A)}] \check{Y}}(t_\mathcal{N}),
\end{equation}
a sum is taken over every permutation of the arguments in the nested commutator, apart from the leftmost one. Therefore, if every argument is connected to the leftmost argument $a_1$ by some chain of direct connections, there will always be at least one term in the sum such that each argument is directly connected to an argument to its left. The retarded composition vanishes, when there is at least one argument in the nested commutator that is not connected to $a_1$, i.e. when the retarded set $\Acal = \{\check{a}_1,\check{a}_2 \cdots \check{a}_A\}$ is disconnected. We therefore have the rule
\begin{equation} \begin{split} \label{eq:6_retarded_set_rule}
\text{A retarded composition vanishes, if it contains a disconnected retarded set}.
\end{split} \end{equation}
As an example, let us take a chain convolution of three functions
\begin{equation}
\bigdiagram{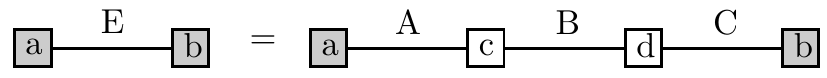} \qquad \con{E}_{ab} = \int_{\gamma'} \con{\bar{E}}_{ab\bar{c}\bar{d}} = \int_{\gamma'} \con{A}_{a\bar{c}} \con{B}_{\bar{c}\bar{d}} \con{C}_{\bar{d}b},
\end{equation}
that appears for example in the Dyson equation. For the component $E^{R(\check{a},\check{b})}_{ab}$ we find the representation
\begin{equation} \label{eq:6_vanishing_sets_example}
\bigdiagram{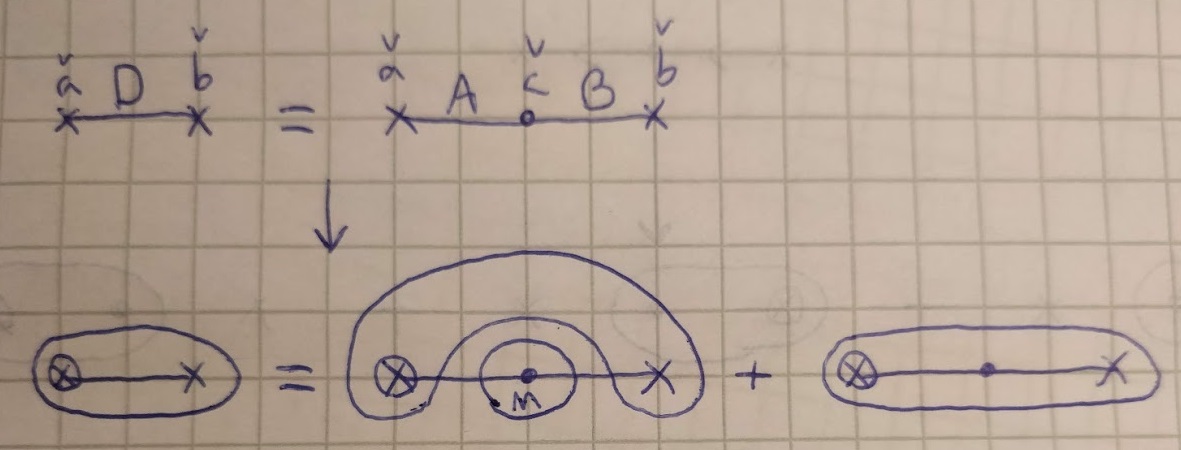}
\end{equation}
Here the first three terms vanish, since $a$ and $b$ are not connected inside the retarded set. Thus a retarded composition of a chain convolution, no matter the length, will never contain any MK components, as it is not possible to place any Matsubara sets without disconnecting the retarded set. From \Eq{eq:6_vanishing_sets_example} we are then left with
\begin{equation} \label{eq:6_vanishing_sets_example_2}
E^{R(\check{a},\check{b})}_{ab} = \int [ \con{A}_{a\bar{c}} \con{B}_{\bar{c}\bar{d}} \con{C}_{\bar{d}b} ]^{R(\ach,\bch\cch\dch)}.
\end{equation}
Expanding the right-hand side of \Eq{eq:6_vanishing_sets_example_2} using \Eq{eq:6_retarded_definition} sums over the six different permutations of $b$, $c$ and $d$. However, out of the six resulting nested commutator expressions five are seen to give zero by applying the rule \eqref{eq:6_nested_commutator_rule}. Since the vertices are connected in a chain, there's only one permutation in which each vertex is directly connected to another to its left, that corresponding to the commutator $[\ach,\cch,\dch,\bch]$, and thus we obtain from \Eq{eq:6_vanishing_sets_example_2} simply
\begin{equation} \label{eq:6_vanishing_sets_example_3}
E^{R(\check{a},\check{b})}_{ab} = \int \Theta_{acdb} [\con{A}_{a\bar{c}} \con{B}_{\bar{c}\bar{d}} \con{C}_{\bar{d}b} ]^{[\ach,\cch,\dch,\bch]}.
\end{equation}
This expression will be simplified further by rules that we will derive later.

\subsection{Separating Retarded Sets}

A diagram in a retarded set representation typically consists of sub-functions in multiple retarded sets that are connected via other sub-functions. Often these diagrams can be split into multiple pieces, so that each retarded set can be handled individually. This is based on the fact that the retarded sets are ordered with respect to each other. If each of the vertices of a sub-function is in a different retarded set, the contour ordering of its vertices is determined, and it will reduce to a single MK component. For example we have
\begin{equation}
[ \con{A}_{abcd} \con{B}_{ghi} \con{C}_{dfg} \con{D}_{ef} ]^{R(\gch,\hch\ich)R(\cch,\ach\bch\dch)R(\ech,\fch)} =  A_{abcd}^{R(\cch,\ach\bch\dch)} B_{ghi}^{R(\gch,\hch\ich)} C_{dfg}^{\gch\dch\fch} D_{ef}^{R(\ech,\fch)}.
\end{equation}
which corresponds to the graph
\begin{equation}
\bigdiagramcenter{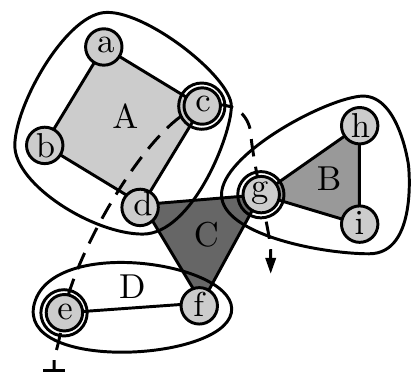} = \bigdiagramcenter{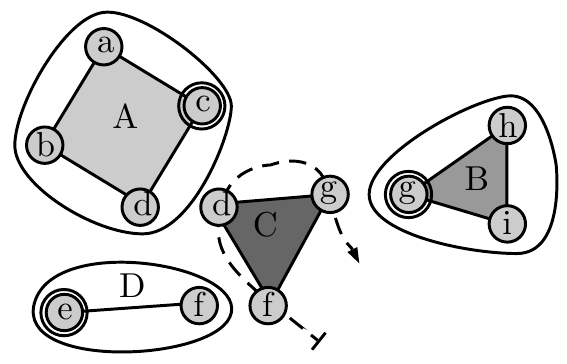},
\end{equation}

To consider this procedure in more detail, let us return to the chain convolution
\begin{equation}
\bigdiagram{6_three_chain_convolution} \qquad \con{E}_{ab} = \int_{\gamma'} \con{\bar{E}}_{ab\bar{c}\bar{d}} = \int_{\gamma'} \con{A}_{a\bar{c}} \con{B}_{\bar{c}\bar{d}} \con{C}_{\bar{d}b}.
\end{equation}
Choosing now to obtain the greater component $E_{ab}^{12}$, we get
\begin{equation} \begin{split} \label{eq:6_example_2}
E^{\ach\bch}_{ab} &= \int \bar{E}^{R(\ach,\cch)R(\bch,\dch)}_{ab\bar{c}\bar{d}} + \int \bar{E}^{R(\ach,\dch)R(\bch,\cch)}_{ab\bar{c}\bar{d}} + \int \bar{E}^{R(\ach,\cch\dch)\bch}_{ab\bar{c}\bar{d}} + \int \bar{E}^{\ach R(\bch,\cch\dch)}_{ab\bar{c}\bar{d}}.
\end{split} \end{equation}

Let us start with the first term with superindex $R(\check{a},\check{c})R(\check{b},\check{d})$. If we expand the retarded sets in terms of commutators, we obtain terms in which the order between $a$ and $c$, as well as $b$ and $d$, varies term by term. However, in every term $c$ is later than $d$ in contour order. Consequently $\con{B}_{cd}$ reduces to the same Keldysh-component $B^{\check{c}\check{d}}_{cd}$ in every term, and we can pull it out of the brackets as a common factor:
\begin{equation} \begin{split} \label{eq:6_separated_ordered_piece}
[\con{A}_{ac} \con{B}_{cd} \con{C}_{db}]^{R(\check{a},\check{c})R(\check{b},\check{d})} &= [\con{A}_{ac}\con{C}_{db}]^{R(\check{a},\check{c})R(\check{b},\check{d})} B^{\check{c}\check{d}}_{cd}.
\end{split} \end{equation}
The crucial point is that all the arguments of function $B$ are in different retarded sets. 
%
%
If we expand the retarded set $R(\check{a},\check{c})$ we obtain
\begin{equation} \begin{split}
[\con{A}_{ac}\con{C}_{db}]^{R(\check{a},\check{c})R(\check{b},\check{d})} &= \theta_{ac} \Big( [\con{A}_{ac}\con{C}_{db}]^{\check{a}\check{c}R(\check{b},\check{d})} - [\con{A}_{ac}\con{C}_{db}]^{\check{c}\check{a}R(\check{b},\check{d})} \Big),
\end{split} \end{equation}
where in both of the terms the contour order of $a$ and $c$ is fixed. Therefore we can now place $A$ in front of the brackets to obtain
\begin{equation} \begin{split} \label{eq:6_separated_retarded_sets}
\theta_{ac} \Big( [\con{A}_{ac}\con{C}_{db}]^{\check{a}\check{c}R(\check{b},\check{d})} - [\con{A}_{ac}\con{C}_{db}]^{\check{c}\check{a}R(\check{b},\check{d})} \big)
&= \theta_{ac} \Big( A^{\ach\cch}_{ac} C_{db}^{R(\check{b},\check{d})} - A^{\cch\ach}_{ac} C_{db}^{R(\check{b},\check{d})} \Big).
\end{split} \end{equation}
We can now factor out $C_{db}^{R(\check{b},\check{d})}$ and obtain
\begin{equation} \begin{split}
[\con{A}_{ac}\con{C}_{db}]^{R(\check{a},\check{c})R(\check{b},\check{d})} 
&= A^{R(\check{a},\check{c})}_{ac} C^{R(\check{b},\check{d})}_{db}.
\end{split} \end{equation}
This result can be represented graphically as
\begin{equation}
\bigdiagram{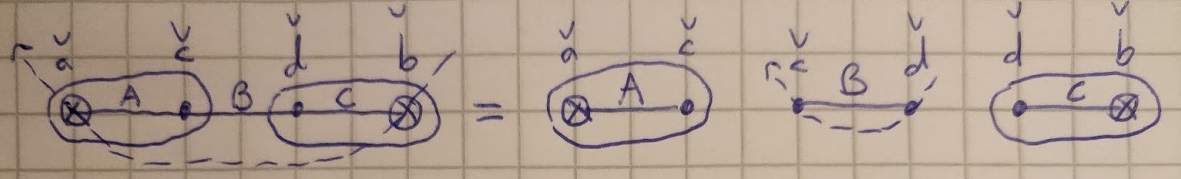}.
\end{equation}
This is a specific instance of a more general rule.
As was discussed, if there is a connecting piece that has multiple vertices inside some retarded sets, it cannot be completely split from a diagram. However, if the connecting piece has a single vertex inside some other retarded sets, those sets can still be split off. For example we have the equation
\begin{equation}
[ \con{A}_{abcd} \con{B}_{ghi} \con{C}_{dfg} \con{D}_{ef} ]^{R(\cch,\ach\bch\dch)R(\ech,\fch\gch\hch\ich)} =  A_{abcd}^{R(\cch,\ach\bch\dch)} [ \con{B}_{ghi} \con{C}_{dfg} \con{D}_{ef}]^{\dch R(\ech,\fch\gch\hch\ich)},
\end{equation}
which corresponds to the graph
\begin{equation}
\bigdiagramcenter{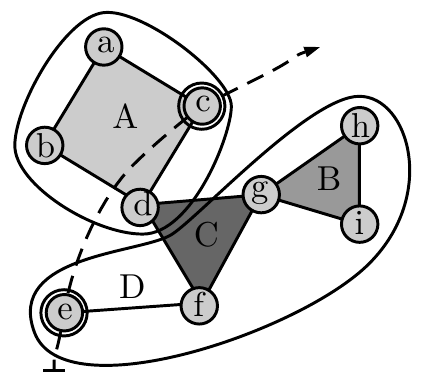} = \bigdiagramcenter{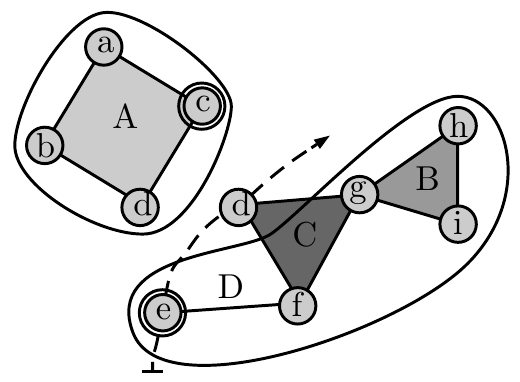}.
\end{equation}

Diagrammatically the above considerations can be condensed to the single statement, that one is allowed to split the diagram by splitting vertices through the process
\begin{equation} \label{eq:6_splitting_vertices}
\bigdiagramcenter{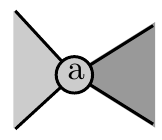} \rightarrow \bigdiagramcenter{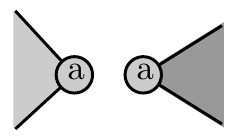}
\end{equation}
whenever this can be done without altering the retarded sets. 
Note that Matsubara sets can be split freely (this statement is merely a diagrammatic equivalent of \Eq{eq:6_matsubara_sets_general}).

To demonstrate this rule, let us now return to \Eq{eq:6_example_2}, the right hand side of which can be written diagrammatically as
\begin{equation} \begin{split}
&\bigdiagramcenter{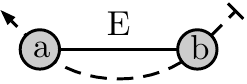} \\
&= \bigdiagramcenter{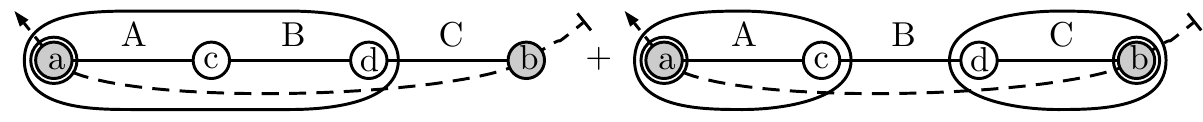} \\
&+ \bigdiagramcenter{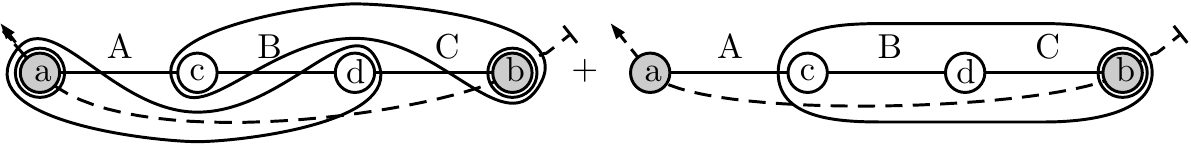}
\end{split} \end{equation}
Note that the third diagram vanishes due to disconnected retarded sets, as per rule \eqref{eq:6_retarded_set_rule}. After using \Eq{eq:6_splitting_vertices} to split the vertices on the edges of retarded sets, we obtain
\begin{equation} \begin{split} \label{eq:6_cutting_example_result}
&\bigdiagramcenter{6_three_chain_convolution_integrand}
= \bigdiagramcenter{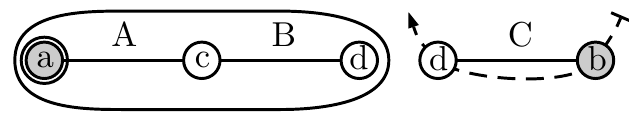} \\
&+ \bigdiagramcenter{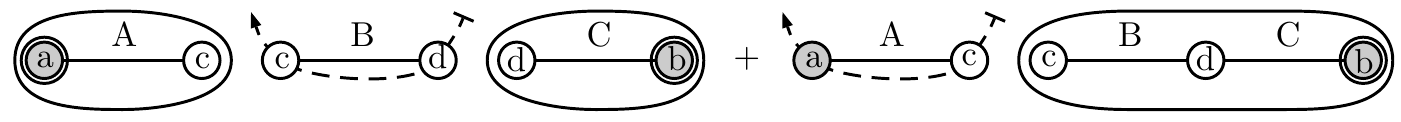}
\end{split} \end{equation}
Converting \Eq{eq:6_cutting_example_result} back into a mathematical expression now yields
\begin{equation} \begin{split} \label{eq:6_example_2_cut}
E^{\ach\bch}_{ab} &=
\int[\con{A}_{a\bar{c}} \con{B}_{\bar{c}\bar{d}}]^{R(\check{a},\check{c}\check{d})} C^{\check{d}\check{b}}_{\bar{d}b} +
\int A^{R(\check{a},\check{c})}_{a\bar{c}} B^{\check{c}\check{d}}_{\bar{c}\bar{d}} C_{\bar{d}b}^{R(\check{b},\check{d})} +
\int A^{\check{a}\check{c}}_{a\bar{c}} [ \con{B}_{\bar{c}\bar{d}} \con{C}_{\bar{d}b}]^{R(\check{b},\check{c}\check{d})}.
\end{split} \end{equation}
However, the square bracketed expressions can not be split further, as cutting between $A$ and $B$ in the first term, for example, splits a retarded set. These terms will be considered more closely in the following sections.

\subsection{Nested Retarded Compositions}

After performing the possible separations of the product, we are generally still left with factors that contain retarded compositions of products of several sub-functions. If the factor is simple enough, one may now expand the retarded composition in terms of nested commutators using \Eq{eq:6_retarded_definition}, and solve each of these one by one. Some of the terms may vanish as per the rule \eqref{eq:6_nested_commutator_rule}.
%
This expansion can always be done, but it easily gets rather cumbersome. It is possible to define an alternative way to expand retarded sets, that opens up more options and thus in many cases allows for a cleaner derivation. Nested retarded compositions are defined by
\begin{equation} \begin{split} \label{eq:nested_retarded_definition}
&\bar{O}^{\check{X} R(R(\check{\Hcal}_1),R(\check{\Hcal}_2) \cdots R(\check{\Hcal}_r)) \check{Y}}(t_\mathcal{N}) = \sum_{P \in S_{r-1}} \Theta_{h_1 h_{P(2)} \cdots h_{P(r)}} \bar{O}^{\check{X} [R(\check{\Hcal}_1),R(\check{\Hcal}_{P(2)}), \ldots, R(\check{\Hcal}_{P(r)})] \check{Y}}(t_\Ncal),
\end{split} \end{equation}
where $\Hcal_i = h_i \cup \Ical_i$, in which $h_i$ is the top element of $\Hcal_i$, such that $R(\Hcal_i) = R(h_i, \Ical_i)$. In the nested commutator the retarded sets $R(\Hcal_i)$ are treated as single elements, so that for example $[R(\Hcal_1), R(\Hcal_2)] = R(\Hcal_1) R(\Hcal_2) - R(\Hcal_2) R(\Hcal_1)$, and for the super-indices we again make use of \Eq{eq:linearCombination}. The resulting multi-retarded compositions are defined as in \Eq{eq:4_retarded_definition}. Let us, for example, take a look at the factor that appears in the first term of \Eq{eq:6_example_2_cut}:
\begin{equation} \begin{split} \label{eq:nested_retarded_example_a}
[\con{A}_{ac} \con{B}_{cd}]^{R(\check{a},\check{c}\check{d})} = \Theta_{acd} [\con{A}_{ac} \con{B}_{cd}]^{[\check{a},\check{c},\check{d}]} + \Theta_{adc} [\con{A}_{ac} \con{B}_{cd}]^{[\check{a},\check{d},\check{c}]}.
\end{split} \end{equation}
Because of the nested commutator structure we can make use of the Jacobi identity
\begin{equation} \label{eq:jacobi_identity}
[[A,B],C] + [[C,A],B] + [[B,C],A] = 0,
\end{equation}
to derive
\begin{equation}
[\con{A}_{ac} \con{B}_{cd}]^{[\check{a},\check{c},\check{d}]} = [\con{A}_{ac} \con{B}_{cd}]^{[\check{a},\check{d},\check{c}]} + [\con{A}_{ac} \con{B}_{cd}]^{[\check{d},\check{c},\check{a}]}.
\end{equation}
This allows us to obtain
\begin{equation} \begin{split} \label{eq:6_nested_retarded_example}
[\con{A}_{ac} \con{B}_{cd}]^{R(\check{a},\check{c}\check{d})} &= \Theta_{acd} \left( [\con{A}_{ac} \con{B}_{cd}]^{[\check{a},\check{d},\check{c}]} + [\con{A}_{ac} \con{B}_{cd}]^{[\check{d},\check{c},\check{a}]} \right) + \Theta_{adc} [\con{A}_{ac} \con{B}_{cd}]^{[\check{a},\check{d},\check{c}]} \\
&= \Theta_{ac} \Theta_{cd} [\con{A}_{ac} \con{B}_{cd}]^{[\check{a},[\check{c},\check{d}]]} + \Theta_{ad} \Theta_{ac} [\con{A}_{ac} \con{B}_{cd}]^{[[\check{a},\check{d}],\check{c}]} \\
&= [\con{A}_{ac} \con{B}_{cd}]^{R(\check{a},R(\check{c},\check{d}))} + [\con{A}_{ac} \con{B}_{cd}]^{R(R(\check{a},\check{d}),\check{c})},
\end{split} \end{equation}
where we can express the result cleanly using nested retarded compositions.

Note that when expanding a retarded set containing nested sets, only the top arguments of the nested sets are included in the sum over permutations, so that for example
\begin{equation}
[\con{A}_{ac} \con{B}_{cd}]^{R(\check{a},R(\check{c},\check{d}))} = \Theta_{ac} [\con{A}_{ac} \con{B}_{cd}]^{[\check{a}, R(\check{c},\check{d})]} = \Theta_{ac} \Theta_{cd} [\con{A}_{ac} \con{B}_{cd}]^{[\check{a}, [\check{c},\check{d}]]}.
\end{equation}
The vanishing rule \eqref{eq:6_retarded_set_rule} still holds, and the retarded composition vanishes if any of the nested sets are disconnected.

Diagrammatically we express the expansion in \Eq{eq:6_nested_retarded_example} as
\begin{equation} \label{fig:6_nested_retarded_example}
\bigdiagramcenter{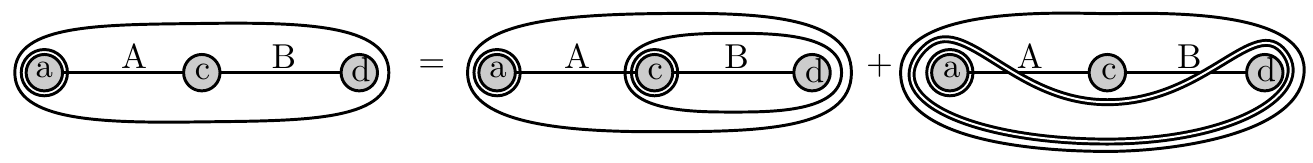},
\end{equation}
Note that a nested retarded set sharing the top argument with the outer retarded set is circled with a double line, to distinguish it from other nested sets on the same level. In figure \ref{fig:6_nested_retarded_example} the second term vanishes, as the retarded set $\{ a,d \}$ is disconnected.

One advantage of the expansion in nested retarded sets, is that since the expansion can be performed with respect to any of the retarded vertices, there is a number of alternative expansions for any retarded set. This allows one to choose the particular expansion that results in the largest number of terms vanishing due to disconnected retarded sets. The various expansions are related by the symmetry of retarded compositions with respect to permutations of the retarded arguments, see \Eq{eq:4_retarded_composition_symmetry}. For example, for the retarded composition in \Eq{eq:nested_retarded_example_a} one has the symmetry $[\cdots]^{R(\check{a},\check{c}\check{d})} = [\cdots]^{R(\check{a},\check{d}\check{c})}$, which allows one to immediately obtain from \Eq{eq:6_nested_retarded_example} the alternative expansion
\begin{equation} \begin{split} \label{eq:6_nested_retarded_alternative}
[\con{A}_{ac} \con{B}_{cd}]^{R(\check{a},\check{c}\check{d})} &= [\con{A}_{ac} \con{B}_{cd}]^{R(\check{a},R(\check{d},\check{c}))} + [\con{A}_{ac} \con{B}_{cd}]^{R(R(\check{a},\check{c}),\check{d})}.
\end{split} \end{equation}
which is expressed diagrammatically as
\begin{equation} \label{fig:6_nested_retarded_alternative}
\bigdiagramcenter{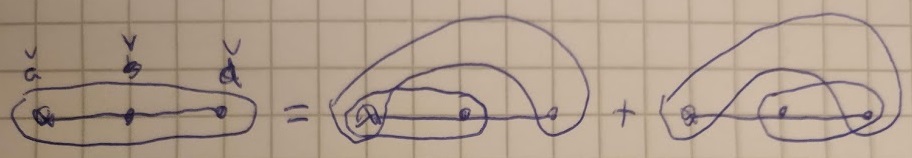}.
\end{equation}
In this case, the expansion in  \Eq{eq:6_nested_retarded_example} and \Eq{fig:6_nested_retarded_example} is the more expedient choice, since in \Eq{fig:6_nested_retarded_alternative} no terms vanish.

The result in \Eq{fig:6_nested_retarded_example} is independent of the sub-functions involved, and can be generalized to (see \Eq{eq:a_nested_retarded_expansion})
\begin{equation} \label{eq:6_nested_retarded_expansion_diagram}
\bigdiagram{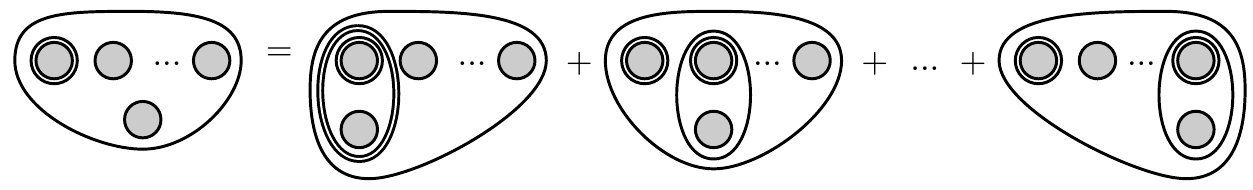}
\end{equation}
Here, the gray dots may represent single vertices or retarded sets. The top element in each retarded set, which may be a retarded set itself, is always designated with a double-lined closed curve. This graphical rule corresponds to \Eq{eq:a_nested_retarded_expansion}, proven in \ref{app:relations_between_retarded_compositions}.
As a further generalization, the outer retarded set in \Eq{eq:6_nested_retarded_expansion_diagram} may itself be inside larger retarded sets (see \Eq{eq:a_nested_retarded_expansionGeneralized2}).

For example, for a four-point function, expanding with respect to argument $4$, we have
\begin{equation}
O^{R(1,234)} = O^{R(R(1,4),23)} + O^{R(1,R(2,4)3)} + O^{R(1,2R(3,4))},
\end{equation}
which we represented diagrammatically as
\begin{equation} \label{eq:6_nested_retarded_expansion_example_a}
\bigdiagramcenter{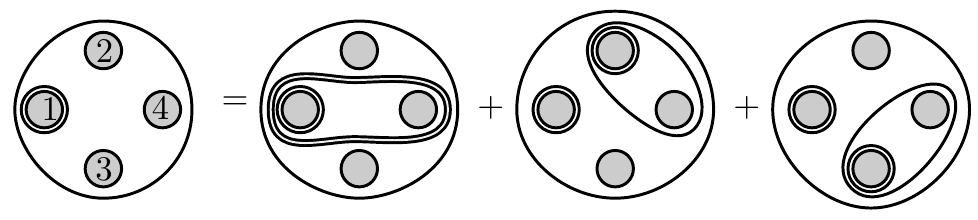}.
\end{equation}

We can expand each term on the right-hand side of \Eq{eq:6_nested_retarded_expansion_example_a} again with respect to vertex $3$ (or the retarded set containing $3$). This leads to
\begin{equation} \begin{split} \label{eq:6_nested_retarded_expansion_example_b}
&\bigdiagramcenter{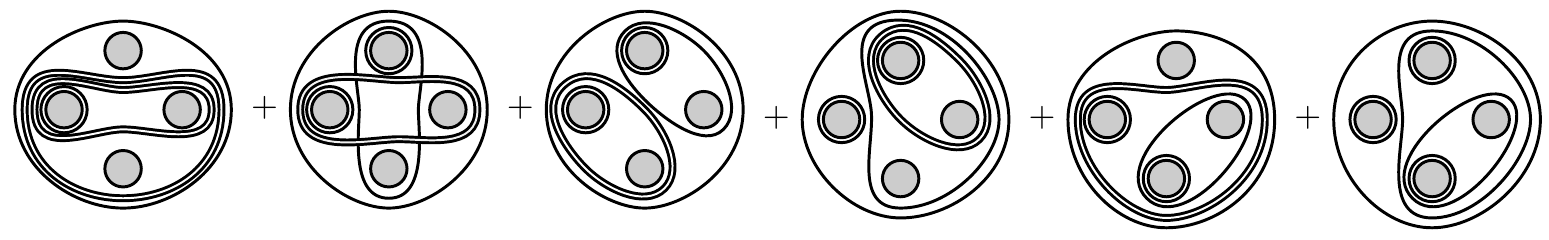}.
\end{split} \end{equation}
We can further apply \Eq{eq:6_nested_retarded_expansion_diagram} in reverse to combine the first and the fifth term, as well as the fourth and the sixth term, in \Eq{eq:6_nested_retarded_expansion_example_b} to obtain
\begin{equation} \begin{split}
\bigdiagramcenter{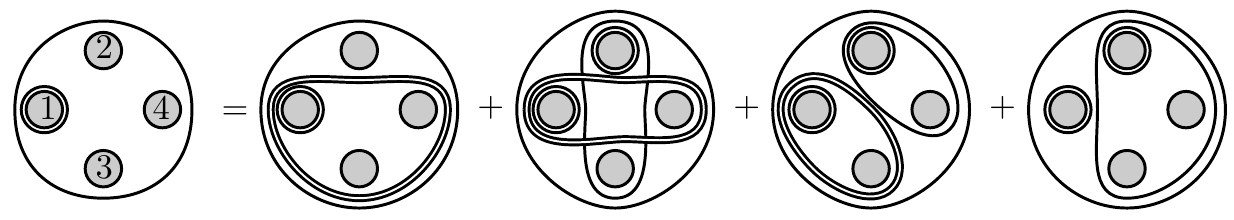}.
\end{split} \end{equation} 
This constitutes a diagrammatic proof of the relation
\begin{equation}
O^{R(1,234)} = O^{R(R(1,34),2)} + O^{R(R(1,4),R(2,3))} + O^{R(R(1,3),R(2,4))} + O^{R(1,R(2,34))}.
\end{equation}

Another useful result relating nested retarded compositions can be derived by considering a retarded composition of the form $[\con{A}_{ab} \con{B}_{b \Ncal}]^{R(\check{a},R(\check{b},L_{\Ncal}))}$, where $L_{\Ncal}$ is an arbitrary string of retarded sets containing the labels $\Ncal = \{ n_1, \ldots, n_N \}$ (such as $L_{\Ncal} = R(n_1,n_2 \cdots n_N)$, to take the simplest option).
Expanding the outermost retarded set in terms of commutators, we obtain
\begin{equation} \begin{split}
[\con{A}_{ab} \con{B}_{b \Ncal}]^{R(\check{a},R(\check{b},L_{\Ncal}))} &= \Theta_{ab} \Big( [\con{A}_{ab} \con{B}_{b \Ncal}]^{\ach R(\check{b},L_{\Ncal})} - [\con{A}_{ab} \con{B}_{b \Ncal}]^{ R(\check{b},L_{\Ncal})\ach} \Big) \\
&= \Theta_{ab} \Big( A^{\ach\bch}_{ab} B_{j \Ncal}^{R(\check{j},L_{\Ncal})} - A^{\bch\ach}_{ab} B_{b \Ncal}^{ R(\check{b},L_{\Ncal})} \Big) \\
&= A^{R(\ach,\bch)}_{ab} B_{b \Ncal}^{R(\check{b},L_{\Ncal})}
\end{split} \end{equation}
Diagrammatically this result can be drawn as
\begin{equation}  \label{eq:6_nested_retarded_result}
\bigdiagramcenter{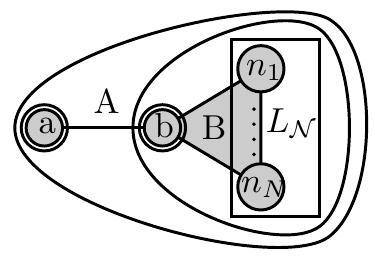} = \bigdiagramcenter{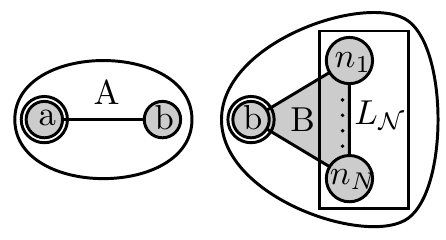}.
\end{equation}

\begin{figure}[ht]
\begin{equation*}
\bigdiagramcenter{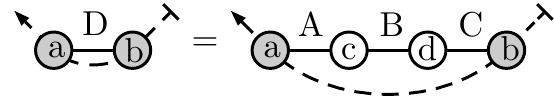}
\end{equation*}
\begin{equation*} \begin{split}
&= \bigdiagramcenter{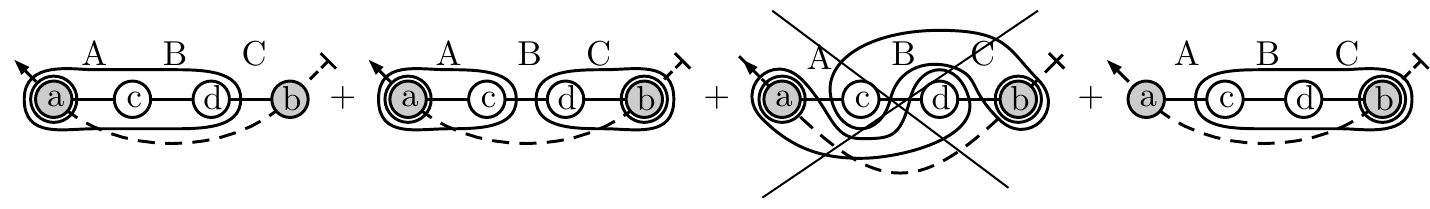} \\
&= \bigdiagramcenter{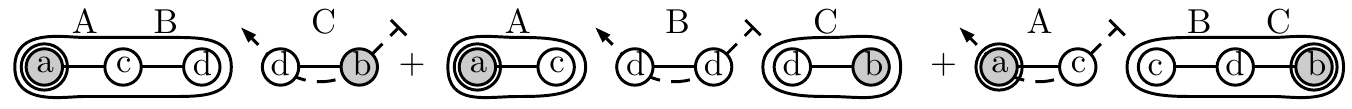} \\
&= \bigdiagramcenter{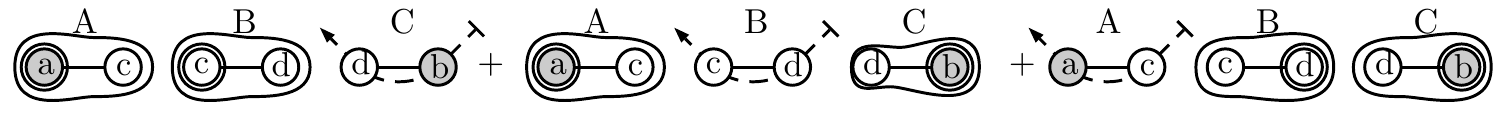}
\end{split} \end{equation*}
\caption{Diagrammatic derivation of the langreth rule \Eq{eq:6_example_2_final} for a chain of two convolutions. We have used the separate results of \Eq{fig:6_nested_retarded_example} and \Eq{6_nested_retarded_split_example} for the first and the last terms in the last step.} \label{fig:chain_convolution_complete_diagrammatic}
\end{figure}

Applying \Eq{eq:6_nested_retarded_result} to the first term in \Eq{fig:6_nested_retarded_example} we have
\begin{equation}\label{6_nested_retarded_split_example}
\bigdiagramcenter{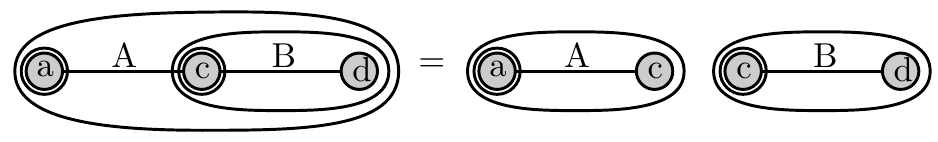},
\end{equation}
which corresponds to
\begin{equation} \label{eq:6_nested_retarded_result_2}
[\con{A}_{ac} \con{B}_{cd}]^{R(\check{a},R(\check{c},\check{d}))} = A^{R(\check{a},\check{c})}_{ac} B^{R(\check{c},\check{d})}_{cd}.
\end{equation}
Using \Eq{eq:6_nested_retarded_result_2}, along with the analogous result for $[ \con{B}_{\bar{c}\bar{d}} \con{C}_{\bar{d}b}]^{R(\check{b},\check{c}\check{d})}$, we can now obtain from \Eq{eq:6_example_2_cut} the Langreth rule
\begin{equation} \begin{split} \label{eq:6_example_2_final}
E^{12}_{ab} &= A^{R(\check{a},\check{c})}_{ac} B^{\check{c}\check{d}}_{cd} C_{db}^{R(\check{b},\check{d})} + A^{R(\check{a},\check{c})}_{ac} B^{R(\check{c},\check{d})}_{cd} C^{\check{d}\check{b}}_{db} + A^{\check{a}\check{c}}_{ac} B^{R(\check{c},\check{d})}_{cd} C^{R(\check{d},\check{b})}_{db}.
\end{split} \end{equation}
The diagrammatic representation of the above equation, including its diagrammatic derivation, is shown in figure \ref{fig:chain_convolution_complete_diagrammatic}.

%
%
%

We now have all the tools to derive the known and extended versions of the Langreth rules. In the following section we will demonstrate the rules laid out above, by going through the process of deriving Langreth rules for two important practical cases, namely the double-triangle graph and the vertex graph.

\section{Langreth Rules for the Double-triangle and Vertex Structures}
\label{sec:7_rules_for_double_triangle}

Let us now apply the extended Langreth rules to two important practical examples introduced in the beginning, that involve two external and two internal arguments:
\begin{equation}
\con{D}_{ab} = \int_{\gamma'} \con{\bar{D}}_{ab\bar{c}\bar{d}} \quad : \quad \bigdiagramcenter{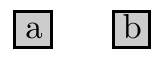} = \bigdiagramcenter{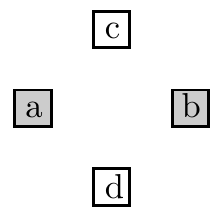}
\end{equation}

We will consider the following two structures:

\begin{enumerate}

\item
The double-triangle structure $\con{\bar{X}}_{abcd} = \con{A}_{ac} \con{B}_{cb} \con{C}_{cd} \con{D}_{ad} \con{E}_{db}$
\begin{equation} \label{eq:7_second_order_exhange_structure}
\con{X}_{ab} = \int_{\gamma'} \con{\bar{X}}_{ab\bar{c}\bar{d}}  \quad : \quad \bigdiagramcenter{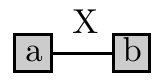} = \bigdiagramcenter{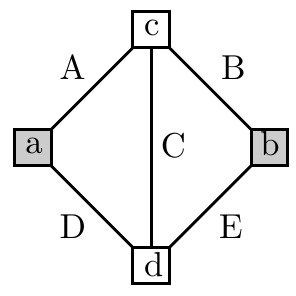}.
\end{equation}

\item
The vertex structure $\con{\bar{H}}_{abcd} = \con{A}_{ac} \con{B}_{ad} \con{C}_{cdb}$
\begin{equation} \label{eq:7_hedin_structure}
\con{H}_{ab} = \int_{\gamma'} \con{\bar{H}}_{ab\bar{c}\bar{d}}  \quad : \quad\bigdiagramcenter{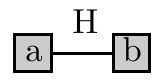} = \bigdiagramcenter{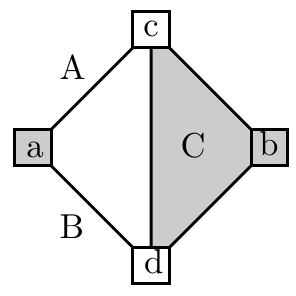}.
\end{equation}

\end{enumerate}

All the Langreth rules for these structures are listed in Table~\ref{tab:double_triangle_rules} and Table~\ref{tab:hedin_rules}. We will provide explicit derivations of a few of the rules.

\subsection{The \texorpdfstring{$D^{M(\check{a}\check{b})}$}{DMab} component.}

For the fully Matsubara restricted component, the retarded set representation is
\begin{equation}
D^{M(\check{a}\check{b})}_{ab} = \bigdiagramcenter{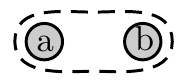} = \bigdiagramcenter{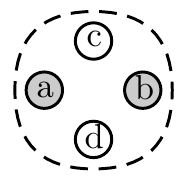}
\end{equation}

\begin{enumerate}

\item
Substituting the double-triangle structure into the above diagram results in:
\begin{equation}
\bigdiagramcenter{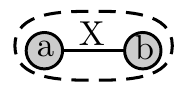} = \bigdiagramcenter{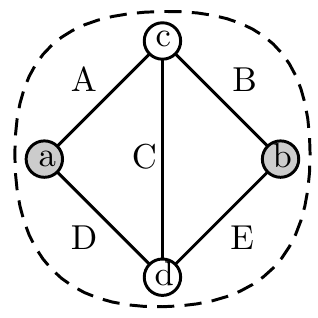} \quad : \quad X^{M(\check{a}\check{b})}_{ab} = \int [ \con{A}_{a\bar{c}} \con{B}_{\bar{c}b} \con{C}_{\bar{c}\bar{d}} \con{D}_{a\bar{d}} \con{E}_{\bar{d}b} ]^{M(\check{a}\check{b}\check{c}\check{d})}
\end{equation}
Since Matsubara sets can be distributed over the sub-functions for the case when no arguments are on the Keldysh branch (see the discussion below \Eq{eq:5_extended_retarded_result}), we find

\begin{equation}
X^{M(\check{a}\check{b})}_{ab} =
\int A^{M(\check{a}\check{c})}_{a\bar{c}} B^{M(\check{c}\check{b})}_{\bar{c}b} C^{M(\check{c}\check{d})}_{\bar{c}\bar{d}} D^{M(\check{a}\check{d})}_{a\bar{d}} E^{M(\check{d}\check{b})}_{\bar{d}b}.
\end{equation}

\item
For the vertex structure we likewise obtain
\begin{equation}
H^{M(\check{a}\check{b})}_{ab} =
\int A^{M(\check{a}\check{c})}_{a\bar{c}} B^{M(\check{a}\check{d})}_{a\bar{d}} C^{M(\check{c}\check{d}\check{b})}_{\bar{c}\bar{d}b}
\end{equation}

\end{enumerate}

\subsection{The \texorpdfstring{$D^{M(\check{a})\check{b}}$}{DMab} component.}

The retarded set representation is
\begin{equation} \label{eq:7_double_triangle_expansion_mixed}
D^{M(\check{a})\check{b}} = \bigdiagramcenter{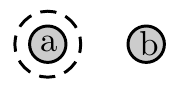} = \bigdiagramcenter{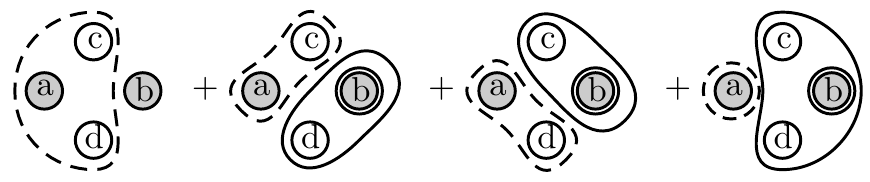}
\end{equation}

\begin{enumerate}

\item
Substituting the double-triangle structure into the above diagram results in:
\begin{equation} \begin{split}
&X^{M(1)2}_{ab} = \bigdiagramcenter{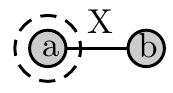} \\
&= \bigdiagramcenter{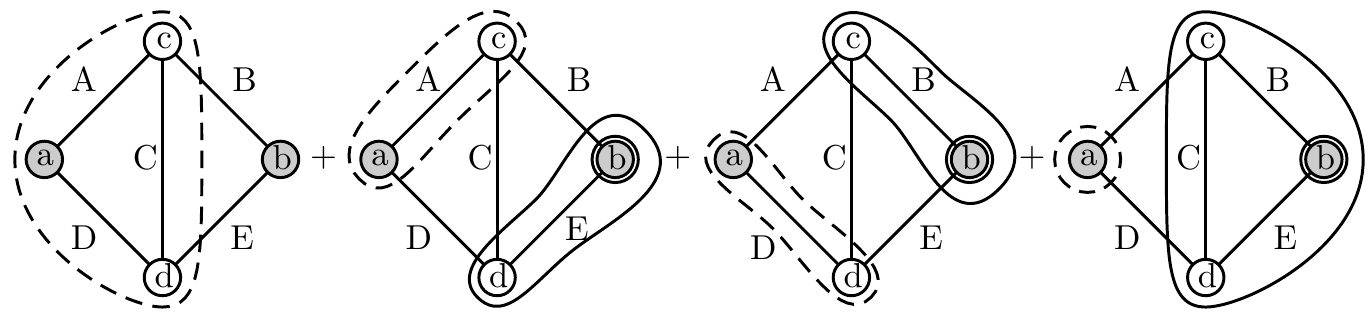} \\
&= \bigdiagramcenter{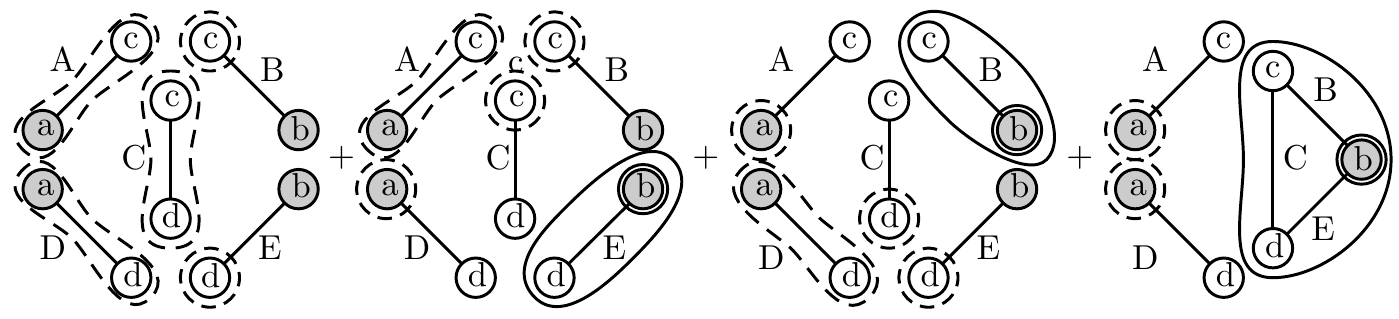} \\
&= \int A^{M(\check{a}\check{c})}_{a\bar{c}} C^{M(\check{c}\check{d})}_{\bar{c}\bar{d}} D^{M(\check{a}\check{d})}_{a\bar{d}} B^{M(\check{c})\check{b}}_{\bar{c}b} E^{M(\check{d})\check{b}}_{\bar{d}b} +
\int A^{M(\check{a}\check{c})}_{a\bar{c}} B^{M(\check{c})\check{b}}_{\bar{c}b} C^{M(\check{c})\check{d}}_{\bar{c}\bar{d}} D^{M(\check{a})\check{d}}_{a\bar{d}} E^{R(\check{b},\check{d})}_{\bar{d}b} \\
&+ \int B^{R(\check{b},\check{c})}_{\bar{c}b} A^{M(\check{a})\check{c}}_{a\bar{c}} C^{M(\check{d})\check{c}}_{\bar{c}\bar{d}} E^{M(\check{d})\check{b}}_{\bar{d}b} D^{M(\check{a}\check{d})}_{a\bar{d}}
+ \int A^{M(\check{a})\check{c}}_{a\bar{c}} D^{M(\check{a})\check{d}}_{a\bar{d}} [ \con{B}_{\bar{c}b} \con{E}_{\bar{d}b} \con{C}_{\bar{c}\bar{d}} ]^{R(\check{b},\check{c}\check{d})}
\end{split} \end{equation}

Here we encounter a retarded composition of a triangle in the term $[ \con{B}_{\bar{c}b} \con{E}_{\bar{d}b} \con{C}_{\bar{c}\bar{d}} ]^{R(\check{b},\check{c}\check{d})}$. Because every vertex connects to every other, there are no symmetries that could be leveraged. Through brute force expansion of the retarded set, either using nested commutators or nested retarded sets, we obtain two forms of a Langreth rule
\begin{equation} \begin{split} \label{eq:6_triangle_langreth_rule}
[ \con{B}_{\bar{c}b} \con{E}_{\bar{d}b} \con{C}_{\bar{c}\bar{d}} ]^{R(\check{b},\check{c}\check{d})} &= B^{R(\check{b},\check{c})}_{\bar{c}b} E^{\check{b}\check{d}}_{\bar{d}b} C^{R(\check{c},\check{d})}_{\bar{c}\bar{d}} + B^{\check{c}\check{b}}_{\bar{c}b} E^{R(\check{b},\check{d})}_{\bar{d}b} C^{R(\check{d},\check{c})}_{\bar{c}\bar{d}} + B^{R(\check{b},\check{c})}_{\bar{c}b} E^{R(\check{b},\check{d})}_{\bar{d}b} C^{\check{d}\check{c}}_{\bar{c}\bar{d}} \\
&= B^{R(\check{b},\check{c})}_{\bar{c}b} E^{\check{d}\check{b}}_{\bar{d}b} C^{R(\check{c},\check{d})}_{\bar{c}\bar{d}} + B^{\check{b}\check{c}}_{\bar{c}b} E^{R(\check{b},\check{d})}_{\bar{d}b} C^{R(\check{d},\check{c})}_{\bar{c}\bar{d}} + B^{R(\check{b},\check{c})}_{\bar{c}b} E^{R(\check{b},\check{d})}_{\bar{d}b} C^{\check{c}\check{d}}_{\bar{c}\bar{d}}.
\end{split} \end{equation}
These two expressions only differ in the change of order of the Keldysh components.

\item
Substituting the vertex structure into \eqref{eq:7_double_triangle_expansion_mixed} leads to
\begin{equation} \begin{split}
&H^{M(1)2}_{ab} = \bigdiagramcenter{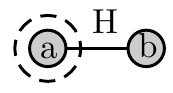} \\
&= \bigdiagramcenter{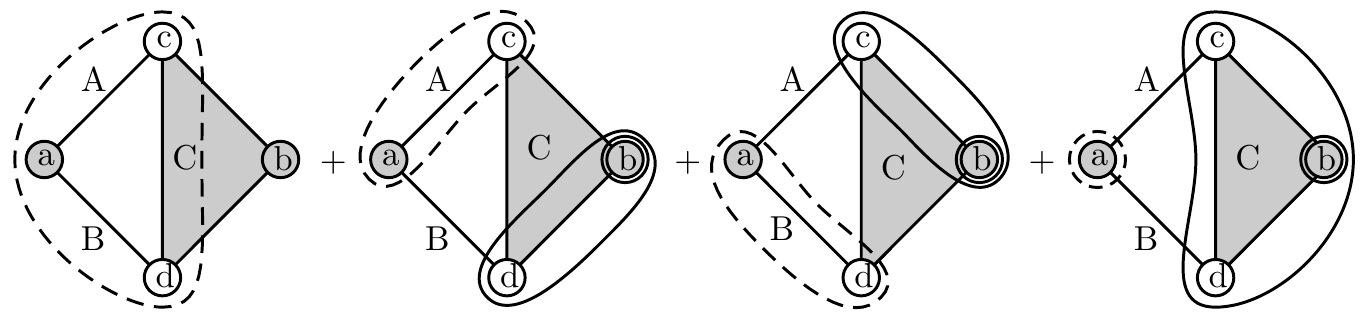} \\
&= \bigdiagramcenter{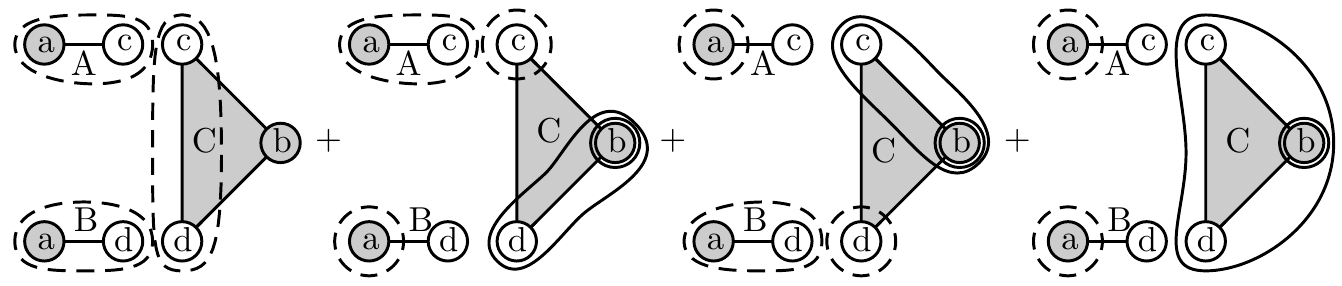} \\
&= \int A^{M(\check{a}\check{c})}_{a\bar{c}} B^{M(\check{a}\check{d})}_{a\bar{d}} C^{M(\check{c}\check{d})\check{b}}_{\bar{c}\bar{d}b} + \int A^{M(\check{a}\check{c})}_{a\bar{c}} B^{M(\check{a})\check{d}}_{a\bar{d}} C^{M(\check{c})R(\check{b},\check{d})}_{\bar{c}\bar{d}b} \\
&+ \int A^{M(\check{a})\check{c}}_{a\bar{c}} B^{M(\check{a}\check{d})}_{a\bar{d}} C^{M(\check{d})R(\check{b},\check{c})}_{\bar{c}\bar{d}b} + \int A^{M(\check{a})\check{c}}_{a\bar{c}} B^{M(\check{a})\check{d}}_{a\bar{d}} C^{R(\check{b},\check{c}\check{d})}_{\bar{c}\bar{d}b}
\end{split} \end{equation}

\end{enumerate}

\subsection{The \texorpdfstring{$D^{\ach\bch}$}{DMab} component.}

The retarded set representation is now
\begin{equation} \begin{split}\label{7_double_triangle_expansion_greater_a}
&D^{\ach\bch}_{ab} = \bigdiagramcenter{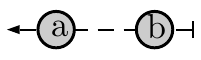} \\
&= \bigdiagramcenter{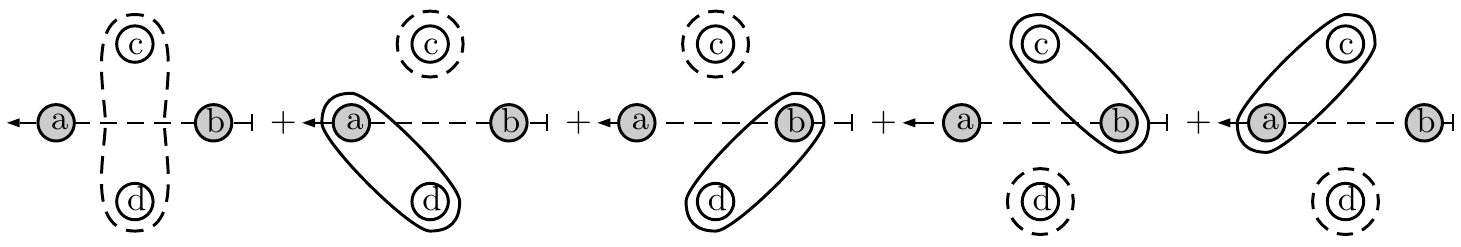} \\
&+ \bigdiagramcenter{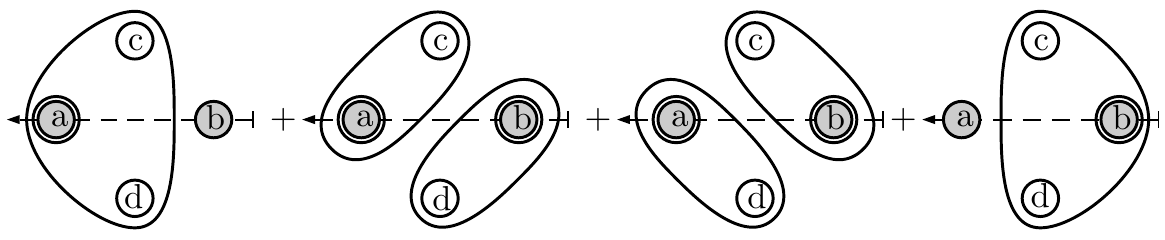}
\end{split} \end{equation}

\begin{enumerate}

\item
After substituting the double-triangle structure we can cut every line in every term, except for the last two terms for which the rule in \Eq{eq:6_triangle_langreth_rule} is needed. The rule is the same for any ordering of the arguments, and therefore by substituting $B_{b\bar{c}} \rightarrow A_{a\bar{c}}$ and $E_{b\bar{d}} \rightarrow D_{a\bar{d}}$ on both sides of \Eq{eq:6_triangle_langreth_rule}, we obtain the rule for $[\con{A}_{a\bar{c}} \con{D}_{a\bar{d}} \con{C}_{\bar{c}\bar{d}}]^{R(\check{a},\check{b}\check{c})}$. The second and third term in \Eq{7_double_triangle_expansion_greater_a} sum up to
\begin{equation}
\bigdiagram{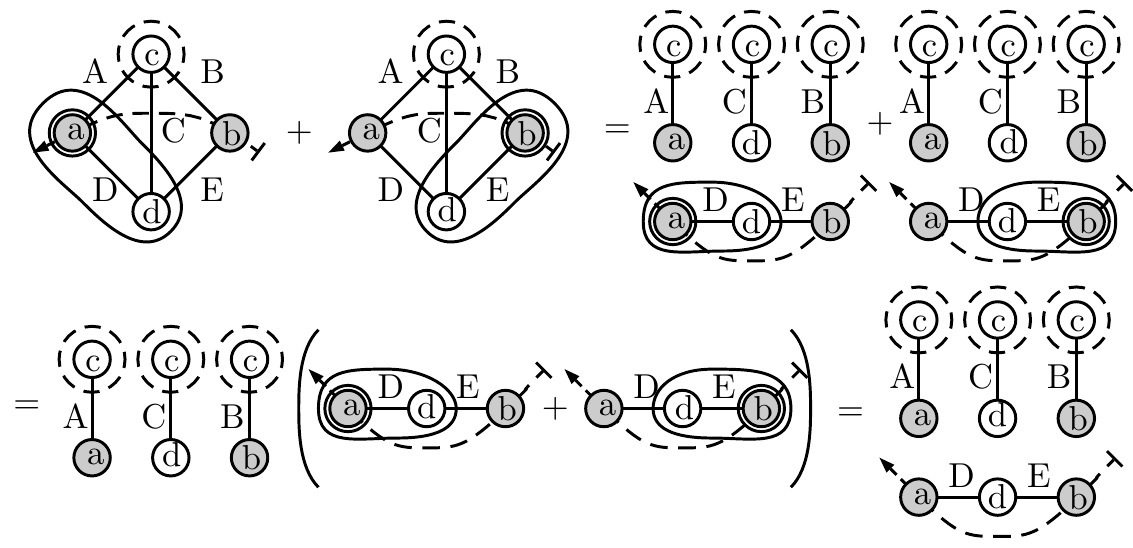}
\end{equation}
and therefore we can express these compactly using a chain-convolution. The same is true for the fourth and fifth terms in \Eq{7_double_triangle_expansion_greater_a}. We obtain
\begin{equation} \begin{split}
X^{\check{a}\check{b}}_{ab} &= \int A^{M(\check{c})\check{a}}_{a\bar{c}} B^{M(\check{c})\check{b}}_{\bar{c}b} C^{M(\check{c}\check{d})}_{\bar{c}\bar{d}} D^{M(\check{d})\check{a}}_{a\bar{d}} E^{M(\check{d})\check{b}}_{\bar{d}b} \\
&+ \int \left[ \con{A}_{a\bar{c}} \con{B}_{\bar{c}b} \right]^{\check{a}\check{b}} C^{M(\check{d})\check{c}}_{\bar{c}\bar{d}} D^{M(\check{d})\check{a}}_{a\bar{d}} E^{M(\check{d})\check{b}}_{\bar{d}b} +
\int A^{M(\check{c})\check{a}}_{a\bar{c}} B^{M(\check{c})\check{b}}_{\bar{c}b} C^{M(\check{c})\check{d}}_{\bar{c}\bar{d}} \left[ \con{D}_{\check{a}\bar{d}} \con{E}_{\bar{d}\check{b}} \right]^{\check{a}\check{b}} \\
&+ \int \left[ \con{A}_{\check{a}\bar{c}} \con{D}_{\check{a}\bar{d}} \con{C}_{\bar{c}\bar{d}} \right]^{R(\check{a},\check{c}\check{d})} B^{\check{c}\check{b}}_{\bar{c}\check{b}} E^{\check{d}\check{b}}_{\bar{d}\check{b}} +
\int A^{R(\check{a},\check{c})}_{\check{a}\bar{c}} B^{\check{c}\check{b}}_{\bar{c}\check{b}} C^{\check{c}\check{d}}_{\bar{c}\bar{d}} D^{\check{a}\check{d}}_{\check{a}\bar{d}} E^{R(\check{b},\check{d})}_{\bar{d}\check{b}} \\
&+ \int A^{\check{a}\check{c}}_{\check{a}\bar{c}} B^{R(\check{b},\check{c})}_{\bar{c}\check{b}} C^{\check{d}\check{c}}_{\bar{c}\bar{d}} D^{R(\check{a},\check{d})}_{\check{a}\bar{d}} E^{\check{d}\check{b}}_{\bar{d}\check{b}} + \int A^{\check{a}\check{c}}_{\check{a}\bar{c}} E^{\check{a}\check{d}}_{\check{a}\bar{d}} \left[ \con{B}_{\bar{c}\check{b}} \con{D}_{\bar{d}\check{b}} \con{C}_{\bar{c}\bar{d}} \right]^{R(\check{b},\check{c}\check{d})}.
\end{split} \end{equation}
After applying the rule in \Eq{eq:6_triangle_langreth_rule}, this can be manipulated into a more compact form, shown in the table \ref{tab:double_triangle_rules}.

\item
After substituting the vertex structure we obtain
\begin{equation} \begin{split}
&H^{12}_{ab} = \bigdiagramcenter{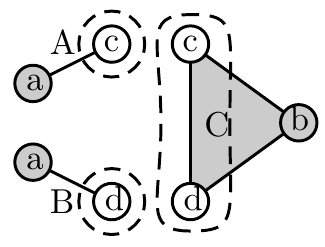} \\
&\bigdiagramcenter{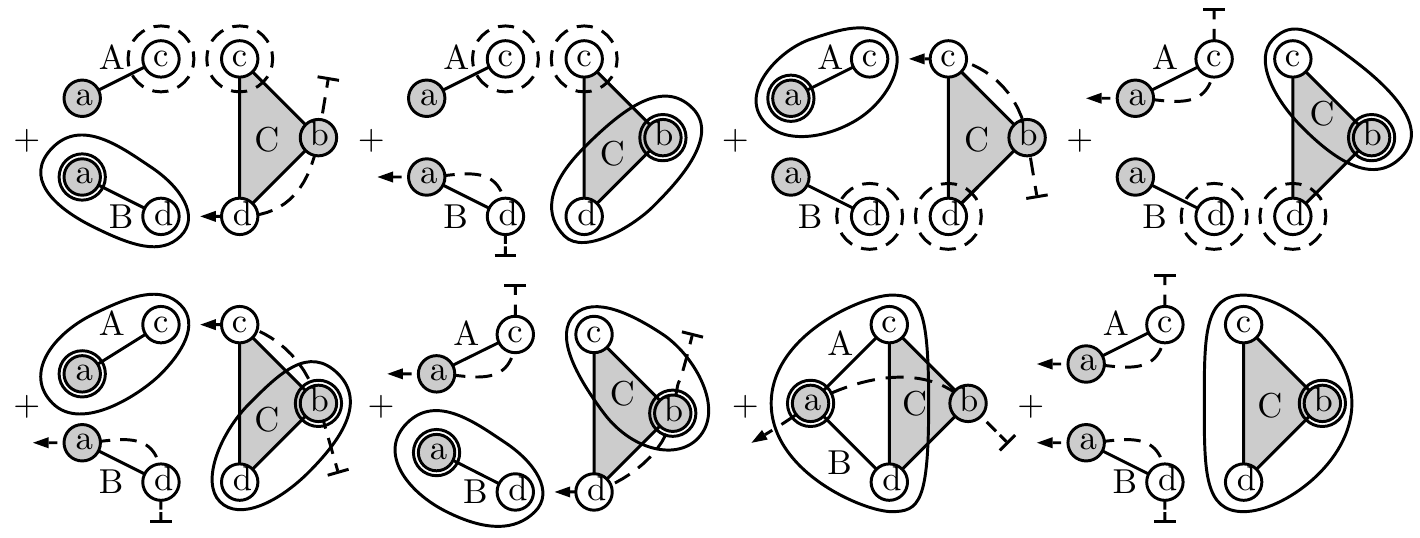}
\end{split} \end{equation}
\begin{equation} \begin{split}
&= \int A^{M(\check{c})\check{a}}_{a\bar{c}} B^{M(\check{d})\check{a}}_{a\bar{d}} C^{M(\check{c}\check{d})\check{b}}_{\bar{c}\bar{d}b} \\
&+ \int A^{M(\check{c})\check{a}}_{a\bar{c}} B^{R(\check{a},\check{d})}_{a\bar{d}} C^{M(\check{c})\check{d}\check{b}}_{\bar{c}\bar{d}b}
+ \int A^{M(\check{c})\check{a}}_{a\bar{c}} B^{\check{a}\check{d}}_{a\bar{d}} C^{M(\check{c})R(\check{b},\check{d})}_{\bar{c}\bar{d}b} \\
&+ \int A^{R(\check{a},\check{c})}_{a\bar{c}} B^{M(\check{d})\check{a}}_{a\bar{d}} C^{M(\check{d})\check{d}\check{b}}_{\bar{c}\bar{d}b}
+ \int A^{\check{a}\check{c}}_{a\bar{c}} B^{M(\check{d})\check{a}}_{a\bar{d}} C^{M(\check{d})R(\check{b},\check{c})}_{\bar{c}\bar{d}b} \\
&+ \int A^{R(\check{a},\check{c})}_{a\bar{c}} B^{\check{a}\check{d}}_{a\bar{d}} C^{\check{c}R(\check{b},\check{d})}_{\bar{c}\bar{d}b}
+ \int A^{\check{a}\check{c}}_{a\bar{c}} B^{R(\check{a},\check{d})}_{a\bar{d}} C^{\check{d}R(\check{b},\check{c})}_{\bar{c}\bar{d}b}
+ \int [ \con{A}_{a\bar{c}} \con{B}_{a\bar{d}} \con{C}_{\bar{c}\bar{d}b} ]^{R(\check{a},\check{c}\check{d})\check{b}}
+ \int A^{\check{a}\check{c}}_{a\bar{c}} B^{\check{a}\check{d}}_{a\bar{d}} C^{R(\check{b},\check{c}\check{d})}_{\bar{c}\bar{d}b}.
\end{split} \end{equation}
Here the second to last term can be handled using the rule \Eq{eq:6_triangle_langreth_rule}, as $b$, being always first in contour order, does not interfere.

\end{enumerate}

\subsection{The \texorpdfstring{$D^{R(\ach,\bch)}$}{DMab} component.}

Performing the retarded set expansion leads to
\begin{equation} \label{eq:7_double_triangle_expansion_retarded}
D^{R(\check{a},\check{b})} = \bigdiagramcenter{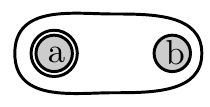} = \bigdiagramcenter{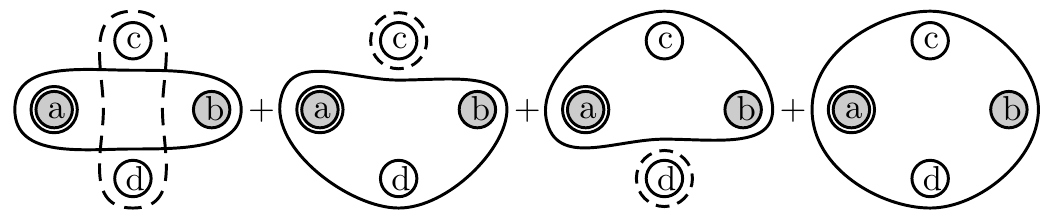}
\end{equation}

\begin{enumerate}

\item
After substituting the double-triangle structure the first term after the last equal sign vanishes, as $a$ and $b$ are not directly connected. In the second and third terms the sub-functions connecting to the Matsubara set can be separated, and the remaining piece is a retarded composition of a convolution (shown in Table \ref{tab:basic_langreth_rules}). The final piece can be worked out by expanding the retarded set, in which the lack of direct connection between $a$ and $b$ leading to cancellations.


\item
The vertex structure can be handled similarly. Starting from \Eq{eq:7_double_triangle_expansion_retarded} the first term after the second equal sign vanishes because $a$ and $b$ are not directly connected. The second and third terms reduce to retarded compositions of convolutions, as the three-point sub-function reduces to an MK function of two contour arguments when one of its vertices is in the Matsubara set.

\end{enumerate}

\subsection{The Tables of Langreth Rules}
\label{sec:tables_of_langreth_rules}

We give in Table \ref{tab:basic_langreth_rules} the known Langreth rules for convolutions and products. The new Langreth rules for the double-triangle structure are shown in Table~\ref{tab:double_triangle_rules}, and the new rules for the vertex structure are shown in Table~\ref{tab:hedin_rules}. For clarity and to conform to common nomenclature, we have used the notation of Langreth for two-point functions:
\begin{equation} \begin{split}
O^{12} = O^>, \quad O^{21} = O^< ,\quad O^{R(1,2)} = O^R, \quad O^{R(2,1)} = O^A \\
O^{M(1)2} = O^\lceil, \quad O^{M(2)1} = O^\rceil, \quad O^{M(12)} = O^M.
\end{split} \end{equation}
We also omit the sub-indices when writing the rules, as these can be read from the contour expression of the structure considered.

Note that in this notation the Langreth rules are specific to a particular ordering of arguments. For example, we have for the chain convolution in the form $\con{C}_{ab} = \int_{\gamma'} \con{A}_{a\bar{c}}\con{B}_{\bar{c}b}$ the rule
\begin{equation}
C^R_{ab} = \int A^R_{a\bar{c}} B^R_{\bar{c}b}.
\end{equation}
If we wish to obtain the rule for a different ordering of arguments, such as $\con{C}_{ab} = \int_{\gamma'} \con{A}_{a\bar{c}}\con{B}_{b\bar{c}}$, we can write the rule above using explicit argument labels as
\begin{equation}
C^{R(\check{a},\check{b})}_{ab} = A^{R(\check{a},\check{c})}_{a\bar{c}} B^{R(\check{c},\check{b})}_{\bar{c}b}.
\end{equation}
In this form the rule remains valid if the argument order is changed, which again indicates the convenience of the háček notation. We can then convert back to the earlier notation to obtain
\begin{equation}
C^{R(\check{a},\check{b})}_{ab} = A^{R(\check{a},\check{c})}_{a\bar{c}} B^{R(\check{c},\check{b})}_{b\bar{c}} \rightarrow C^R_{ab} = A^R_{a\bar{c}} B^A_{\bar{c}b}.
\end{equation}

We have made use of the simpler Langreth rules to simplify the notation for the more complex ones. Thus there appears for example $\left[ A_{a\bar{c}} B_{\bar{c}b} D_{a\bar{d}} E_{\bar{d}b} \right]^R_{ab}$, which is a product of two convolutions and can be worked out using the rules in Table \ref{tab:basic_langreth_rules} to give
\begin{equation} \begin{split}
\left[ A_{a\bar{c}} B_{\bar{c}b} D_{a\bar{d}} E_{\bar{d}b} \right]^R_{ab} &= \left[ A_{a\bar{c}} B_{\bar{c}b} \right]^R_{ab} \left[ D_{a\bar{d}} E_{\bar{d}b} \right]^<_{ab} + \left[ A_{a\bar{c}} B_{\bar{c}b} \right]^<_{ab} \left[ D_{a\bar{d}} E_{\bar{d}b} \right]^A_{ab} \\
&= A^R_{a\bar{c}} B^R_{\bar{c}b} \left( D^R_{a\bar{d}} E^<_{\bar{d}b} + D^<_{a\bar{d}} E^A_{\bar{d}b} \right) + \left( A^R_{a\bar{c}} B^<_{\bar{c}b} + A^<_{a\bar{c}} B^A_{\bar{c}b} \right) D^A_{a\bar{d}} E^A_{\bar{d}b}.
\end{split} \end{equation}
Likewise there are chains of three two-point sub-functions, such as
\begin{equation}
[ D_{a\bar{d}} C_{\bar{c}\bar{d}} B_{\bar{c}b} ]^R_{ab} = [ D_{a\bar{d}} C_{\bar{c}\bar{d}} ]^R_{a\bar{d}} B^R_{\bar{c}b} = D^R_{a\bar{d}} C^A_{\bar{c}\bar{d}} B^R_{\bar{c}b}.
\end{equation}

\begin{table}
\centering
\begin{tabular}{@{}lcl@{}}
\toprule
$\con{D}_{ab} = \int_{\gamma'} \con{A}_{a\bar{c}} \con{B}_{\bar{c}b}$ && $\con{D}_{ab} = \con{A}_{ab} \con{B}_{ba}$ \\
\cmidrule{1-1} \cmidrule{3-3}
$D^> = \int A^R B^> + \int A^> B^A + \int A^\rceil B^\lceil$ && $D^> = A^> B^<$ \\
$D^< = \int A^R B^< + \int A^< B^A + \int A^\rceil B^\lceil$ && $D^< = A^< B^>$ \\
$D^R = \int A^R B^R$ && $D^R = \left\{ \begin{array}{ll} A^R B^< + A^< B^A \\ A^R B^> + A^> B^A \end{array} \right.$ \\
$D^A = \int A^A B^A$ && $D^A = \left\{ \begin{array}{ll} A^A B^< + A^< B^R \\ A^A B^> + A^> B^R \end{array} \right.$ \\
$D^\rceil = \int A^\rceil B^M + \int A^R B^\rceil$ && $D^\rceil = A^\rceil B^\lceil$ \\
$D^\lceil = \int A^\lceil B^A + \int A^M B^\lceil$ && $D^\lceil = A^\lceil B^\rceil$ \\
$D^M = \int A^M B^M$ && $D^M = A^M B^M$ \\
\bottomrule
\end{tabular}
\caption{The Langreth rules for convolutions (left) and products (right). See section \ref{sec:tables_of_langreth_rules} for explanation of the notation. \label{tab:basic_langreth_rules}}
\end{table}

\begin{table}
\centering
\begin{tabular}{@{}l@{}}
\toprule
$\con{G}_{ab} = \int_{\gamma'} \con{\bar{G}}_{ab\bar{c}\bar{d}} = \int_{\gamma'} \con{A}_{a\bar{c}} \con{B}_{\bar{c}b} \con{C}_{\bar{c}\bar{d}} \con{D}_{a\bar{d}} \con{E}_{\bar{d}b}$ \\
\midrule
$G^> = \begin{array}{l}
\int A^\rceil B^\lceil C^M D^\rceil E^\lceil + \int \left[ A B \right]^> C^\rceil D^\rceil E^\lceil + \int A^\rceil B^\lceil C^\lceil \left[ D E \right]^> \\
+ \int \left[ A B \right]^> C^> \left[ D E \right]^> + \int \left[ A B \right]^> C^R ( D^< - D^A) E^> + \int A^> ( B^< + B^R ) C^A \left[ D E \right]^> \end{array}$ \\ \addlinespace
$G^< = \begin{array}{l}
\int A^\rceil B^\lceil C^M D^\rceil E^\lceil + \int \left[ A B \right]^< C^\rceil D^\rceil E^\lceil + \int A^\rceil B^\lceil C^\lceil \left[ D E \right]^< \\
+ \int \left[ A B \right]^< C^> \left[ D E \right]^< + \int \left[ A B \right]^< C^R D^< ( E^> - E^R ) + \int ( A^> + A^A ) B^< C^A \left[ D E \right]^< \end{array}$ \\ \addlinespace
$G^R = \begin{array}{l} \int \left[ A B \right]^R C^\rceil D^\rceil E^\lceil + \int A^\rceil B^\lceil C^\lceil \left[ D E \right]^R \\
+ \int \left[ A B D E \right]^R C^> + \int \left[ A B \right]^R C^R D^< E^> + \int A^> B^< C^A \left[ D E \right]^R \\
+ \int A^> [ D C B ]^R E^> + \int B^< [ A C E ]^R D^< \end{array} $ \\ \addlinespace
$G^A = \begin{array}{l} \int \left[ A B \right]^A C^\rceil D^\rceil E^\lceil + \int A^\rceil B^\lceil C^\lceil \left[ D E \right]^A \\
+ \int \left[ A B D E \right]^A C^> + \int \left[ A B \right]^A C^R D^< E^> + \int A^> B^< C^A \left[ D E \right]^A \\
+ \int A^> [ D C B ]^A E^> + \int B^< [ A C E ]^A D^< \end{array} $ \\ \addlinespace
$G^\rceil = \int A^\rceil B^M C^M D^\rceil E^M + \int A^\rceil B^M C^\lceil D^R E^\rceil + \int A^R B^\rceil C^\rceil D^\rceil E^M + \int [ A D C ]^{1} B^\rceil E^\rceil $ \\ \addlinespace
$G^\lceil = \int A^M B^\lceil C^M D^M E^\lceil + \int A^M B^\lceil C^\lceil D^\lceil E^A + \int A^\lceil B^A C^\rceil D^M E^\lceil + \int A^\lceil D^\lceil [ C  B E ]^{1} $ \\ \addlinespace
$G^M = \int A^M B^M C^M D^M  E^M $ \\
\midrule
$\con{F}_a = \int_{\gamma'} \con{\bar{F}}_{a\bar{b}\bar{c}} = \int_{\gamma'} \con{A}_{a\bar{b}} \con{B}_{a\bar{c}} \con{C}_{\bar{b}\bar{c}}$ \\
\midrule
$F^1 = \int A^\rceil B^\rceil C^M + \int A^\rceil B^R C^\lceil + \int A^R B^\rceil C^\rceil + \int \bar{F}^{R(\check{a},\check{b}\check{c})}_{a\bar{b}\bar{c}} $ \\
$\bar{F}^{R(\check{a},\check{b}\check{c})}_{abc} = \left\{ \begin{array}{ll} A^R B^> C^R + A^< B^R C^A + A^R B^R C^< \\ A^R B^< C^R + A^> B^R C^A + A^R B^R C^>  \end{array} \right.$ \\
\bottomrule
\end{tabular}
\caption{Langreth rules for the double-triangle structure. See section \ref{sec:tables_of_langreth_rules} for explanation of the notation. \label{tab:double_triangle_rules}}
\end{table}

\begin{table}
\centering
\begin{tabular}{@{}l@{}}
\toprule
$\con{H}_{ab} = \int_{\gamma'} \con{\bar{H}}_{ab\bar{c}\bar{d}} = \int_{\gamma'} \con{A}_{a\bar{c}} \con{B}_{a\bar{d}} \con{C}_{\bar{d}b\bar{c}} $ \\
\midrule
$H^> = \begin{array}{l}
\int A^\rceil B^\rceil C^{M(\check{c}\check{d})\check{b}} \\
+ \int A^> B^\rceil C^{M(\check{d})R(\check{b},\check{c})} + \int A^R B^\rceil C^{M(\check{d})\check{c}\check{b}} + \int A^\rceil B^> C^{M(\check{c})R(\check{b},\check{d})} + \int A^\rceil B^R C^{M(\check{c})\check{d}\check{b}} \\
+ \int A^> B^R C^{R(\check{d},\check{c})\check{b}} + \int A^R B^< C^{R(\check{c},\check{d})\check{b}} + \int A^R B^> C^{\check{c}R(\check{b},\check{d})} + \int A^> B^R C^{\check{d}R(\check{b},\check{c})} \\
+ \int A^R B^R C^{\check{c}\check{d}\check{b}} + \int A^> B^> C^{R(\check{b},\check{c}\check{d})} \end{array}$ \\ \addlinespace
$H^< = \begin{array}{l}
\int A^\rceil B^\rceil C^{M(\check{c}\check{d})\check{b}} \\
+ \int A^< B^\rceil C^{M(\check{d})R(\check{b},\check{c})} + \int A^R B^\rceil C^{M(\check{d})\check{b}\check{c}} + \int A^\rceil B^< C^{M(\check{c})R(\check{b},\check{d})} + \int A^\rceil B^R C^{M(\check{c})\check{b}\check{d}} \\
+ \int A^> B^R C^{\check{b}R(\check{d},\check{c})} + \int A^< B^R C^{R(\check{b},\check{c})\check{d}} + \int A^R B^< C^{\check{b}R(\check{c},\check{d})} + \int A^R B^< C^{R(\check{b},\check{d})\check{c}} \\
+ \int A^R B^R C^{\check{b}\check{c}\check{d}} + \int A^< B^< C^{R(\check{b},\check{c}\check{d})} \end{array}$ \\ \addlinespace
$H^R = \begin{array}{l} \int A^R B^\rceil C^{M(\check{d})R(\check{c},\check{b})}
+ \int A^\rceil B^R C^{M(\check{c})R(\check{d},\check{b})} \\
+ \int A^> B^R C^{R(\check{d},\check{b}\check{c})} + \int A^R B^< C^{R(\check{c},\check{b}\check{d})} + \int A^R B^R \big( C^{\check{c}R(\check{d},\check{b})} + \int C^{R(\check{c},\check{b})\check{d}} \big) \end{array} $ \\ \addlinespace
$H^A = \begin{array}{l} \int A^A B^\rceil C^{M(\check{d})R(\check{b},\check{c})}
+ \int A^\rceil B^A C^{M(\check{c})R(\check{b},\check{d})} \\
+ \int \big( A^> B^A + A^A B^< \big) C^{R(\check{b},\check{c}\check{d})} + \int A^R B^A C^{\check{c}R(\check{b}\check{d})} + \int A^A B^R C^{R(\check{b},\check{c})\check{d}} \end{array} $ \\ \addlinespace
$H^\rceil = \begin{array}{l} \int A^\rceil B^\rceil C^{M(\check{b}\check{c}\check{d})} + \int A^\rceil B^R C^{M(\check{b}\check{c})\check{d}} + \int A^R B^\rceil C^{M(\check{b}\check{d})\check{c}} \\
+ \int A^R B^< C^{M(\check{b})R(\check{c},\check{d})} + \int A^> B^R C^{M(\check{b})R(\check{d},\check{c})} + \int A^R B^R C^{M(\check{b})\check{c}\check{d}} \end{array} $ \\ \addlinespace
$H^\lceil = \begin{array}{l} \int A^M B^M C^{M(\check{c}\check{d})\check{b}} + \int A^M B^\lceil C^{M(\check{c})R(\check{b},\check{d})} + \int A^\lceil B^M C^{M(\check{d})R(\check{b},\check{c})}
+ \int A^\lceil B^\lceil C^{R(\check{b},\check{c}\check{d})} \end{array} $ \\ \addlinespace
$H^M = \int A^M B^M C^{M(\check{b}\check{c}\check{d})} $ \\
\bottomrule
\end{tabular}
\caption{Langreth rules for the vertex structure. See section \ref{sec:tables_of_langreth_rules} for explanation of the notation. \label{tab:hedin_rules}}
\end{table}

\section{Conclusions}\label{sec:conclusions}

Non-equilibrium Green's function methods require a translation between contour quantities, well suited for representing the abstract theory, and real-time quantities, well suited for numerical calculations. In this paper we have provided general rules to perform this translation effectively.
%
%

We have constructed a diagrammatic recipe to straightforwardly obtain generalized Langreth rules for the important cases of the double-triangle structure and the vertex diagram.
%
%
Our diagrammatic recipe can be applied to other structures of interest, such as the Hedin equation for the vertex function, and for the Bethe-Salpeter equation.
%
The general rules laid out in this paper will make this possible.

Apart from deriving Langreth rules, the results derived in this paper are of use in other contexts. As an example, we recently showed that an expression for the self-energy, that yields a positive semi-definite spectral function, can be derived in non-equilibrium situations by making use of generalized retarded compositions of half-diagrams \cite{Hyrkas2018}.

\ack
D.K. acknowledges the Academy of Finland for funding under Project
No. 308697. M.H. thanks the Finnish Cultural Foundation for support. R.v.L. acknowledges the Academy of Finland for funding under Project No. 317139.

\appendix

\section{Decomposition of Step Functions}
\label{app:decomposition_of_step_functions}

In this section, we will show how to write products of two step functions as sums in terms of permutations of a single step function. This joining of step functions is useful when discussing the retarded composition. As an example, we consider the multiplication of two step functions containing two times:
\begin{equation} \begin{split} \label{eq:exampleTheta}
 \Theta(t_1,t_2) \Theta(t_3,t_4) &=
  \Theta(t_1,t_2,t_3,t_4) +
  \Theta(t_1,t_3,t_2,t_4) +
  \Theta(t_1,t_3,t_4,t_2) \\ &+
  \Theta(t_3,t_1,t_2,t_4) +
  \Theta(t_3,t_1,t_4,t_2) +
  \Theta(t_3,t_4,t_1,t_2).
\end{split} \end{equation}
The sum contains all six permutations for which $t_1$ is to the left of $t_2$, and $t_3$ to the left of $t_4$.
%
This example can be generalized. Let us consider two sets of time variables, $t_1,t_2,\ldots, t_k$ and $t_{k+1},t_{k+2},\ldots, t_m$, and the product $ \Theta(t_1, t_2, \ldots, t_k) \Theta(t_{k+1}, t_{k+2}, \ldots, t_m)$. The multiplication can be written as a sum over permutations $R \in \Rcal_{m,k}$ of the time arguments $t_1,\ldots,t_m$ in a single step function,
\begin{equation} \label{thetafunctions}
 \Theta(t_1, t_2, \ldots, t_k) \Theta(t_{k+1}, t_{k+2}, \ldots, t_m) = \sum_{R \in \Rcal_{m,k}} \Theta \left(t_{R(1)},t_{R(2)},\ldots,t_{R(m)}\right).
\end{equation}
where we sum over all permutations $R \in \Rcal_{m,k}$ that retain the relative ordering among $t_1,\ldots, t_k$  and $t_{k+1},\ldots,t_m$, separately, as imposed by the original step functions on the left-hand side. That is, the permutation $R$ orders the times such that the position of $t_1$ is to the left of $t_2$, and so on until $t_k$, and the same for the  times $t_{k+1}, \ldots, t_m$. In case a step function contains only one or zero arguments, we define that step function to yield 1.

Equation \eqref{eq:exampleTheta} is a special case of \Eq{thetafunctions} with $m=4$ and $k=2$, for which there are a six permutations that keep the original time ordering. We write these permutations as
\begin{equation}\label{eq:exampleThetaPermutation}
 \sum_{R \in \Rcal_{4,2}} R(1234) = 1234 + 1324 + 1342 + 3124 + 3142 + 3412.
\end{equation}
The set $\Rcal_{m,k}$ is a subset of the symmetric group $\Scal_m$. The set $\Rcal_{m,k}$ is, however, not a group itself, since inverse permutations are not always included. As an example, we can take $\Rcal_{4,2}$, and the permutation $R(1234) = 1342$ from \Eq{eq:exampleThetaPermutation}. The inverse is $R^{-1}(1234) = 1423$, which is not in $\Rcal_{4,2}$.

The total number of permutations $|\Rcal_{m,k}|$, i.e. the size of the set $\Rcal_{m,k}$, is given by the following combinatorical argument. Let us assume that $t_{k+1},\cdots,t_m$ are ordered by the permutation in the correct order. The number of ways we can place the $k$ time arguments $t_1,\cdots,t_k$ among $t_{k+1},\cdots,t_m$ is given by $m(m-1)(m-2)\cdots (m-k) = \frac{m!}{(m-k)!}$. Of all these permutations, one out of every $k!$ permutations has the arguments $t_1,\cdots,t_k$ in the right order. Dividing by $k!$ gives the size of the set $\Rcal_{m,k}$ as
\begin{equation}
 |\Rcal_{m,k}| = \frac{m!}{(m-k)!k!} = \binom{m}{k}.
\end{equation}

The set $\Rcal_{m,k}$ can be defined in a compact manner. We note that the argument $t_{R(i)}$ is at position $i$ in the step function on the right hand side of \Eq{thetafunctions}. Writing $l=R(i)$, the argument $t_l$ is at position $R^{-1}(l)$. Thus, the set
$\Rcal_{m,k}$ can be defined as the set of permutations $R$ that satisfy
\begin{align}
\begin{split}\label{eq:definingR2}
 (i < j) \text{ and } (i,j \leq k)) \Rightarrow R^{-1}(i) < R^{-1}(j) \\
 (i < j) \text{ and } (i,j > k)) \Rightarrow R^{-1}(i) < R^{-1}(j).
 \end{split}
\end{align}
Writing $R^{-1}(i) = l$ in \Eq{eq:definingR2}, we can equivalently define the set $\Rcal_{m,k}$ to contain the permutations that satisfy
\begin{align}
\begin{split}\label{eq:definingR}
 (R(i) < R(j)) \text{ and } (R(i),R(j) \leq k)) \Rightarrow i < j \\
 (R(i) < R(j)) \text{ and } (R(i),R(j) > k)) \Rightarrow i < j.
 \end{split}
\end{align}

It turns out that we will need one more result, which is the multiplication of two step functions containing $t_1,\cdots,t_k$ and $t_{k+1},\cdots,t_m$ separately, when the order of the first set of times is reversed:
\begin{equation}
 \Theta(t_k, t_{k-1}, \cdots, t_1) \Theta(t_{k+1}, t_{k+2}, \cdots, t_m) = \sum_{T \in \Tcal_{m,k}} \Theta \left(t_{T(1)},t_{T(2)},\cdots,t_{T(m)}\right). \label{thetafunctionsReversed}
\end{equation}
The set $\Tcal_{m,k}$ is closely related to $\Rcal_{m,k}$. The size of the set is the same, $|\Tcal_{m,k}| = \binom{m}{k}$, and it can be defined by either of the two relations
\begin{align}
\begin{split}\label{eq:definingTcal1}
(i > j) \text{ and } (i,j \leq k )\Rightarrow T^{-1}(i) < T^{-1}(j) \\
 (i < j) \text{ and } (i,j > k) \Rightarrow T^{-1}(i) < T^{-1}(j).
\end{split}
 \end{align}
or
\begin{align}
\begin{split}\label{eq:definingTcal2}
 (T(i) > T(j)) \text{ and } (T(i),T(j) \leq k )\Rightarrow i < j \\
 (T(i) < T(j)) \text{ and } (T(i),T(j) > k) \Rightarrow i < j.
 \end{split}
\end{align}


The multiplication of step functions are related to the structure of nested commutators, as we show in the next appendix.

\section{The Structure of Nested Commutators}
\label{app:structure_of_nested_commutators}

In this section, we elucidate the structure of the nested commutator and derive some useful results. Our discussion on structure follows a similar one in Ref.~\cite{Zagier1970}.
%

The nested commutator of $m+1$ objects, $[x,1,2,\cdots m]$, where each number $x,1,\cdots,m$ represents a different object, is defined as
\begin{equation}\label{eq:nestedCommutator}
 [x,1,2,\cdots m] = [\cdots[[x,1],2],\cdots,m].
\end{equation}
The nested commutator in \Eq{eq:nestedCommutator} has $2^{m}$ terms, of which one half has a prefactor $+1$ and one half a prefactor $-1$. It is convenient to group the terms according to the number of objects, $k$, that appear to the left of $x$ in each term. The sign of the terms is given by $(-1)^k$. Let us take the example of $m=4$ and $k=2$:
\begin{equation}\label{eq:exampleCommutator}
 [x,1,2,3,4]_2 = 21x34 + 31x24 + 41x23 + 32x14 + 42x13 + 43x12
\end{equation}
where $_2$ denotes that two objects are to the left of $x$, and therefore all terms have a positive sign. In \Eq{eq:exampleCommutator}, the numbers to the left of $x$ are in decreasing order, while the numbers to the right are in increasing order.

In general, we write a nested commutator as
\begin{equation}\label{eq:commutatorWithk}
 [x,1,2,\cdots,m] = \sum_k [x,1,2,\cdots,m]_k,
\end{equation}
where the terms $[x,1,2,\cdots,m]_k$, with $k$ numbers to the left of $x$, can be written as
\begin{equation}\label{eq:defNestedCommutatorQ}
 [x,1,2,\cdots,m]_k = (-)^k \sum_{Q \in \Qcal_{m,k}}Q(1,2,\cdots,k)x Q(k+1,\cdots,m),
\end{equation}
where we sum over all permutations $Q \in \Qcal_{m,k}$ such that the numbers to the left of $x$ are in decreasing order, and the numbers to the right of $x$ in increasing order. In the example above, \Eq{eq:exampleCommutator}, we have
\begin{equation}
 [x,1,2,3,4]_2 = \sum_{Q \in \Qcal_{4,2}}Q(12)x Q(34)
\end{equation}
with
\begin{equation}
 \sum_{Q \in \Qcal_{4,2}} Q(1234) = 2134 + 3124 + 4123 + 3214 + 4213 + 4312.
\end{equation}
We see that $Q(1) > Q(2)$, while $Q(3) < Q(4)$.

The set $\Qcal_{m,k}$ can be compactly defined as containing those permutations $Q$ that fulfill
\begin{align}
\begin{split}\label{eq:definingQcal}
 & (i > j) \text{ and } (i,j \leq k )\Rightarrow  Q(i) < Q(j)\\
 & (i < j) \text{ and } (i,j > k) \Rightarrow Q(i) < Q(j).
 \end{split}
\end{align}
Comparing with \Eq{eq:definingTcal1}, we see that the set $\Qcal_{m,k}$ contain exactly the inverses of the set $\Tcal_{m,k}$.


We will now prove some relations between nested commutators, that are useful when working with retarded compositions. Let us now denote a nested commutator of $n$ objects by
\begin{equation}
A_n = [1,\ldots,n].
\end{equation}
To begin with we observe that
\begin{align} \label{eq:b_nested_commutator_result_0}
[1,\ldots,n] &= [A_{i-1},i,\ldots,n].
\end{align}

We will then prove that
\begin{align}\label{eq:b_nested_commutator_result_2}
\begin{split}
[\ldots,i-1,[i,x],i+1,\ldots] &= [\ldots,i-1,i,x,i+1,\ldots]
\\
&- [\ldots,i-1,x,i,i+1,\ldots] \qquad \text{when } i>1.
\end{split}
\end{align}
For \Eq{eq:b_nested_commutator_result_2}, we obtain, from \Eq{eq:b_nested_commutator_result_0},
\begin{equation} \begin{split} \label{eq:b_nested_commmutator_3}
[\ldots,i-1,[i,x],i+1,\ldots] &= [A_{i-1},[i,x],i+1,\ldots] \\
&= [[[A_{i-1},[i,x]],i+1],\ldots]
\end{split} \end{equation}
Using the Jacobi identity, \Eq{eq:jacobi_identity}, we find
\begin{equation}
[A_{i-1},[i,x]] = [[A_{i-1},i],x] - [[A_{i-1},x],i].
\end{equation}
Inserting the Jacobi identity into \Eq{eq:b_nested_commmutator_3} leads to
\begin{equation} \begin{split}
&[[[[A_{i-1},i],x],i+1],\ldots] - [[[[A_{i-1},x],i],i+1],\ldots] \\
&= [A_{i-1},i,x,i+1,\ldots] - [A_{i-1},x,i,i+1,\ldots] \\
&= [\ldots,i-1,i,x,i+1,\ldots] - [\ldots,i-1,x,i,i+1,\ldots],
\end{split} \end{equation}
which proves \Eq{eq:b_nested_commutator_result_2}.

Finally we will show that
\begin{equation} \begin{split} \label{eq:b_nested_commutator_result_3}
&[\ldots,i,x,i+1,\ldots] \\
&= [[1,x],2,\ldots] + [1,[2,x],3,\ldots] + \ldots + [\ldots,i-1,[i,x],i+1,\ldots]
\end{split} \end{equation}

This result can be proven by induction, with the base step
\begin{equation} \begin{split}
&[[1,x],2,\ldots] + [1,[2,x],3,\ldots] \\
&= [1,x,2,\ldots] + [1,2,x,3,\ldots] - [1,x,2,3,\ldots] \\
&= [1,2,x,3,\ldots],
\end{split} \end{equation}
and the induction step
\begin{equation} \begin{split}
&[\ldots,i,x,i+1,\ldots] + [\ldots,i,[i+1,x],i+2,\ldots] \\
&= [\ldots,i,x,i+1,\ldots] + [\ldots,i,i+1,x,i+2,\ldots] - [\ldots,i,x,i+1,i+2,\ldots] \\
&= [\ldots,i+1,x,i+2,\ldots]
\end{split} \end{equation}

\section{Expansion of Retarded Compositions in Terms of Nested Retarded Compositions}
\label{app:relations_between_retarded_compositions}
%

A general retarded composition can be defined as
\begin{equation} \label{eq:a_retarded_definition}
O^{\check{X} R(Z_1,Z_2 \cdots Z_r) \check{Y}} = \sum_{P \in S_{r-1}} \Theta_{h_1 h_{P(2)} \cdots h_{P(r)}} O^{\check{X} [Z_1,Z_{P(2)}, \ldots, Z_{P(r)}] \check{Y}},
\end{equation}
where $Z_i$ is some object with a defined a top element $h_i$. In the simplest case of $Z_i = \check{h}_i$ for all $i$, \Eq{eq:a_retarded_definition} reduces to the definition of a simple retarded composition in \Eq{eq:4_fully_retarded_definition}. If $Z_i$, for some $i$, is taken to be a retarded set, for example $Z_i = R(\Hcal_i)$ for some $\Hcal_i = h_i \cup I_i$, \Eq{eq:a_retarded_definition} defines a nested retarded composition. Finally $Z_i$ itself can be nested retarded set, for example $Z_i = R(R(\Hcal_i),R(\Hcal_r))$. In this case the top element $h_i$ is the top element of $\Hcal_i$.

Here we will use the relations between nested commutators derived in \ref{app:structure_of_nested_commutators} to derive the result expressed diagrammatically in \Eq{eq:6_nested_retarded_expansion_diagram}.
We will show that a nested retarded composition can be expanded as a sum of nested retarded compositions as
\begin{equation} \label{eq:a_nested_retarded_expansion}
\bar{O}^{\check{X} R(Z_1,Z_2 \cdots Z_r) \check{Y}}(t_\mathcal{N}) = \sum_{i = 1}^{r-1} \bar{O}^{\check{X} R(Z_1,Z_2 \cdots Z_{i-1}  R(Z_i,Z_r)Z_{i-1} \cdots Z_{r-1}) \check{Y}}(t_\mathcal{N}).
\end{equation}
Here we have expanded with respect to $Z_r$. Note that since the left hand side is symmetric with respect to permutations of the indices $2\cdots r$, the expansions with respect to any of the objects $Z_2,\ldots,Z_r$ give the same result.

To keep the presentation cleaner we suppress $X$ and $Y$ as well as the time-arguments. For the right hand side of \Eq{eq:a_nested_retarded_expansion} we define a new set of $Z'$ objects as
\begin{equation} \begin{split}
Z'_i &= R(Z_i, Z_r) \\
Z'_k &= Z_k \quad k \neq i.
\end{split} \end{equation}
Note that the top element of $Z'_n$ is the same as for $Z_n$, i.e. $h_n$, for all $n$.
Using this definition the right hand side of \Eq{eq:a_nested_retarded_expansion} can be expanded using \Eq{eq:a_retarded_definition}. For each $i$ we can restrict our attention to the term that contain the unit permutation, as the other terms can be obtained by subsequent permutations at the end of the derivation. This gives
\begin{equation} \begin{split} \label{eq:a_nested_retarded_example_3}
\sum_{i = 1}^{r-1} \Theta_{h_1 \cdots h_{r-1}} \bar{O}^{[Z'_1,Z'_2  \cdots Z'_{i-1} Z'_i Z_{i+1} \cdots Z'_{r-1}]} &= \sum_{i = 1}^{r-1} \Theta_{h_1 \cdots h_{r-1}} \bar{O}^{[Z_1,Z_2 \cdots Z_{i-1} R(Z_i,Z_r)Z_{i+1} \cdots Z_{r-1}]} \\
&= \sum_{i = 1}^{r-1} \Theta_{h_1 \cdots h_{r-1}} \Theta_{h_i h_r} \bar{O}^{[Z_1,Z_2  \cdots Z_{i-1} [Z_i,Z_r] Z_{i+1}\cdots Z_{r-1}]}.
\end{split} \end{equation}

The product of step functions can be written as a sum of step functions using
\begin{equation} \label{eq:a_step_function_combination}
\Theta_{h_1 \cdots h_{r-1}} \Theta_{h_i h_r} = \sum_{j = i}^{r-1} \Theta_{h_1 \cdots h_j h_r  h_{j+1} \cdots h_{r-1}},
\end{equation}
We then reorder the sums
\begin{equation}
\sum_{i = 1}^{r-1} \sum_{j = i}^{r-1} \rightarrow \sum_{j = 1}^{r-1} \sum_{i = 1}^j,
\end{equation}
which leads to
\begin{equation} \begin{split} \label{eq:a_nested_retarded_example_4}
\sum_{i = 1}^{r-1} \Theta_{h_1 \cdots h_{r-1}} \bar{O}^{[Z'_1,Z'_2 \cdots Z'_i \cdots Z'_{r-1}]} &= \sum_{j = 1}^{r-1} \Theta_{h_1 \cdots h_j h_r  h_{j+1} \cdots h_{r-1}} \sum_{i = 1}^j \bar{O}^{[Z_1,Z_2 \cdots Z_{i-1} [Z_i,Z_r] Z_{i+1} \cdots Z_{r-1}]} \\
&= \sum_{j = 1}^{r-1} \Theta_{h_1 \cdots h_j h_r  h_{j+1} \cdots h_{r-1}} \bar{O}^{[Z_1,Z_2, \ldots, Z_{j},Z_r, Z_{j+1}, \ldots, Z_{r-1}]},
\end{split} \end{equation}
where on the last line we have used the relation between nested commutators shown in \Eq{eq:b_nested_commutator_result_3}. Summing over all the permutations of $2 \cdots r-1$ in \Eq{eq:a_nested_retarded_example_4} yields the right-hand side of \Eq{eq:a_nested_retarded_expansion} as a consequence of the definition of a retarded composition, \Eq{eq:a_retarded_definition}. The sum over permutations yields
\begin{equation} \label{eq:a_nested_retarded_example_5}
\sum_{P \in S_{r-2}} \sum_{j = 1}^{r-1} \Theta_{h_1 \cdots h_{P(j)} h_r  h_{P(j+1)} \cdots h_{P(r-1)}} \bar{O}^{[Z_1,Z_{P(2)}, \ldots, Z_{P(j)},Z_r, Z_{P(j+1)}, \ldots, Z_{P(r-1)}]}
\end{equation}
Here for each permutation of $2,\ldots,r-1$, one sums over all positions of $r$ in the sequence. The effect is the same as summing over all permutations of $2,\ldots,r$. \Eq{eq:a_nested_retarded_example_5} then becomes equal to the definition of the retarded composition $\bar{O}^{R(Z_1,Z_2 \cdots Z_r)}$, as given by \Eq{eq:a_retarded_definition}.
This completes the proof of \Eq{eq:a_nested_retarded_expansion}.

As a useful example of the expansion in \Eq{eq:a_nested_retarded_expansion}, we consider a case in which all objects $Z_i$ only contain single elements, that is $Z_i = \check{h}_i$. The expansion formula in this case is
\begin{equation} \label{eq:a_nested_retarded_expansionExampleSimple}
\bar{O}^{\check{X} R(\check{h}_1,\check{h}_2 \cdots \check{h}_r) \check{Y}}(t_\mathcal{N}) = \sum_{i = 1}^{r-1} \bar{O}^{\check{X} R(\check{h}_1,\check{h}_2 \cdots \check{h}_{i-1} R(\check{h}_i,\check{h}_r) \check{h}_{i+1} \cdots \check{h}_{r-1}) \check{Y}}(t_\mathcal{N}).
\end{equation}

Finally we will consider the case of expanding one of the nested sets. For example, suppose that we have $O^{\check{X} R(Z_1,Z_2 \cdots Z_r) \check{Y}}(t_\mathcal{N})$ where $Z_k = R(Z'_1, Z'_2 \cdots Z'_s)$. In this case it is possible to apply \Eq{eq:a_nested_retarded_expansion} on $Z_k$ to obtain
\begin{equation} \label{eq:a_nested_retarded_expansionGeneralized2}
\begin{split}
\bar{O}^{\check{X} R(Z_1,Z_2 \cdots Z_{k-1} R(Z'_1, Z'_2 \cdots Z'_s) Z_{k+1} \cdots Z_r) \check{Y}}(t_\mathcal{N})
\\
=\sum_{i = 1}^{s-1} \bar{O}^{\check{X} R(Z_1,Z_2 \cdots Z_{k-1} R(Z'_1, Z'_2 \cdots  R(Z'_i, Z'_s)  \cdots Z'_{s-1})Z_{k+1} \cdots Z_r) \check{Y}}(t_\mathcal{N}).
\end{split}
\end{equation}
To prove \Eq{eq:a_nested_retarded_expansionGeneralized2} we expand the outermost retarded set in nested commutators. After expanding also the nested commutators thus generated, one obtains a sum in which terms of the form considered above in \Eq{eq:a_nested_retarded_expansion} appear. \Eq{eq:a_nested_retarded_expansion} can then be applied to each of these terms to expand $Z_k$. Because in the sum of terms generated by \Eq{eq:a_nested_retarded_expansion} the top element of the expanded retarded set remains the same in each term, we can reconstruct the outermost retarded set, which leads to \Eq{eq:a_nested_retarded_expansionGeneralized2}. Note that expanding a set deeper in the nested structure raises no further issues. In each such case expanding the outer retarded sets allows one to eventually reach a situation in which \Eq{eq:a_nested_retarded_expansion} can be applied.

\section{Retarded-set Representation for Retarded Compositions}
\label{app:expansion_for_retarded_compositions}

Here we will outline the proof of \Eq{eq:4_retarded_result}, which states that for two Keldysh-functions $\con{O}$ and $\con{\bar{O}}$ related by
\begin{equation} \label{eq:a_integral_equation}
\con{O}(z_\mathcal{E}) = \int_\gamma \dif z_\mathcal{I} \, \con{\bar{O}}(z_\mathcal{N}),
\end{equation}
the retarded compositions are related by
\begin{equation} \label{eq:a_retarded_composition_result}
O^{R(\check{h}_1,\check{\mathcal{H}}_1)\cdots R(\check{h}_H,\check{\mathcal{H}}_H)}(t_\mathcal{E}) = \int_{t_0}^\infty \dif t_\mathcal{I} \sum_\mathcal{I} \bar{O}^{R(\check{h}_1,\check{\mathcal{H}}_1 \cup \check{\mathcal{I}}_1) \cdots R(\check{h}_H,\check{\mathcal{H}}_H \cup \check{\mathcal{I}}_H)}(t_\mathcal{N}).
\end{equation}

We first prove the case for a single internal argument $z_i$, for which
\begin{equation}
\con{O}(z_\mathcal{E}) = \int_\gamma \dif z_i \, \con{\bar{O}}(z_\Ncal),
\end{equation}
and
\begin{equation} \label{eq:a_retarded_preliminary_result}
O^{R(\check{e}_1,\check{e}_2 \cdots \check{e}_E)}(t_\mathcal{E}) = \int_{t_0}^\infty \dif t_i \, \bar{O}^{R(\check{e}_1,\check{e}_2 \cdots \check{e}_E \check{i})}(t_\Ncal).
\end{equation}
%
Suppressing the time-arguments, the proof of \Eq{eq:a_retarded_preliminary_result} proceeds as follows:

\begin{enumerate}

\item
We expand the left-hand side of \Eq{eq:a_retarded_preliminary_result} in terms of nested commutators using \Eq{eq:4_retarded_definition}. This gives
\begin{equation} \label{eq:a_retarded_example_2}
O^{R(\check{e}_1,\check{e}_2 \cdots \check{e}_E)} = \sum_{P \in S_{E-1}}\Theta_{e_1 e_{P(2)} \cdots e_{P(E)}} O^{[\check{e}_1,\check{e}_{P(2)},\ldots,\check{e}_{P(E)}]}.
\end{equation}

\item
Each Keldysh component of $\con{O}$ can be written in terms of retarded compositions of $\con{\bar{O}}$ using \Eq{eq:4_1_multi_parameter_result}. This gives for example for the Keldysh component  $O^{\check{e}_1 \cdots \check{e}_E}$
\begin{equation} \label{eq:a_danielewicz_rule}
O^{\check{e}_1 \cdots \check{e}_E} = \int_{t_0}^\infty \dif t_i \sum_{j = 1}^E \bar{O}^{\check{e}_1 \cdots R(\check{e}_j, \check{i}) \cdots \check{e}_E},
\end{equation}
and similarly for each permutation of $e_1,\ldots,e_E$. Thus if in \Eq{eq:a_retarded_example_2} we expand the nested commutators, apply \Eq{eq:a_danielewicz_rule} to each Keldysh component, and then reconstruct the nested commutator, we obtain
\begin{equation} \begin{split} \label{eq:a_retarded_example_3}
O^{R(\check{e}_1,\check{e}_2 \cdots \check{e}_E)} &= \int_{t_0}^\infty \dif t_i \sum_{j = 1}^E \sum_{P \in S_{E-1}}\Theta_{e_1 e_{P(2)} \cdots e_{P(E)}} O^{[\check{e}_1,\check{e}_{P(2)},\ldots,R(\check{e}_j, \check{i}),\ldots,\check{e}_{P(E)}]} \\
&= \int_{t_0}^\infty \dif t_i \sum_{j = 1}^E O^{R(\check{e}_1,\check{e}_2 \cdots R(\check{e}_j, \check{i}) \cdots \check{e}_E)},
\end{split} \end{equation}
where on the second line we have used the definition of a nested retarded composition, \Eq{eq:a_retarded_definition}.

\item
We can now apply \Eq{eq:a_nested_retarded_expansionExampleSimple} derived in \ref{app:relations_between_retarded_compositions} to the right hand side of \Eq{eq:a_retarded_example_3} to obtain \Eq{eq:a_retarded_preliminary_result}, which was to be proven.

\end{enumerate}

Multi-retarded compositions can be obtained by going through the same steps as above. To demonstrate this we will consider an example with four external arguments and take the component
\begin{equation} \label{eq:a_retarded_ab_cd}
O^{R(\check{a},\check{b})R(\check{c},\check{d})} = \Theta_{ab} \Theta_{cd} O^{[\check{a},\check{b}][\check{c},\check{d}]}.
\end{equation}

In the second step we again write the Keldysh components of $\con{O}$ in terms of retarded compositions of $\con{\bar{O}}$ using \Eq{eq:4_1_multi_parameter_result}. For four arguments we have
\begin{equation}
O^{abcd} = \int_{t_0}^\infty \dif t_i \, \Big[ \bar{O}^{R(\check{a},\check{i})\check{b}\check{c}\check{d}} + \bar{O}^{\check{a}R(\check{b},\check{i})\check{c}\check{d}} + \bar{O}^{\check{a}\check{b}R(\check{c},\check{i})\check{d}} + \bar{O}^{\check{a}\check{b}\check{c}R(\check{d},\check{i})} \Big],
\end{equation}
etc. Substituting these into \Eq{eq:a_retarded_ab_cd} leads to
\begin{equation} \begin{split} \label{eq:d_result_1}
&\Theta_{ab} \Theta_{cd} O^{[\check{a},\check{b}][\check{c},\check{d}]} \\
&= \int_{t_0}^\infty \dif t_i \, \Theta_{ab} \Theta_{cd} \Big( \bar{O}^{[R(\check{a},\check{i}),\check{b}][\check{c},\check{d}]} + \bar{O}^{[\check{a},R(\check{b},\check{i})][\check{c},\check{d}]} + \bar{O}^{[\check{a},\check{b}][R(\check{c},\check{i}),\check{d}]} + \bar{O}^{[\check{a},\check{b}][\check{c},R(\check{d},\check{i})]} \Big) \\
&= \int_{t_0}^\infty \dif t_i \, \Big( \bar{O}^{R(R(\check{a},\check{i}),\check{b})R(\check{c},\check{d})} + \bar{O}^{R(\check{a},R(\check{b},\check{i}))R(\check{c},\check{d})} + \bar{O}^{R(\check{a},\check{b})R(R(\check{c},\check{i}),\check{d})} + \bar{O}^{R(\check{a},\check{b})R(\check{c},R(\check{d},\check{i}))} \Big).
\end{split} \end{equation}
Applying \Eq{eq:a_nested_retarded_expansion} in step iii) now leads to
\begin{equation}
\Theta_{ab} \Theta_{cd} O^{[\check{a},\check{b}][\check{c},\check{d}]} = \int_{t_0}^\infty \dif t_i \, \Big[ \bar{O}^{R(\check{a},\check{b}\check{i})R(\check{c},\check{d})} + \bar{O}^{R(\check{a},\check{b})R(\check{c},\check{d}\check{i})} \Big].
\end{equation}
For an arbitrary retarded composition $O^{R(\check{h}_1,\check{\mathcal{H}}_1)\cdots R(\check{h}_H,\check{\mathcal{H}}_H)}$, performing these steps leads to
\begin{equation} \label{eq:a_retarded_preliminary_result_multi}
O^{R(\check{h}_1,\check{\mathcal{H}}_1)\cdots R(\check{h}_H,\check{\mathcal{H}}_H)}(t_\mathcal{E}) = \int_{t_0}^\infty \dif t_i \, \sum_{j=1}^H \bar{O}^{R(\check{h}_1,\check{\mathcal{H}}_1)\cdots R(\check{h}_j,\check{\mathcal{H}}_j \cup \check{i}) \cdots R(\check{h}_H,\check{\mathcal{H}}_H)}(t_\mathcal{N}),
\end{equation}
in which the internal argument is added to each retarded set in turn.

Having considered the case of a single internal argument, the more general situation can be handled simply by applying \Eq{eq:a_retarded_preliminary_result_multi} repeatedly for each integral in \Eq{eq:a_integral_equation}. In this way every internal argument gets added to each retarded set, and the end result is a sum in which the internal arguments are distributed in every possible way among the retarded sets. In other words we obtain \Eq{eq:a_retarded_composition_result} which was to be proven.

\section*{References}
\bibliographystyle{iopart-num}
\bibliography{aForPapers-aPaperMarkku}


\end{document}

%% file: newcommands.tex
\newcommand{\Sigmacal}{\mathit{\Sigma}}

\renewcommand{\S}{\Sigmacal}

\newcommand{\Ical}{\mathcal{I}}

\newcommand{\Kcal}{\mathcal{K}}

\newcommand{\Uh}{\hat{U}}

\newcommand{\Hh}{\hat{H}}

\newcommand{\Oh}{\hat{O}}

\newcommand{\rhoh}{\hat{\rho}}

\newcommand{\Nh}{\hat{N}}

\newcommand{\Vh}{\hat{V}}

\newcommand{\dif}{\text{d}}

\newcommand{\trace}[1]{\Tr\left [ #1 \right ]}

\newcommand{\Eq}[1]{Eq.~\eqref{#1}}

\newcommand{\zb}{\bar{z}}

\newcommand{\tb}{\bar{t}}

\newcommand{\Scal}{{\mathcal{S}}}

\newcommand{\ext}[1]{#1'}
\newcommand{\con}[1]{\mathscr{#1}}

\newcommand{\diagram}[1]{\includegraphics[scale=0.5]{figures/#1.pdf}}
\newcommand{\bigdiagram}[1]{\includegraphics[scale=1.0]{figures/#1.pdf}}
\newcommand{\diagramcenter}[1]{\vcenter{\hbox{\diagram{#1}}}}
\newcommand{\bigdiagramcenter}[1]{\vcenter{\hbox{\bigdiagram{#1}}}}

\newcommand{\Ob}{\bar{O}}

\newcommand{\Ncal}{\mathcal{N}}
\newcommand{\Tcal}{\mathcal{T}}
\newcommand{\Rcal}{\mathcal{R}}
\newcommand{\Qcal}{\mathcal{Q}}
\newcommand{\Mcal}{\mathcal{M}}
\newcommand{\Acal}{\mathcal{A}}
\newcommand{\Ecal}{\mathcal{E}}
\newcommand{\Hcal}{\mathcal{H}}

\newcommand{\ach}{\check{a}}
\newcommand{\bch}{\check{b}}
\newcommand{\cch}{\check{c}}
\newcommand{\dch}{\check{d}}
\newcommand{\ech}{\check{e}}
\newcommand{\fch}{\check{f}}
\newcommand{\gch}{\check{g}}
\newcommand{\hch}{\check{h}}
\newcommand{\ich}{\check{i}}